\pdfoutput=1
\documentclass[acmsmall,nonacm]{acmart}
\AtBeginDocument{%
  \providecommand\BibTeX{{%
    \normalfont B\kern-0.5em{\scshape i\kern-0.25em b}\kern-0.8em\TeX}}}

\setcopyright{acmcopyright}
\copyrightyear{2024}
\acmYear{2024}
\acmDOI{XXXXXXX.XXXXXXX}

\acmConference[PODC 2024]{}{}{}
\acmBooktitle{PODC 2024} 

\usepackage[ruled]{algorithm}
\usepackage{algpseudocode, amsfonts}
\usepackage{xcolor}
\usepackage{tcolorbox}

\usepackage{ioa_code}
\usepackage{scalerel,stackengine}
\newtheorem{theorem}{Theorem}
\newcommand{\prf}[1]{{}}

\newtheorem{property}{Property}

\newtheorem{Def}{Definition}[section]

\newcommand{\sacode}[5]
{ \vspace{.06in} \hrule \vspace{.06in} 
 {\setlength{\parindent}{0cm}\bf #1}: \\
 \footnotesize {\setlength{\parindent}{0cm}\bf Signature:}\B \nobreak
 \normalsize \begin{quote} \nobreak #2 \end{quote}
 \footnotesize {\setlength{\parindent}{0cm}\bf States:}\B \nobreak
 \begin{quote} \nobreak #3 \end{quote}
 {\setlength{\parindent}{0cm}\bf Transitions:} \nobreak
 \vspace{-.2in} \nobreak
 \normalsize #4
 \vspace{-.06in} \hrule \vspace{.06in} 
}

\newcommand{\act}[1]{%
    \relax\ifmmode
        \mathord{\mathcode`\-="702D\sf #1\mathcode`\-="2200}%
    \else
        $\mathord{\mathcode`\-="702D\sf #1\mathcode`\-="2200}$%
    \fi
}

\newcommand{\tup}[1]{%
    \relax\ifmmode
      \langle #1 \rangle%
    \else
        $\langle$#1$\rangle$%
    \fi
}

\newcommand{\seq}[1]{%
    \relax\ifmmode
      \langle \! \langle #1 \rangle \! \rangle%
    \else
        $\langle \! \langle$ #1 $\rangle \! \rangle$%
    \fi
}

\newcommand{\B}{\vspace*{-\smallskipamount}}

\newcommand{\FF}{\vspace*{\medskipamount}}

\newcommand{\ms}[1]{%
    \relax\ifmmode
        \mathord{\mathcode`\-="702D\it #1\mathcode`\-="2200}%
    \else
        {\it #1}%
    \fi
}

\newcommand{\lit}[1]{%
    \relax\ifmmode
        \mathord{\mathcode`\-="702D\sf #1\mathcode`\-="2200}%
    \else
        {\it #1}%
    \fi
}

\newcommand{\XDK}[1]{}
\newcommand{\remove}[1]{} 
\newcommand{\proofremove}[1]{} 
\newcommand{\uselater}[1]{} 

\newcommand{\case}[1]{
        {\vspace{1em}{\setlength{\parindent}{0cm}\bf Case #1:}}}

\makeatletter
\def\mainlistofsymbols{
  \normalsize
  \vspace*{1.5 em}
  \@starttoc{los}
}

\def\partonelistofsymbols{
  \normalsize
  \vspace*{1.5 em}
  \@starttoc{p1los}
}

\def\parttwolistofsymbols{
  \normalsize
  \vspace*{1.5 em}
  \@starttoc{p2los}
}

\def\l@symbol#1#2{\addpenalty{-\@highpenalty} \vskip 4pt plus 2pt
{\@dottedtocline{0}{0em}{8em}{#1}{#2}}}
\makeatother

\newcommand{\newhiddensym}[2]{%
}

\newcommand{\algIOA}[2]{\ifmmode{\text{#1}_{#2}}\else{$\text{#1}_{#2}$}\fi}

\newcommand{\EX}{\ifmmode{\xi}\else{$\xi$}\fi}
\newcommand{\EXF}{\ifmmode{\phi}\else{$\phi$}\fi}

\renewcommand{\state}{\sigma}
\newcommand{\st}{\sigma}

\newcommand{\quo}[1]{Q_{#1}}
\newcommand{\inter}[1]{
	\ifmmode{\left(\bigcap_{\mathcal{Q}\in#1}\mathcal{Q}\right)}
	\else{$\left(\bigcap_{\mathcal{Q}\in#1}\mathcal{Q}\right)$}
	\fi
}

\newcommand{\idSet}{\mathcal{I}}
\newcommand{\wSet}{\mathcal{W}}
\newcommand{\rdSet}{\mathcal{R}}
\newcommand{\recSet}{\mathcal{G}}
\newcommand{\srvSet}{\mathcal{S}}

\newcommand{\vSet}{\mathcal{V}}

\newcommand{\cSet}{\mathcal{I}}

\newcommand{\confSet}{\mathcal{C}}
\newcommand{\servers}[1]{ #1.Servers}
\newcommand{\quorums}[1]{ #1.Quorums}
\newcommand{\consensus}[1]{ #1.Con}

\newcommand{\op}{\pi}

\newcommand{\rd}{\rho}
\newcommand{\wrt}{\omega}

\mathchardef\mhyphen="2D

\newcommand{\pr}{p}
\newcommand{\rdr}{r}

\newcommand{\bef}{\rightarrow}

\newcommand{\vid}[1]{\ifmmode{\nu_{#1}}\else{$\nu_{#1}$}\fi}

\newcommand{\seen}{\ifmmode{seen}\else{$seen$}\fi}

\newcommand{\ares}{{\sc Ares}}

\newcommand{\valSet}{{\mathcal V}}

\newcommand{\tsSet}{{\mathcal T}}

\newcommand{\tg}[1]{\tau_{#1}}
\newcommand{\tgc}[1]{\tau_{\text{c}}}
\newcommand{\vc}[1]{v_{\text{c}}}
\newcommand{\tgs}[1]{\tg{}s}
\newcommand{\tgvs}[1]{\tg{}$-$vs}
\newcommand{\maxts}[1]{\ifmmode{maxTS_{#1}}\else{$maxTS_{#1}$}\fi}
\newcommand{\maxtag}[1]{\ifmmode{maxTag_{#1}}\else{$maxTag_{#1}$}\fi}
\newcommand{\maxpair}[1]{\ifmmode{maxMPair_{#1}}\else{$maxMPair_{#1}$}\fi}
\newcommand{\mintag}[1]{\ifmmode{minTag_{#1}}\else{$minTag_{#1}$}\fi}
\newcommand{\maxps}{\ifmmode{maxPS}\else{$maxPS$}\fi}
\newcommand{\conftg}[1]{\ifmmode{confirmed_{#1}}\else{$confirmed_{#1}$}\fi}
\newcommand{\maxconftag}{\ifmmode{\ms{maxCT}}\else{$maxCT$}\fi}

\usepackage[binary-units=true]{siunitx}
\usepackage[normalem]{ulem}

\usepackage{arydshln}

\newcommand{\myemph}[1]{{\it #1}}

\newcommand{\myparagraph}[1]{\setlength{\parindent}{0cm}\textbf{#1}}
\newcommand{\WRP}{\par\qquad\(\hookrightarrow\)\enspace}

\newcommand{\cvec}[2]{\mathbf{c}^{#1}_{#2}}
\newcommand{\atT}[2]{#1|_{#2}}
\newcommand{\status}[1]{#1.status}
\newcommand{\config}[1]{#1.cfg}
\newcommand{\configPB}[1]{\PBcolor{#1}.cfg}

\newcommand{\Coded}{code\act{-}elems}

\newcommand{\dap}[1]{{DAP(#1)}}

\algblockdefx[Operation]{Operation}{EndOperation}%
[2]{{\bf operation} $\act{#1}$(#2)}%
{{\bf end operation}}
\algblockdefx[Procedure]{Procedure}{EndProcedure}%
[2]{{\bf procedure} $\act{#1}$(#2)}%
{{\bf end procedure}}
\algblockdefx[Receive]{Receive}{EndReceive}%
[2]{{\bf Upon receive} (#1)$_{\text{ #2 }}${\bf from} $q$}%
{{\bf end receive}}

\newcommand\wwidehat[1]{%
\savestack{\tmpbox}{\stretchto{%
  \scaleto{%
    \scalerel*[\widthof{\ensuremath{#1}}]{\kern-.6pt\bigwedge\kern-.6pt}%
    {\rule[-\textheight/2]{1ex}{\textheight}}
  }{\textheight}%
}{0.5ex}}%
\stackon[1pt]{#1}{\tmpbox}%
}

\usepackage{url}

\usepackage{graphicx}
\usepackage{multicol,amsthm}
\usepackage{subcaption}
\usepackage{adjustbox}

\newcommand{\coARES}{{\sc Co\ares{}}}
\newcommand{\fcoARES}{{\sc \coARES F}}
\newcommand{\frfs}{{\sc CoBFS}}

\newcommand{\daputdata}[2]{ {#1}.{\act{put-data}(#2)}}
\newcommand{\dagetdata}[1]{ {#1}.{\act{get-data}()}}
\newcommand{\dagettag}[1]{ {#1}.{\act{get-tag}()}}

\newcommand{\rambo}{{\sc RAMBO}}
\newcommand{\smStore}{{\sc SM-Store}}
\newcommand{\dynaStore}{{\sc DynaStore}}
\newcommand{\SpSnStore}{{\sc SpSnStore}}

\newcommand{\abd}{{\sc ABD}}
\newcommand{\ec}{{\sc EC}}

\newcommand{\abdbased}{{\sc ABD-based}}
\newcommand{\ecbased}{{\sc EC-based}}

\newcommand{\ARESabd}{{\sc \ares \abd{}}}
\newcommand{\ARESec}{{\sc \ares \ec{}}}

\newcommand{\coARESabd}{{\sc \coARES \abd{}}}
\newcommand{\coARESec}{{\sc \coARES \ec{}}}

\newcommand{\fARESabd}{{\sc \coARESabd F}}
\newcommand{\fARESec}{{\sc \coARESec F}}

\newcommand{\ARESopt}{{\sc \ares{} II}}
\newcommand{\ecdapII}{\ecdap{} II}
\newcommand{\sequencetraversalII}{{\sc sequence traversal II}}

\usepackage{changepage}

\newcommand{\abddap}{\abd-DAP}
\newcommand{\abddapII}{\abddap{} II}
\newcommand{\ecdap}{\ec-DAP}
\newcommand{\ecdapopt}{\ecdap{}opt}

\newcommand{\myparagraphitalic}[1]{\setlength{\parindent}{0cm}\textbf{\textit{#1}}}

\newcommand{\PBcolor}[1]{\textcolor{red}{#1}}
\newcommand{\GCcolor}[1]{\textcolor{blue}{#1}}
\newcommand{\RECONcolor}[1]{\textcolor{mydarkgreen}{#1}}

\scriptsize
\definecolor{mydarkgreen}{RGB}{40, 100, 0}
\definecolor{mydarkorange}{RGB}{204, 85, 0}

\newcommand{\PBbox}[1]{
    \fcolorbox{red}{white}{
        \begin{varwidth}{\linewidth}
         #1 
        \end{varwidth}
    }%
}

\newcommand{\PBboxParam}[2]{%
    \begin{varwidth}{\linewidth}
    \begin{tcolorbox}[colback=white,colframe=red, sharp corners, left=0pt, boxsep=0pt, boxrule=0.5pt, width=#1\linewidth-2\fboxsep-2\fboxrule]
         #2  
    \end{tcolorbox}
    \end{varwidth}
}

\newcommand{\GCbox}[1]{%
    \fcolorbox{blue}{white}{%
        \begin{varwidth}{\linewidth}
            #1
        \end{varwidth}%
    }%
}

\newcommand{\GCboxParam}[2]{%
    \begin{varwidth}{\linewidth}
    \begin{tcolorbox}[colback=white,colframe=blue, sharp corners, left=0pt, boxsep=0pt, boxrule=0.5pt, width=#1\linewidth-2\fboxsep-2\fboxrule]
         #2  
    \end{tcolorbox}
    \end{varwidth}
}

\newcommand{\RECONbox}[1]{%
    \fcolorbox{mydarkgreen}{white}{%
        \begin{varwidth}{\linewidth}
            #1
        \end{varwidth}%
    }%
}

\usepackage{todonotes}

\begin{document}

\title{\ARESopt{}: Tracing the Flaws of a (Storage) God}


\author{Chryssis Georgiou}
\affiliation{%
  \institution{University of Cyprus}
  \city{Nicosia}
  \country{Cyprus}}
\email{chryssis@ucy.ac.cy}

\author{Nicolas Nicolaou}
\affiliation{%
  \institution{Algolysis Ltd}
  \city{Limassol}
  \country{Cyprus}}
\email{nicolas@algolysis.com}

\author{Andria Trigeorgi}
\affiliation{%
  \institution{University of Cyprus}
  \city{Nicosia}
  \country{Cyprus}}
\affiliation{%
  \institution{Algolysis Ltd}
  \city{Limassol}
  \country{Cyprus}}
\email{atrige01@ucy.ac.cy}

\renewcommand{\shortauthors}{}


\begin{abstract}
\ares{} is a modular framework, designed to implement
dynamic, reconfigurable, fault-tolerant, read/write and strongly consistent distributed shared memory objects.
Recent enhancements of the framework have realized the efficient implementation of large objects, by introducing versioning and data striping techniques. In this work, we identify performance bottlenecks of the \ares's variants by utilizing distributed tracing, a popular technique for monitoring and profiling distributed systems. We then propose optimizations across all versions of \ares, aiming in overcoming the identified flaws, while preserving correctness. 
We refer to the optimized version of \ares{} as \ARESopt, which now features a piggyback mechanism, a garbage collection mechanism, and a batching reconfiguration technique for improving the performance and storage efficiency of the original \ares{}. We rigorously prove the correctness of \ARESopt, and we demonstrate the performance improvements by an experimental comparison (via distributed tracing) of the \ARESopt{} variants with their original counterparts.\smallskip

\vspace{-.5em}
\end{abstract}

\remove{
\begin{abstract}
The \ares{} algorithm is a dynamic (reconfigurable) storage, allowing the set of servers to be modified while keeping the service available.
Numerous variations of the \ares{} algorithm have introduced novel approaches for effectively managing dynamic reconfigurations and large objects. 
The latest version, \fARESec{}, incorporates reconfigurations and improves performance by combining a fragmentation approach and a second level of striping with erasure coding, 
making storage efficient on servers.  
The fragmentation approach is achieved through \frfs{}, a framework of a DSS (Distributed Storage System) designed to boost concurrent access to large shared data objects (such as files).
Despite these advancements, optimizing system effectiveless and performance is still crucial. Adoption of distributing tracing techniques such as OpenTelemetry is necessary for accurate bottleneck detection, as traditional performance analysis approaches could miss underlying bottlenecks. Once bottlenecks are detected in any \ares{} version, we propose solutions to address them, applicable across all versions.
Our optimized algorithm, \ARESopt{}, features a piggyback mechanism, garbage collection, and batching reconfiguration to improve performance and storage efficiency of \ares{}.
We demonstrate that these optimizations maintain the correctness conditions of the original \ares{}.
Finally, we again use distributed tracing to compare the performance of all \ARESopt{} variations with their original counterparts.
\end{abstract}
}

\remove{

}
\maketitle

\section{Introduction}
\myparagraph{Distributed Shared Memory Emulation.}
In an era where data is being generated at an unprecedented rate, effectively dealing with Big Data challenges has become a critical endeavor. To manage this large amount of data, organizations are increasingly adopting large-scale systems known as Distributed Storage Systems (DSS) which can divide the data across multiple servers for high availability, data redundancy, and recovery purposes. One of the fundamental
structures to implement a DSS is a Distributed Shared Memory (DSM) emulation.

For three decades, a series of works 
(e.g., \cite{ref_article_ABD, ref_article_MWMRABD, ref_article_ERATO, ref_article_fastRead, ref_article_semifast}) proposed
solutions for building DSM emulations,
allowing data to be shared concurrently offering
basic memory elements, i.e. registers, with 
strong consistency guarantees. Linerazibility (atomicity) \cite{HW90} is the 
most challenging, yet intuitive consistency guarantee that such solutions provide.
Attiya, Bar-Noy and Dolev~\cite{ref_article_ABD} present the first fault-tolerant emulation of atomic shared memory in an asynchronous message passing system, also known as \abd{}. Subsequent algorithms (eg.,~\cite{ref_article_ERATO,ref_article_semifast,ref_article_fastRead}) have built upon \abd{}, aiming to reduce communication overhead while imposing minimal computation overhead.
The problem of keeping copies consistent becomes even more challenging when the set of servers need to be modified, leading to the development of  dynamic solutions and reconfiguration services.
Examples of reconfigurable storage algorithms
are \rambo{}~\cite{rambo}, \dynaStore{}~\cite{dynastore}, \smStore{}~\cite{smartmerge}, \SpSnStore{}~\cite{SpSnStore}, and \ares~\cite{ARES}. \smallskip

\myparagraph{ARES and its Extensions.}
\ares{} is a modular framework, designed to implement
reconfigurable, fault-tolerant, read/write distributed linearizable (atomic) shared memory objects.
Unlike other reconfigurable algorithms, \ares{} does not
define the exact methodology to access the object replicas. Rather, it relies on \emph{data access primitives} (DAPs), which are used for expressing  
the data access strategy (i.e., how they retrieve and update the object
data) of different shared memory algorithms (e.g., ABD). One DAP implementation supports Erasure Coding, making \ares{} the first reconfigurable DSM providing a level of data striping. 
%
Recently, in~\cite{FragARES}, two extensions of \ares{} have been proposed. The first, called \coARES{}, extends \ares{} by providing {\em coverability}~\cite{NFG16}. Coverability 
extends linearizability by ensuring that a write operation is performed on the latest ``version'' of the object (see Section~\ref{sec:model}). \fcoARES{} extends \coARES{} by utilizing the fragmentation (striping) strategy of \frfs~\cite{SIROCCO_2021}, making \ares{} suitable for handling large objects (such as files). Section~\ref{sec:ARESnCo} overviews these three variants of \ares. Many versions of \ares{} are obtained, when these variants are used with different DAP implementations. Experimental evaluations~\cite{FragARES,ref_article_ARESvsCommercial} of the various versions of \ares, while revealing interesting trade-offs, they also suggested that there is still room for improvement, especially when compared to commercial DSSs. An initial speculation was hindering that the reconfiguration mechanism of \ares, which is common to all versions, causes redundant communication. Thus, there was a need for a more thorough investigation of the communication deficiency and overall performance of \ares. \smallskip

\remove{
A significant drawback of \fcoARES{} is its sequential handling of current configurations (proposed or installed through reconfiguration operations) in each read/write/reconfig operation. This involves reading the configuration prior to each invocation and transferring messages, either for retrieving/putting information for an object from/to replicas, from one configuration to the next.}

\myparagraph{Distributed Tracing.}
Traditionally, in distributed shared memory emulations (e.g.~\cite{do-rambo,ref_article_blobseer,spiegelman2016space,jehl_et_al:LIPIcs:2017:7100,berger2018integrated,spiegelman2017dynamic,HNS17,ref_article_ERATO,Self-stabilizationOverhead9,ref_article_ARESvsCommercial}), performance analysis and bottleneck detection have often focused on measuring the overall latency of operational or computational time, rather than examine the latency of individual components in detail. This approach provides a high-level view of the system's performance but may lack the level of detail required to precisely identify the underlying causes of the bottlenecks.  

In recent years, the realm of distributed systems has witnessed the emergence of innovative software monitoring tools. These tools, often referred to as tracing tools, are designed to trace and visualize the interactions within a distributed system. The technique behind the tools, known as {\em Distributed Tracing}, has gained significant traction and has been implemented by most of the major actors of Cloud-Computing for their own monitoring needs, e.g., Google~\cite{sigelman2010dapper}, Twitter~\cite{aniszczyk2012zipkin}, Uber~\cite{uberjaeger}, Sigelman~\cite{opentracing,ref_url_opentelemetry}.
Unlike logging, which merely records events and data, distributed tracing offers a comprehensive perspective on how requests flow through interconnected components, enabling a better understanding of performance and dependencies in distributed systems.\smallskip

\myparagraph{Contributions.}
In this work we bring distributed tracing into the realm of DSM and demonstrate its usefulness by turning the identified flaws of \ares{} into optimizations, yielding \ARESopt.

\myparagraphitalic{Distributed tracing.} We identify performance bottlenecks in different versions of \ares{} using OpenTelemetry~\cite{opentelemetry-python}, an open-source tool for distributed tracing. We seamlessly integrate this tool with other observability tools, such as Jaeger and Grafana Jaeger, enabling us to combine its tracing data with metrics, logs, and visualizations. This integration allows us to trace requests as they traverse various components, providing a holistic view of the algorithms' behavior. 

\myparagraphitalic{ARES II.} Once bottlenecks are detected, we develop optimizations that ensure a performance boost while preserving the correctness conditions of the original algorithms, leading to \ARESopt{}. The optimizations mainly concern the reconfiguration mechanism, and are as follows: $(i)$ to expedite configuration discovery we introduce piggy-back data on read/write messages; $(ii)$ for service
longevity and to expedite configuration discovery, we introduce a garbage collection mechanism
that removes obsolete configurations and updates older configurations with newly established ones; and $(iii)$ to expedite reconfiguration we introduce a batching mechanism where a single configuration is applied not on a single but multiple objects concurrently; this is particularly useful for the fragmented version of \ARESopt. We rigorously prove the correctness of \ARESopt, and we demonstrate the performance improvements by
an experimental comparisons  of the \ARESopt{} versions with their original counterparts.

\remove{
In contrast to its predecessor, this algorithm employs a different approach to handling current configurations. Instead of processing them in a separate communication round, the new algorithm utilizes a piggy-back technique. Additionally, the new algorithm initiates a reconfiguration operation on the fragmented object (\ares{}) as opposed to issuing a series of reconfiguration operations on the blocks of the fragmented object. Up to this point, these improvements have been focused on reducing communication latency. For storage efficiency, we introduce a garbage collection (GC) mechanism that operates in the background to remove older configurations.
}

\section{System Settings and Definitions}
\label{sec:model}

We explore an asynchronous message-passing system with processes communicating through reliable point-to-point channels, allowing potential message reordering. 
\smallskip
 
\myparagraph{Clients and servers.}
The system is a collection of crash-prone, asynchronous processors with unique identifiers (ids) from a totally-ordered set, composed of two main disjoint sets of processes: 
(a) a set $\cSet{}$ of client processes ids that may perform operations on a replicated object,
and 
(b) a set $\srvSet$ of server processes ids;  
servers host and maintain replicas of shared data. 
A \textit{quorum} is defined as a subset of $\srvSet$. A \textit{quorum system}~\cite{QuorumSystems} is a collection of pair-wise intersecting quorums.

In this work, we deal with dynamic environments, where the {\em configuration} of the system may dynamically change over time due to servers removal or addition. A configuration is a data type that describes the service setup, i.e., the finite set of servers, their grouping into intersecting sets (i.e., quorum system), and various parameters needed for the implementation of the atomic storage service (see Section~\ref{sec:ARESnCo} for the precise contains of \ares's configurations).  Dynamic addition and removal of servers may
lead the system from one configuration to another, with different set of servers and parameters. {\em Reconfiguration} is the process responsible to migrate the system from one configuration to the next.
%
There are three distinct sets of client processes: a set $\wSet$ of writers, a set $\rdSet$ of readers, and a set $\recSet$ of reconfiguration clients. 
Each writer is allowed to modify the value of a shared object, and each reader is allowed to obtain 
the value of that object. Reconfiguration clients attempt to introduce new configurations 
to the system in order to mask transient {server errors or include new servers} and to ensure the longevity of the service.\smallskip

\remove{
\myparagraph{Configurations.} 
A \textit{configuration} $c \in \confSet$,~$\confSet$ being a set of unique identifiers, comprises:
$(i)$ $\servers{c}\subseteq\srvSet$: a set of server identifiers; 
$(ii)$ $\quorums{c}$: the set of quorums on $\servers{c}$, s.t. $\forall Q_1,Q_2\in\quorums{c}, Q_1,Q_2\subseteq\servers{c}$ and $Q_1\cap Q_2\neq \emptyset$; 
$(iii)$ $\dap{c}$: the set of primitives that clients in $\idSet$ may invoke on $\servers{c}$; 
and $(iv)$ $\consensus{c}$: a consensus instance with the values from $\confSet$, 
implemented on servers in $\servers{c}$.}

\myparagraph{Executions, histories and operations.} 
An execution $\xi$ of a distributed algorithm $A$ is a sequence of states and actions reflecting real-time evolution. The history $H_\xi$ is the subsequence of actions in $\xi$. An operation $\pi$ is invoked in $\xi$ when its invocation appears in $H_\xi$, and it responds when the matching response is present. An operation is \emph{complete} in $\xi$ when both its invocation and matching response appear in $H_\xi$ in order. $H_\xi$ is sequential if it starts with an invocation, with each invocation immediately followed by its matching response; otherwise, it is concurrent. $H_\xi$ is complete if every invocation has a matching response, ensuring that each operation in $\xi$ is complete. $\pi$ \textit{precedes} (or \textit{succeeds}) $\pi'$ in real-time in $\xi$, denoted $\pi\rightarrow \pi'$, if $\pi$'s response appears before $\pi'$'s invocation in $H_\xi$. Two operations are concurrent if neither precedes the other.\smallskip

\myparagraph{Consistency.} We consider \textit{linearizability}~\cite{HW90} and \textit{coverability}~\cite{NFG16}  of 
R/W objects.
A complete history $H_\xi$ is \textit{linearizable} if there exists some total order on the
operations in $H_\xi$ s.t. 
it respects the real-time order $\rightarrow$ of operations, and is 
consistent with the semantics of operations.
\textit{Coverability} extends linearizability by ensuring that a write operation is performed on the latest \textit{version} of the object. In particular, coverability is defined over a \textit{totally ordered} set of {versions}, and introduces the notion of \textit{versioned (coverable) objects}. Per~\cite{NFG16}, a coverable object is a type of a R/W object where each value written is assigned with a version, and a write succeeds only if its associated version is the latest one; otherwise, the write becomes a read operation.

\myparagraph{Fragmented Objects and Fragmented coverability.} 
As defined in~\cite{SIROCCO_2021}, a
\textit{block object} is a concurrent R/W object with a bounded value domain. 
A \textit{fragmented object} is a totally ordered sequence of \textit{block} objects, initially containing an empty block. 
\textit{Fragmented coverability}~\cite{SIROCCO_2021} is a consistency property defined over fragmented objects. It guarantees that concurrent write operations on different coverable blocks of the fragmented object would \textit{all} prevail (as long as each write is tagged with the latest version of each block), whereas only one write on the same block eventually prevails (all other concurrent writes on the same block would become read operations).
Thus, a fragmented object implementation satisfying this property may lead to higher access concurrency~\cite{SIROCCO_2021}.\smallskip

\myparagraph{Tags.}
We use logical tags $\tg{}$ as pairs $(ts, wid)$ to order operations, where $ts \in \mathbb{N}$ is a timestamp, and $wid \in \mathcal{W}$ is a writer ID. Let $\mathcal{T}$ be the set of all tags. For any $\tg{1}, \tg{2} \in \mathcal{T}$, $\tg{2} > \tg{1}$ if $(i)$ $\tg{2}.ts > \tg{1}.ts$ or $(ii)$ $\tg{2}.ts = \tg{1}.ts$ and $\tg{2}.wid > \tg{1}.wid$. Each tag is associated with a value of the object.

\section{Overview of \ares{} and its Extensions}
\label{sec:ARESnCo}


\ares{}~\cite{ARES} is a reconfigurable algorithm, designed as a modular framework to implement dynamic, fault-tolerant, read/write distributed linearizable (atomic) shared memory objects. We first present \ares, and then we overview two recently proposed extensions of \ares.

\myparagraph{Main components.} 
\ares{} consists of three major components: $(i)$ Reconfiguration Protocol: Handles the introduction and installation of new configurations. Reconfiguration clients propose and install new configurations via the reconfig  operation.
$(ii)$ Read/Write Protocol: Executes read and write operations invoked by readers and writers, respectively.
$(iii)$ DAP Implementation: Implements Distributed Atomic Primitives (DAPs) for each installed configuration. These primitives respect certain properties and are used by reconfig, read, and write operations (more below).\vspace{.2em}

\myparagraph{Configurations in \ares.}
A \textit{configuration} $c \in \confSet$,~$\confSet$ being a set of unique identifiers, comprises:
$(i)$ $\servers{c}\subseteq\srvSet$: a set of server identifiers; 
$(ii)$ $\quorums{c}$: the set of quorums on $\servers{c}$, s.t. $\forall Q_1,Q_2\in\quorums{c}, Q_1,Q_2\subseteq\servers{c}$ and $Q_1\cap Q_2\neq \emptyset$;
$(iii)$ $\dap{c}$: the set of primitives that clients in $\idSet$ may invoke on $\servers{c}$; and 
$(iv)$ $\consensus{c}$: a consensus instance with the values from $\confSet$, 
implemented on servers in $\servers{c}$.
We refer to a server $s \in \servers{c}$ as a \myemph{member} of  configuration $c$.\vspace{.2em}


\myparagraph{DAPs.} \ares{} allows for reconfiguration between completely different protocols in principle, as long as they can be expressed using three DAPs: $(i)$ the \act{get-tag}, which returns the tag of an object, 
$(ii)$ the \act{get-data},
which returns a $\tup{tag, value}$ pair, and 
$(iii)$ the \act{put-data}($\tup{tag, value}$), which accepts a $\tup{tag, value}$ as an argument. 

For the DAPs to be useful, they need to satisfy a property, referred in~\cite{ARES} as {\bf Property 1}, which involves two conditions: {\bf (C1)} if a \act{put-data}($\tup{\tg, v})$
precedes a \act{get-data} {(or \act{get-tag})} operation that returns $\tg{}'$, then $\tg{}'\geq\tg{}$, and {\bf (C2)} if a
\act{get-data} returns $\tup{\tg{}',v'}$ then there exists \act{put-data}($\tup{\tg{}', v'})$ that precedes or is concurrent to the get-data operation.

In~\cite{ARES}, two different atomic shared R/W algorithms were expressed in terms of 
DAPs. These are the DAPs for the  \abd{} algorithm~\cite{ref_article_ABD},
and the DAPs for an erasure coded based approach presented for the first time in~\cite{ARES}. Both DAPs were shown to satisfy Property 1. In the rest of the manuscript we refer to the two DAP implementations as \abddap{} and \ecdap{}, respectively.  
\abddap{} focuses on data consistency through replication, while \ecdap{} offers fault tolerance and data protection by dividing data into smaller fragments with redundancy.\vspace{.2em}


\myparagraph{Configuration sequence and sequence traversal.}
A configuration sequence $cseq$ in \ares{} is defined as a sequence 
of pairs $\tup{c, status}$ where $c\in\confSet$, and $status\in\{P, F\}$, where $P$ stands for pending and $F$ for finalized. 
Configuration sequences are constructed and stored in clients, while each
server in a configuration $c$ only maintains the configuration that follows $c$ in 
a local variable $nextC\in\confSet\cup\{\bot\}\times\{P, F\}$. {This way, \ares{} attempts to construct a global distributed sequence (or list) of configurations $\mathcal{G}_L$.}

Any \act{read/write/reconfig} operation utilizes the \myemph{sequence traversal} mechanism
which consists of three actions: 
$(i)$ $\act{get-next-config}()$, to discover the next configuration, 
$(ii)$ $\act{put-config}()$, which writes back $nextC$ to a quorum of servers, to ensure that a state is discoverable by any subsequent operation,
and  $(iii)$ $\act{read-config}()$, which finally returns the updated configuration sequence. \vspace{.2em}

\myparagraph{Reconfiguration operation.} To perform a reconfiguration operation $\act{reconfig}(c)$, a client $r$ follows 4 steps: $(i)$ It executes a sequence traversal to discover the latest configuration sequence $cseq$. $(ii)$ It attempts to add $\tup{c,P}$ at the end of $cseq$ by proposing $c$ to a consensus mechanism. The outcome of the consensus may be a configuration $c'$ (possibly different than $c$) 
proposed by some reconfiguration client.  
$(iii)$ The client determines the maximum tag-value pair of the object, say $\tup{\tg,v}$
by executing \act{get-data}  
and transfers the pair to $c'$ 
by performing $\act{put-data}(\tup{\tg,v})$ on $c'$. $(iv)$ Once the update of the value is complete, $r$ {\em finalizes} the proposed configuration by setting 
$nextC=\tup{c', F}$ in a quorum of servers of the last configuration in its $cseq$.

In~\cite{ARES}, it was shown that this reconfiguration procedure 
guarantees that if $cseq_p$ and $cseq_q$ are configuration sequences obtained by  
any two clients $p$ and $q$, then either 
$cseq_p$ is a prefix of $cseq_q$, or vice versa.\vspace{.2em}

\myparagraph{Read/Write operations.}
A \act{write} (or \act{read}) operation $\op$ by a client $p$ is executed by performing the following actions: 
$(i)$ 
 $\op$ invokes a $\act{read-config}$ action to obtain the latest configuration sequence $cseq$, 
$(ii)$ $\op$ invokes a $\act{get-tag}$ (if a write) or $\act{get-data}$ (if a read)
in each configuration, starting from the last finalized to the last configuration
in $cseq$, and discovers the maximum $\tg{}$ or $\tup{\tg{},v}$ pair respectively,
and 
$(iii)$ repeatedly invokes $\act{put-data}(\tup{\tg{}',v'})$, where $\tup{\tg{}',v'}=\tup{\tg{}+1,v'}$ if $\op$ is a write and $\tup{\tg{}',v'}=\tup{\tg,v}$  if $\op$ is a read in the last configuration in $cseq$, and $\act{read-config}$
to discover any new configuration, 
until no additional configuration is observed.\vspace{.3em}

\remove{
\myparagraph{Read/Write operations.}
The read and write operations are composed of two phases: discovery ($Phase1$) and propagation ($Phase2$). Additionally, there are two instances of the $\act{read-config}$ action involved in the phases:

$(i)$ Discovery Phase: During the discovery phase of read and write operations, a client performs the $\act{read-config}$ action to obtain the latest configuration information and invokes $\act{get-tag}$ (if a write) or $\act{get-data}$ (if a read). Finally, it discovers the maximum $tag$ or $\tup{tag, value}$ pair, respectively.

$(ii)$ Propagation Phase: Once clients have discovered the latest configurations and tags, they enter the propagation phase. In this phase, clients propagate new $\tup{tag, value}$ pairs using the $\act{put-data}$ primitive to ensure that the latest data is distributed across the system. Clients continue to perform $\act{read-config}$ actions to check for any new configurations introduced while propagating data.

\myparagraphitalic{Reconfiguration operation.}
To perform a reconfiguration operation $\act{recon}(c)$, a client $r$ goes through four steps:
$(i)$ $r$ discovers the latest configuration sequence; $(ii)$ $r$ proposes a new configuration to be added at the end of the sequence. The client uses a consensus mechanism to agree on this new configuration, which might end up being different from the one initially proposed.
EISAI SE LA8oS SHMEIO. AYTO TO KEIMENO EINAI removed. EINAI PIO PANW TO KEIMENO POY EMFANIZETAI!!!ok
GIA AYTO DEN EMFANIZETAI H ALLAGH :-)
nia ;p
$(iii)$ $r$ retrieves the latest $tag-value$ pair using $get-data$. Then $r$ adds this pair to the new configuration using $\act{put-data}$; $(iv)$ $r$ finalizes the new configuration on a set of servers responsible for the last configuration, ensuring agreement on the sequence.
}

\myparagraph{Extension 1:}  \coARES{} is a coverable version of \ares{}, introduced in~\cite{FragARES}.
Recall from Section~\ref{sec:model} that 
coverability guarantees that writing to an object succeeds when connecting the new value with the ``current" version of the object; otherwise, the write turns into a read operation that provides the latest version and its associated value.
\coARES{} uses an {\em optimized} DAPs implementation presented in~\cite{FragARES}. 
This optimization reduces unnecessary data transfers by having servers send only the $\tup{tag, value}$ pairs with higher tags than the client's tag in \act{get-data}. Also, \act{put-data} only occurs if the maximum tag exceeds their local one.\vspace{.3em}

\myparagraph{Extension 2:} \fcoARES, introduced in~\cite{FragARES}, is the fragmented version of \coARES{} that is suitable for fragmented objects and guarantees fragmented coverability (cf. Section~\ref{sec:model}). It is obtained by the integration of \coARES{} with the \frfs{} framework presented in~\cite{SIROCCO_2021}. 
As a fragmented object is a sequence of blocks (fragments), write operations can modify individual blocks, leading to higher access concurrency. 
\frfs{} is composed of two modules: $(i)$ a Fragmentation Module (FM), and $(ii)$ a Distributed Shared Memory Module (DSMM). In brief, the FM implements the fragmented object,  while the DSMM implements an interface to a shared memory service that allows operations on individual block objects. To this respect, \frfs{} is flexible enough to utilize any underlying DSM implementation, including \coARES{}. 
When \coARES{} is used as the DSMM with \ecdap{}, we obtain a two-level striping (one level from fragmentation and one from Erasure Coding) reconfigurable DSM providing strong consistency and high access concurrency for large (fragmented) objects.\vspace{.3em} 

\myparagraphitalic{Remark:} We note that both extensions of \ares{} mainly involve the read/write operations (applied on coverable or fragmented objects). All three versions share the same reconfiguration mechanism, thus any optimization introduced on the reconfiguration operation of \ares, follows naturally on its extensions as well (with some minor modifications). Thus, in Section~\ref{sec:optimizations}, for simplicity of presentation, we present optimizations on the {reconfiguration mechanism of the} original \ares.

\section{Tracing Bottlenecks}
\label{sec:TracingBottlenecks}
We first provide some additional background on distributed tracing and then we proceed to provide the details on how we have applied distributed tracing on the different version of \ares.\smallskip

{\bf Distributed Tracing} is a technique used to monitor and profile distributed systems by tracing individual requests or transactions as they move across multiple components and systems. It involves adding code to the distributed system to gather detailed data on the flow of requests and the behavior of each component. This data is then combined and analyzed to gain a better understanding of the system's overall performance and to identify any problems or bottlenecks.
In particular, the distributed tracing method creates \emph{traces}, which are records of the activity of individual requests as they pass through various microservices in a distributed system. These traces can be used to diagnose and debug problems in the system, as well as to gain insights into how the system is operating. Also, the individual units of work within a trace are called \emph{spans}. Each span corresponds to a specific piece of work that is performed as a request passes through a microservice. Spans can be used to track the performance of individual components of a system and to identify bottlenecks or other issues that may be impacting system performance.

Distributed tracing is implemented by various open-source and commercial tools. Some of the popular implementations and tools for distributed tracing include:
OpenTelemetry~\cite{ref_url_opentelemetry}, Datadog~\cite{datadog}, Google Cloud Trace~\cite{googlecloudtrace} etc.
OpenTelemetry is a popular open-source project maintained by Cloud Native Computing Foundation(CNCF) which resulted from merging 
two mature technologies: OpenTracing and OpenCensus. It is an observability framework that provides a standard way to collect telemetry data from distributed systems, including tracing, metrics, and logs. It provides integration with various backends, such as Jaeger~\cite{uberjaeger} and Zipkin~\cite{ref_url_zipkin}. Both collectors (Jaeger and Zipkin) can store the collected data in-memory, or persistently with a supported backend such as Apache Cassandra or Elasticsearch.
For the visualization of collected data into meaningful charts and plots, a proposed tool  \emph{Grafana Jaeger}~\cite{ref_url_grafana}. 
Grafana is a popular open-source platform for visualizing and analyzing data. It has built-in support for Jaeger, allowing for visualization of Jaeger tracing data through Grafana.

\subsection{EXPERIMENTAL EVALUATION SETUP}

In this section, we describe the setup of the distributed tracing used in this work. 

\paragraph{OpenTelemetry embedding and Jaeger setup.} 
Incorporating OpenTelemetry, we trace read/write/reconfig requests, track timing, and interactions within our implementation, exporting collected data to Jaeger~\cite{uberjaeger}. Our Jaeger setup uses Cassandra as backend storage with an external lab URL accessible via web browsers. We also visualize traces through Grafana Jaeger integration and extend Jaeger's TTL from 2 days to ``forever" to retain traces indefinitely in Cassandra storage, preventing automatic deletion.

\paragraph{Evaluated Algorithms.} We have evaluated the performance of the following algorithms:

\begin{itemize}
    \item \ARESabd{}: This is the version of \ares{} 
    that uses the \abddap{} implementation. 
    \item \coARESabd{}: This is the version of  
    \coARES{} that uses the optimized 
    \abddap{} implementation (see Section~\ref{sec:ARESnCo}).
    \item \fARESabd{}: This is the version of \fcoARES{} that used the optimized \abddap{} implementation; i.e., it is the fragmented version of \coARESabd{}.
    \item \ARESec{}: This is a version of \ares{} that uses the \ecdap{} implementation. 
    \item \coARESec{}: This is \coARES{} that uses the optimized \ecdap{} implementation. 
    \item \fARESec{}: This is the two-level data striping  algorithm obtained when \fcoARES{} is used with the optimized  \ecdap{} implementation; i.e., it is the fragmented version of \coARESec{}.
\end{itemize}

\remove{
\begin{itemize}
    \item \ARESabd{}. This is a version of \ares{}~\cite{ARES} that uses the \abddap{} implementation. 
    \item \coARESabd{}. This is a coverable version of \ares{} (\coARES{}), presented in~\cite{FragARES}, that uses the \abddap{} implementation~\cite{ARES}. The \abddap{} implementation is optimized similarly to the \ecdapopt{} presented in~\cite{FragARES}.
    \item \fARESabd{}. This combines \fcoARES{}~\cite{FragARES} with the \abddap{}, creating the fragmented version of \coARESabd{}.
    \item \ARESec{}. This is a version of \ares{} that uses the \ecdap{} implementation~\cite{ARES}. 
    \item \coARESec{}. This is a coverable version of \ares{} (\coARES{})~\cite{FragARES} that uses the \ecdapopt{} presented in~\cite{FragARES}. 
    \item \fARESec{}. This is the two-level striping  algorithm presented in~\cite{FragARES} when used with the \ecdapopt{} implementation~\cite{FragARES}, i.e., it is the fragmented version of \ARESec{}.
\end{itemize}
}

\begin{figure}[t]
\includegraphics[width=0.5\textwidth]{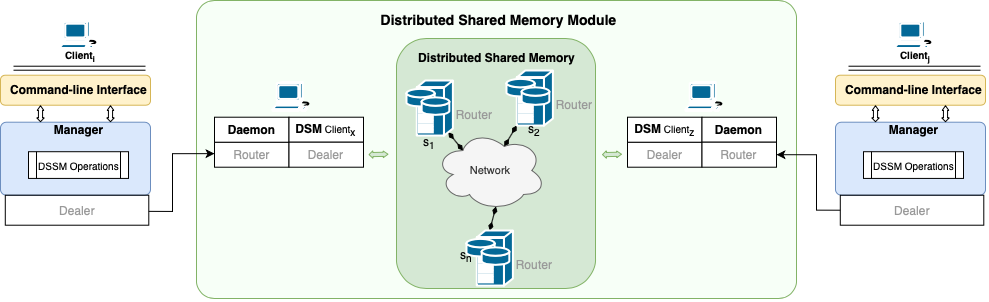}
\caption{The architecture of our implementation.} \label{architecture-basic}
\end{figure}

We have implemented all these algorithms using the same code and communication libraries, based on the architecture in Fig.~\ref{architecture-basic}. The system consists of two main modules: (i) a Manager (User Layer), and (ii) a Distributed Shared Memory Module (DSMM Layer). The Manager provides a client interface (CLI) for DSM access, with each client having its manager handling commands. In this setup, clients access the DSMM via the Manager, while servers maintain shared objects through the DSMM. The Manager uses the DSMM as an external service for read and write operations, allowing flexibility in utilizing various DSM algorithms. The algorithms are all written in Python, and we achieve asynchronous communication between layers using DEALER and ROUTER sockets from the ZeroMQ library~\cite{ref_url_zmq}.

\renewcommand{\abdbased}{ABD-based}
\renewcommand{\ecbased}{EC-based}

In the remainder, for simplicity, we will refer to \ARESabd{}, \coARESabd{} and \fARESabd{} as \abdbased{} algorithms and \ARESec{}, \coARESec{} and \fARESec{} as \ecbased{} algorithms.

\paragraph{Procedures of interest.} 
Using OpenTelemetry Python Library~\cite{opentelemetry-python}, we monitor the \emph{communication} and \emph{computational} overheads of read, write and reconfig operations in both User and DSMM debug levels. 

\paragraph{Distributed Experimental Setup on Emulab.}
We used 39 physical machines on a LAN with no delays or packet loss. The nodes had 2.4 GHz 64-bit Quad Core Xeon E5530 ``Nehalem" processor and 12 GB RAM. A physical controller node orchestrated the experiments, while servers were on different machines. Clients were deployed in a round-robin fashion, with some machines hosting multiple client instances. For instance, with 38 machines, 11 servers, 5 writers, and 50 readers, servers used the first 11 machines, and the rest hosted readers, with 5 of them also running as writers.

\paragraph{Parameters of algorithms:} 
In \ecbased{} algorithms, quorum size is determined by $\left\lceil \frac{n+k}{2} \right\rceil$, while in \abdbased{} ones, it is $\left\lfloor \frac{n}{2} \right\rfloor+1$. Here, $n$ is the total server count, $k$ is the number of encoded data fragments, and $m$ represents parity fragments (i.e., $n-k$). In \ecbased{} algorithms, higher $k$ increases quorum size but results in smaller coded elements, while a high $k$ and low $m$ mean less redundancy and lower fault tolerance. When $k=1$, it is equivalent to replication. The parameter $\delta$ in \ecbased{} algorithms signifies the maximum concurrent $\act{put-data}$ operations, i.e., the number of writers.

\subsection{EXPERIMENTAL SCENARIOS AND RESULTS}
\label{appendix:scenarios+results}
We outline the scenarios and their settings. In each scenario, a writer initializes the system by creating a text file with a specific initial size. As writers update the file, its size grows. While these experiments use random byte strings in text files, our implementations accommodate various file types.
 
Readers and writers use stochastic invocation with random times in intervals of $[1...rInt]$ and $[1..wInt]$ (where $rInt, wInt = 3sec$). Each read/write client performs 50 operations, and reconfigurers, if present, perform 15 reconfigurations at intervals of $15sec$.

\paragraph{We present three types of scenarios:}
\begin{itemize}
    \item File Size: examine performance when using
  different initial file sizes. 
    \item Participation Scalability: examine performance as the number of service participants increases. 
    \item Block Sizes: examine performance under different block sizes (only for fragmented algorithms).
     \item \ec{} Parameter $k$: examine performance with varying $k$ encoded data fragments (for \ecdap{} algorithms).
     \item Longevity: examine performance with reconfigurers switching between DAPs and random server changes under various reconfigurer counts.
\end{itemize}

\subsection{Results and Findings}

Below, we provide Grafana Jaeger-trace graphs, displaying sample traces with average duration for each scenario. Note that all scenarios underwent at least three executions for reliable results.

\subsubsection{File Size}
The scenario is made to measure the performance of algorithms when we vary the size of the shared object, $f_{size}$, from \SI{1}{\mega\byte} to \SI{512}{\mega\byte} by doubling the size in each simulation run.
The maximum, minimum and average block sizes (\emph{rabin fingerprints} parameters), for the fragmented algorithms, were set to \SI{1}{\mega\byte}, \SI{512}{\kilo\byte} and \SI{512}{\kilo\byte} respectively.For \ecbased{} algorithms we used parity $m=5$ yielding quorum sizes of $9$ and for \abdbased{} algorithms we used quorums of size $6$. We fixed the number of concurrent participants to $|\wSet|=5, |\rdSet|=5, |\srvSet|=11$.

\begin{figure*}[htbp]
  \centering
  \begin{subfigure}{0.4\textwidth}
  \includegraphics[width=1\textwidth]{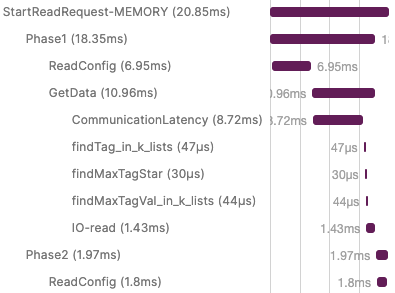}
  \caption{\footnotesize algorithm:\fARESec{}}
  \label{fig:filesize:READ - CoARES_EC-F - DSMM}
  \end{subfigure}
  \hfill
  \begin{subfigure}{0.4\textwidth}
  \includegraphics[width=1\textwidth]{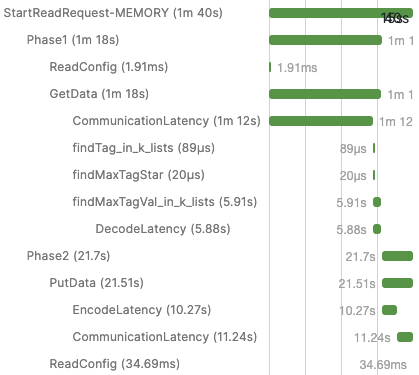}
  \label{fig:filesize:READ - ARES_EC - DSMM}
  \caption{\footnotesize algorithm:\ARESec{}}
  \end{subfigure}
    \caption{\footnotesize READ Operation - S:11, W:5, R:5, fsize: 512MB}
    \label{fig:filesize:READ - ARES_EC & CoARES_EC-F - DSMM}
\end{figure*}

\myparagraph{Results:} The read latency in \fARESec{} involves multiple \emph{block read} requests, fragmenting the file into blocks and executing each block read on shared memory (DSMM).
Fig.~\ref{fig:filesize:READ - ARES_EC & CoARES_EC-F - DSMM} shows the time it takes for two 
read operations to complete on the shared memory (one block in \fARESec{} and the whole object in \ARESec{}).
From Fig.~\ref{fig:filesize:READ - ARES_EC & CoARES_EC-F - DSMM} we observe that the communication and computation latencies vary, while the latency of the $\act{read-config}$ operation incurs a stable overhead regardless of the object size. 
It is worth noting that in this experiment the configuration remains unchanged, and thus time spent for configuration discovery is unnecessary. 
Due to $\act{read-config}$, fragmented algorithms also suffer a stable overhead for each block of the file. This overhead increases when handling 
large objects split into many blocks, ultimately defeating the purpose of fragmentation.

\subsubsection{Participation Scalability}

This scenario is constructed to analyze the read and write 
latency of the algorithms, as the number of readers, writers and servers increases. 
We varied the number of readers $|R|$ from the set $\{5,15,50\}$ and the number of writers from the set $\{5,10,15,20\}$, while the number of servers $|S|$ varies from $3$ to $11$. 
We calculate all possible combinations of readers, writers and servers where the number of readers or writers is kept to $5$. 
The size of the file used is \SI{4}{\mega\byte}.
The maximum, minimum and average block sizes 
were set to \SI{1}{\mega\byte}, \SI{512}{\kilo\byte} and \SI{512}{\kilo\byte} respectively. 
To match the fault-tolerance of \abdbased{} algorithms, we used a
different parity for \ecbased{} algorithms (except in the 
case of 3 servers to avoid replication). 
With this, the \ec{} client has to wait for responses from a larger quorums. The parity value of the \ecbased{} algorithms is set to $m$=1 for $|\srvSet|=3$ and $m$=5 for $|\srvSet|=11$. 

\begin{figure}
    \centering
    \begin{minipage}{0.45\textwidth}
    \centering
    \includegraphics[width=1\linewidth]{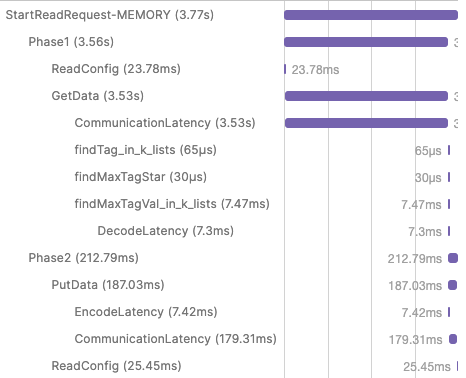}
    \caption{\footnotesize READ Operation - algorithm: \ARESec{}, S:3, W:5, R:50, fsize:4MB, Debug Level:DSMM}
    \label{fig:scalability:READ - ARES_EC - 3_5_50 - DSMM}
    \end{minipage}
    \hspace{10mm}
    \begin{minipage}{0.45\textwidth}
    \centering
    \includegraphics[width=1\linewidth]{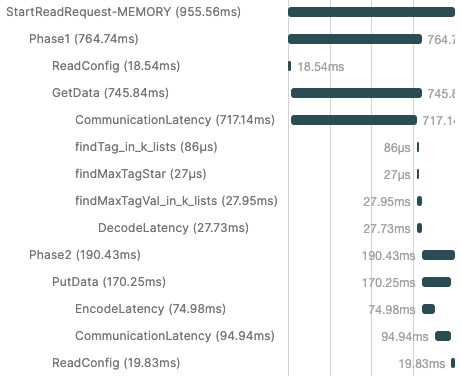}
    \caption{\footnotesize READ Operation - algorithm: \ARESec{}, S:11, W:5, R:50, fsize:4MB, Debug Level:DSMM}
    \label{fig:scalability:READ - ARES_EC - 11_5_50 - DSMM}
    \end{minipage}
\end{figure}

\myparagraphitalic{Results:} In Figs.~\ref{fig:scalability:READ - ARES_EC - 3_5_50 - DSMM} and~\ref{fig:scalability:READ - ARES_EC - 11_5_50 - DSMM}, we observe \ARESec{} read operations in DSMM with varying server counts. A significant difference is seen between $3$ and $11$ servers, with the latter having a shorter read time due to reduced communication latency in $Phase1$. This is because more servers lead to smaller message sizes. However, $DecodeLatency$ and $EncodeLatency$ are slightly longer with $11$ servers. \coARESec{} shows a similar pattern but with smaller latencies due to optimization.

In contrast, the read latency of \fARESec{} in the USER level remains consistent with increasing server count since the object is already divided at the USER level, and DSMM further divides each block, reducing data transfers and improving read latency.

\subsubsection{Block Sizes}
This scenario provides insights into the performance of the read and write operations at different block sizes, highlighting the impact of block sizes  variations on the overall duration of the operations.
We varied the minimum and average $b_{sizes}$ from \SI{2}{\mega\byte} to \SI{64}{\mega\byte} and the maximum $b_{size}$ from \SI{4}{\mega\byte} to \SI{128}{\mega\byte}. The number of servers $|S|$ is fixed to 11, the number of writers $|W|$ and readers $|R|$ to $5$. 
The parity value of \fARESec{} is set to $m=5$.
The size of the initial file used was set to \SI{512}{\mega\byte}.  

\begin{figure}
    \centering
    \centering
    \begin{minipage}{0.25\textwidth}
    \centering
    \includegraphics[width=1\linewidth]{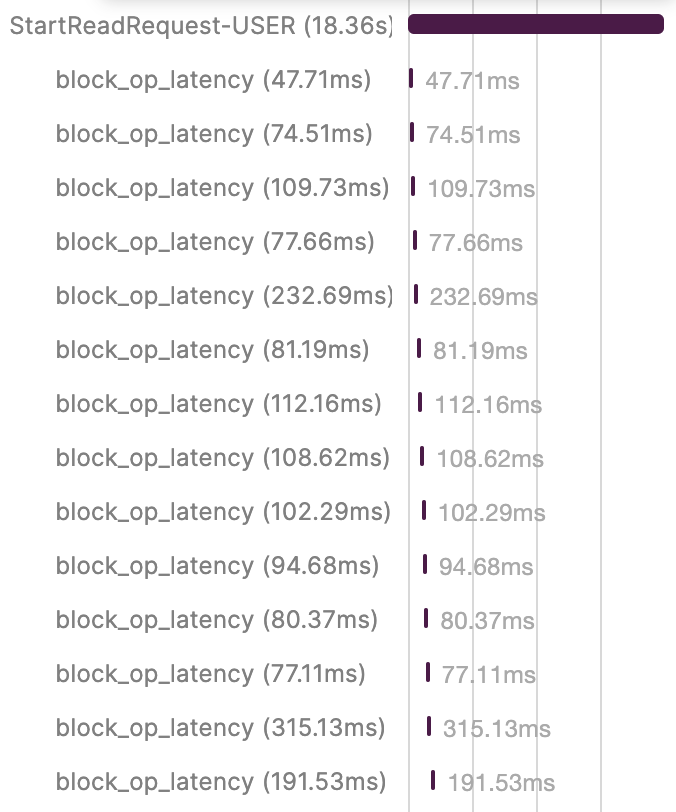}
    \caption{\footnotesize READ Operation - algorithm: \fARESec{}, S:11, W:5, R:5, fsize:512MB, Min/Avg Block Size:2MB, max Block Size:4MB, Debug Level:USER}
    \label{fig:blocksize:READ - CoARES_EC-F - 11_5_5 - 2_4 - USER}
    \end{minipage}
    \hspace{8mm}
    \begin{minipage}{0.25\textwidth}
    \centering
    \includegraphics[width=1\linewidth]{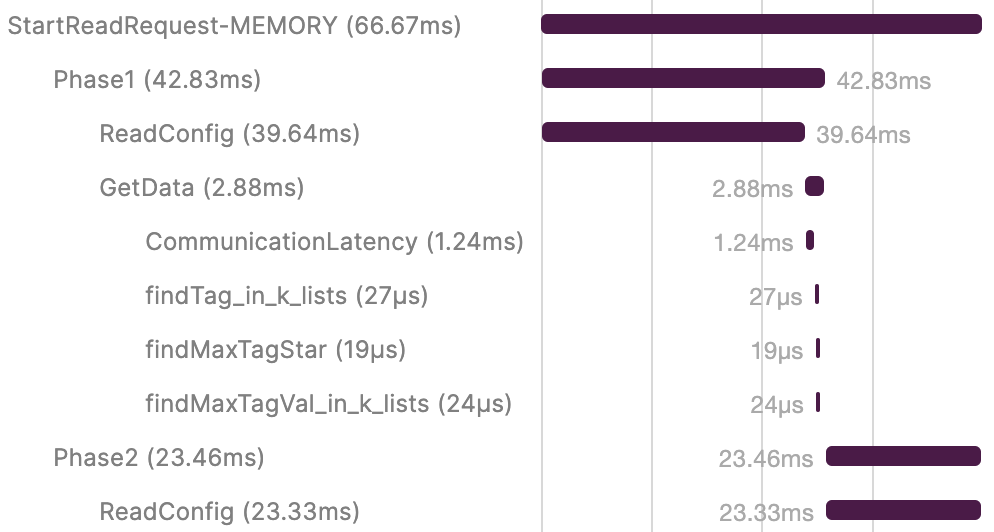}
    \caption{\footnotesize READ Operation - algorithm: \fARESec{}, S:11, W:5, R:5, fsize:512MB, Min/Avg Block Size:2MB, max Block Size:4MB, Debug Level:DSMM}
    \label{fig:blocksize:READ - CoARES_EC-F - 11_5_5 - 2_4 - DSMM}
    \end{minipage}
    \hspace{8mm}
    \begin{minipage}{0.25\textwidth}
    \centering
    \includegraphics[width=1\linewidth]{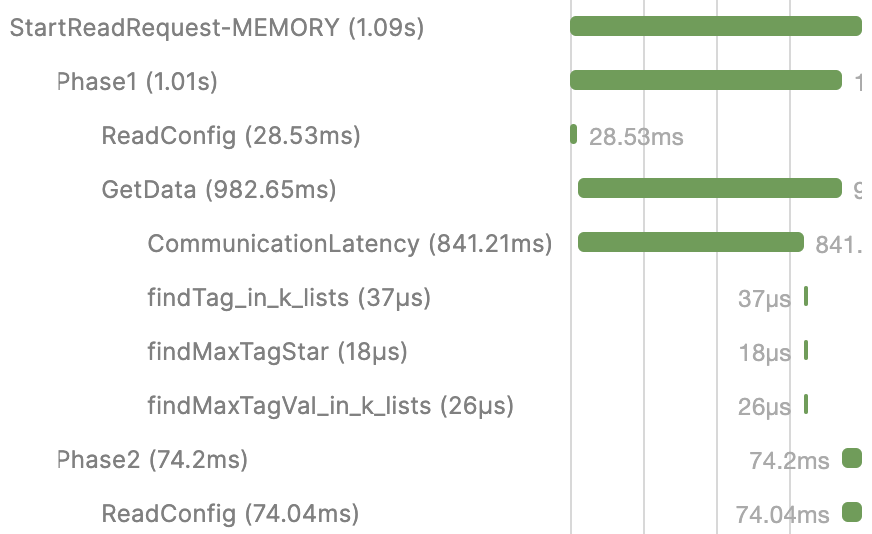}
    \caption{\footnotesize READ Operation - algorithm: \fARESec{}, S:11, W:5, R:5, fsize:512MB, Min/Avg Block Size:64MB, max Block Size:128MB, Debug Level:DSMM}
    \label{fig:blocksize:READ - CoARES_EC-F - 11_5_5 - 64_128 - DSMM}
    \end{minipage}
\end{figure}

\myparagraphitalic{Results:} In previous experiments, we observed that \fcoARES{} experiences higher read and write latencies with larger block sizes. The reason becomes clear when examining the details. For instance, in the read latency with minimal block sizes (Fig.~\ref{fig:blocksize:READ - CoARES_EC-F - 11_5_5 - 2_4 - USER}), the reader must retrieve numerous small blocks, each taking a short time. In contrast, the read latency with maximum block sizes involves fewer but larger blocks, impacting the $CommunicationLatency$, $EncodeLatency$, and $DecodeLatency$ in DSMM (not depicted due to fast reads).

\subsubsection{\ec{} Parameter $k$}
This scenario applies only to \ecbased{} algorithms since we examine how the read and write latencies are affected as we modify the erasure-code fragmentation parameter $k$ (a parameter of Reed-Solomon). We assume $11$ servers and we increase $k$ from $2$ to $10$. The number of writers (and hence the value of $\delta$) are set to $5$. The number of readers is fixed to $15$. The size of the object used is \SI{4}{\mega\byte}.
The maximum, minimum and average block sizes 
were set to \SI{1}{\mega\byte}, \SI{512}{\kilo\byte} and \SI{512}{\kilo\byte} respectively. 

\begin{figure}
    \centering
    \centering
    \begin{minipage}{0.45\textwidth}
    \centering
    \includegraphics[width=1\linewidth]{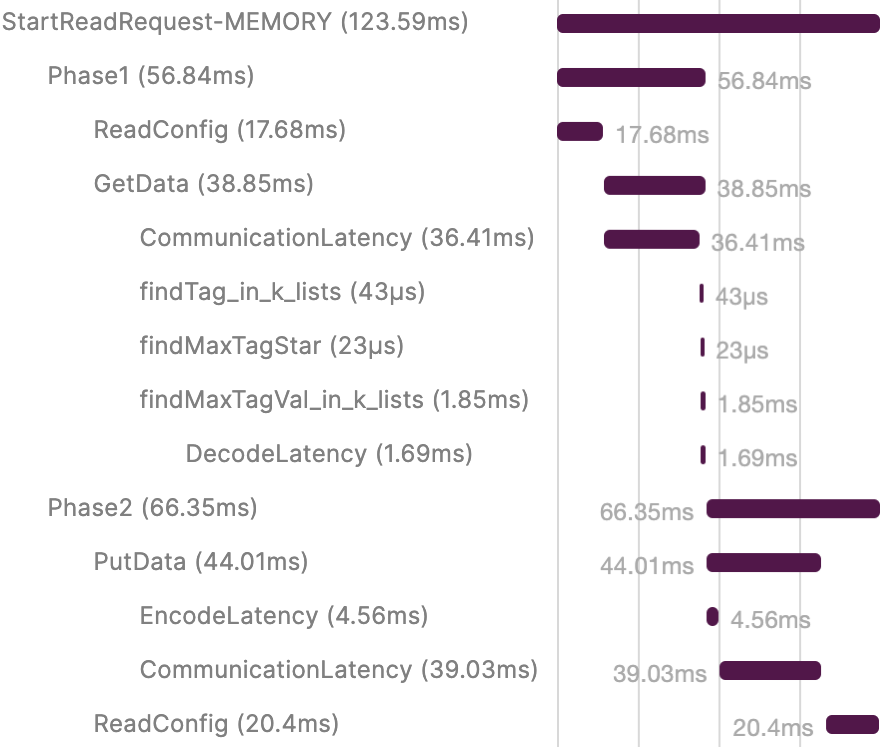}
    \caption{\footnotesize READ Operation - algorithm: \fARESec{}{}, S:11, k:1, W:5, R:5, fsize:4MB, Debug Level:DSMM}
    \label{fig:differentk:READ - CoARES_EC-F - k=1 - DSMM}
    \end{minipage}
    \begin{minipage}{0.45\textwidth}
    \centering
    \includegraphics[width=1\linewidth]{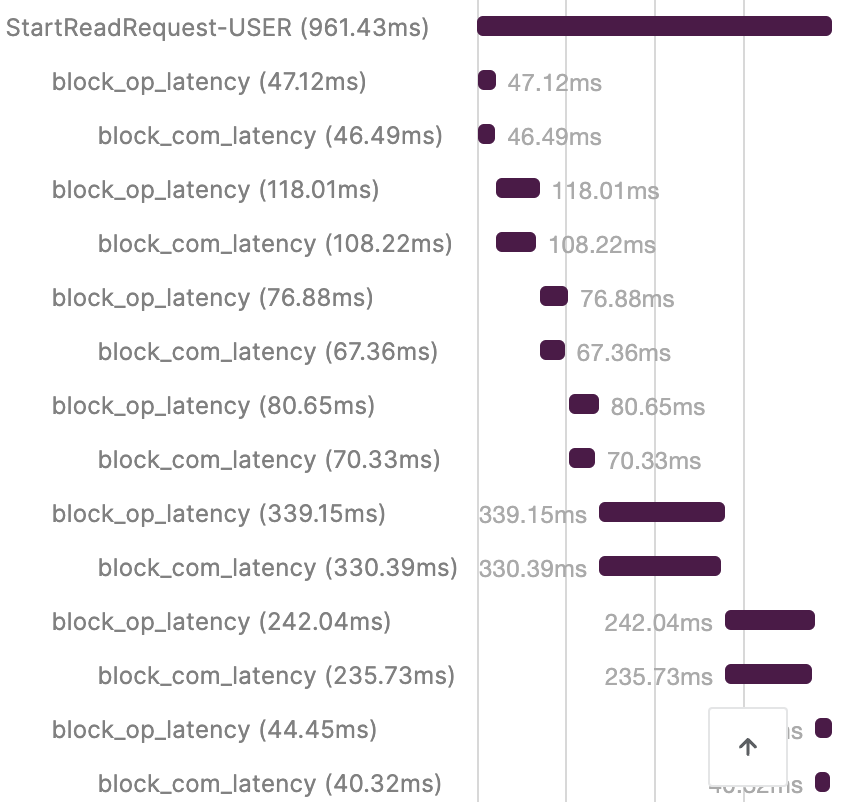}
    \caption{\footnotesize READ Operation - algorithm: \fARESec{}{}, S:11, k:10, W:5, R:5, fsize:4MB, Debug Level:DSMM}
    \label{fig:differentk:READ - CoARES_EC-F - k=10 - DSMM}
    \end{minipage}
    \hspace{15mm}
    \begin{minipage}{0.45\textwidth}
    \centering
    \includegraphics[width=1\linewidth]{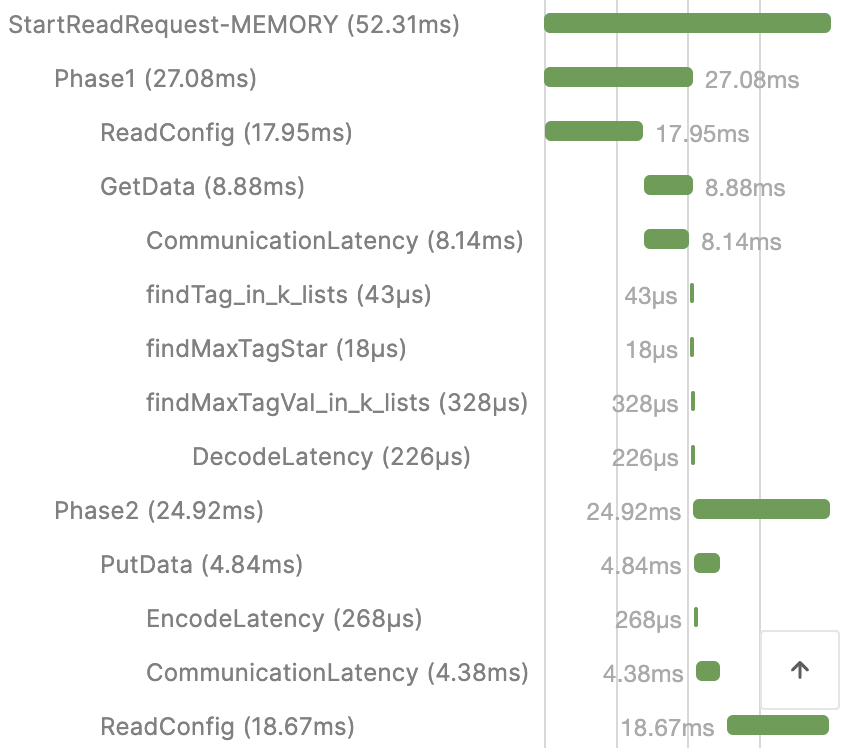}
    \caption{\footnotesize READ Operation - algorithm: \fARESec{}, S:11, k:1, W:5, R:5, fsize:4MB, Debug Level:USER}
    \label{fig:differentk:READ - CoARES_EC-F - k=1 - USER}
    \end{minipage}
    \begin{minipage}{0.45\textwidth}
    \centering
    \includegraphics[width=1\linewidth]{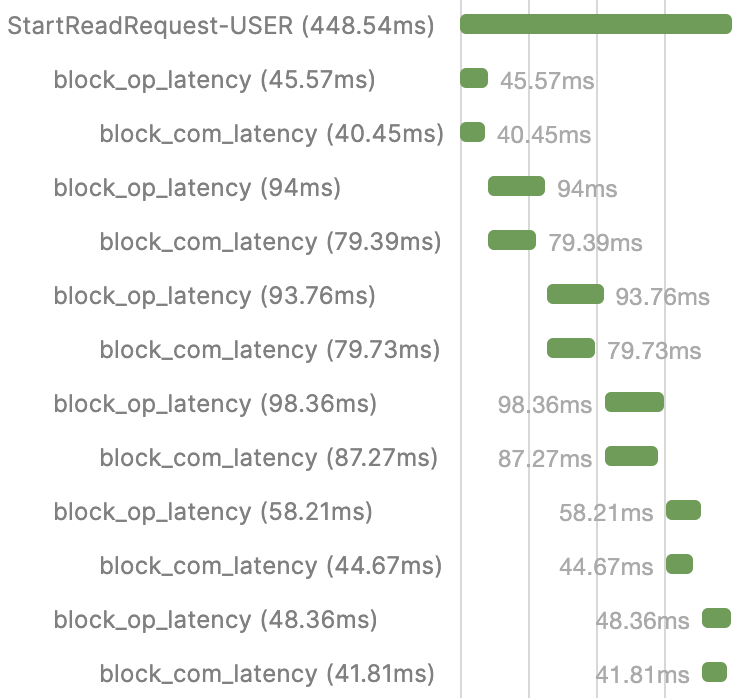}
    \caption{\footnotesize READ Operation - algorithm: \fARESec{}, S:11, k:10, W:5, R:5, fsize:4MB, Debug Level:USER}
    \label{fig:differentk:READ - CoARES_EC-F - k=10 - USER}
    \end{minipage}
\end{figure}

\myparagraphitalic{Results:} In DSMM (Figs.~\ref{fig:differentk:READ - CoARES_EC-F - k=1 - DSMM},~\ref{fig:differentk:READ - CoARES_EC-F - k=10 - DSMM}), we observe reduced communication latency at $k=10$. A higher $k$ reduces read/write latency, while a lower $k$ increases redundancy and fault tolerance at the cost of performance.

At $k=1$ (Fig.~\ref{fig:differentk:READ - CoARES_EC-F - k=1 - DSMM}), encoding latency ($EncodeLatency$) is significantly higher than decode latency due to more parity fragments ($m$) with lower $k$, increasing redundancy. Conversely, at $k=10$ (Fig.~\ref{fig:differentk:READ - CoARES_EC-F - k=10 - USER}) with $m=1$, encoding and decoding are faster with simpler calculations.

\subsubsection{Longevity}

These scenarios examine the performance and verify the correctness of \ares{} when reconfigurations coexist with read/write operations. The reconfigurers changes the DAP alternatively and choose servers randomly between $[3,5,7,9,11]$ servers. The parity value of the \ec{} algorithm is set to $m=1$ for $|S|=3$, $m=2$ for $|S|=5$, $m=3$ for $|S|=7$, $m=4$ for $|S|=9$ and $m=5$ for $|S|=11$. We varied the number of reconfigurers from $1$ to $5$. The numbers of writers and readers are fixed to $5$ and $15$ respectively. The size of the file used is \SI{4}{\mega\byte}. We used the external implementation of Raft consensus algorithms, which was used for the service reconfiguration and was deployed on $6$ physical nodes on Emulab.

\begin{figure}
    \centering
    \begin{minipage}{0.25\textwidth}
    \centering
    \includegraphics[width=1\linewidth]{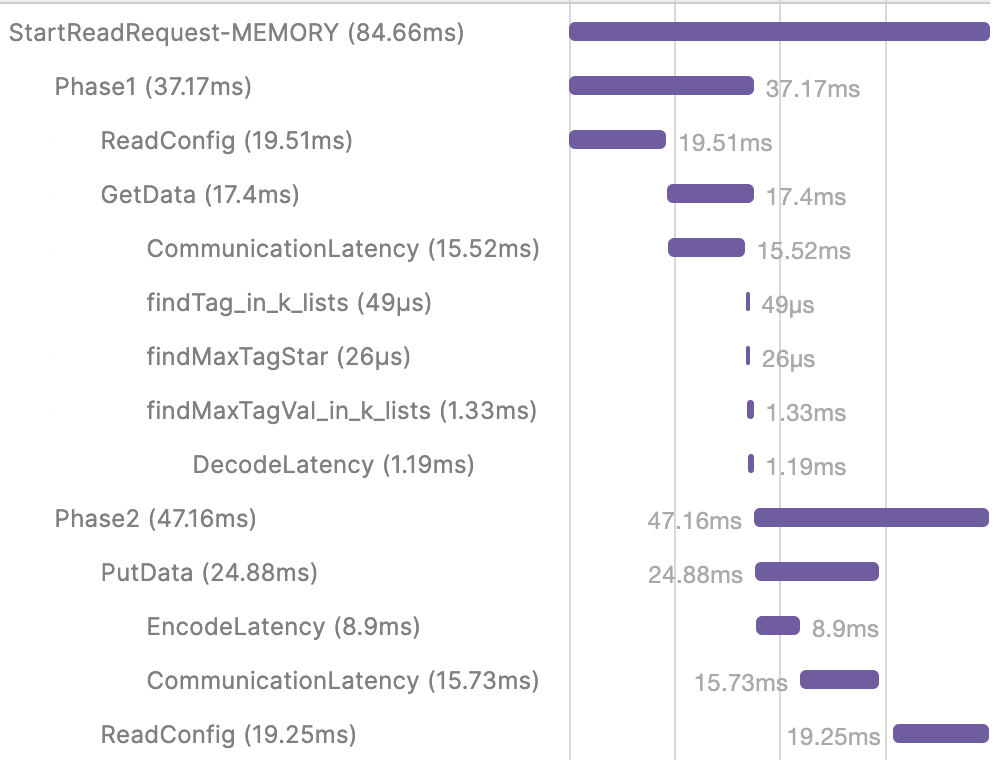}
    \caption{\footnotesize READ Operation - Algorithm:\fcoARES{}, S:11, W:5, R:15, G=1, fsize:4MB, Debug Level:DSMM}
    \label{fig:lonevity:READ - G=1 - DSMM}
    \end{minipage}
    \hspace{8mm}
    \begin{minipage}{0.25\textwidth}
    \centering
    \includegraphics[width=1\linewidth]{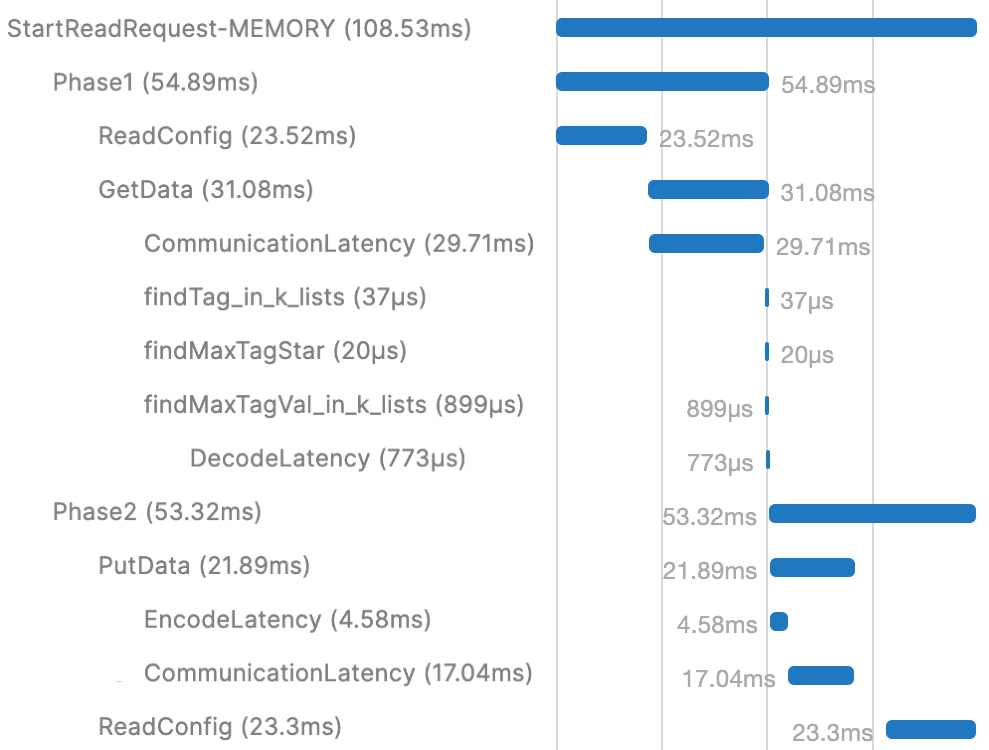}
    \caption{\footnotesize READ Operation - Algorithm:\fcoARES{}, S:11, W:5, R:15, G=5, fsize:4MB, Debug Level:DSMM}
    \label{fig:lonevity:READ - G=5 - DSMM}
    \end{minipage}
    \hspace{8mm}
    \begin{minipage}{0.25\textwidth}
    \centering
    \includegraphics[width=1\linewidth]{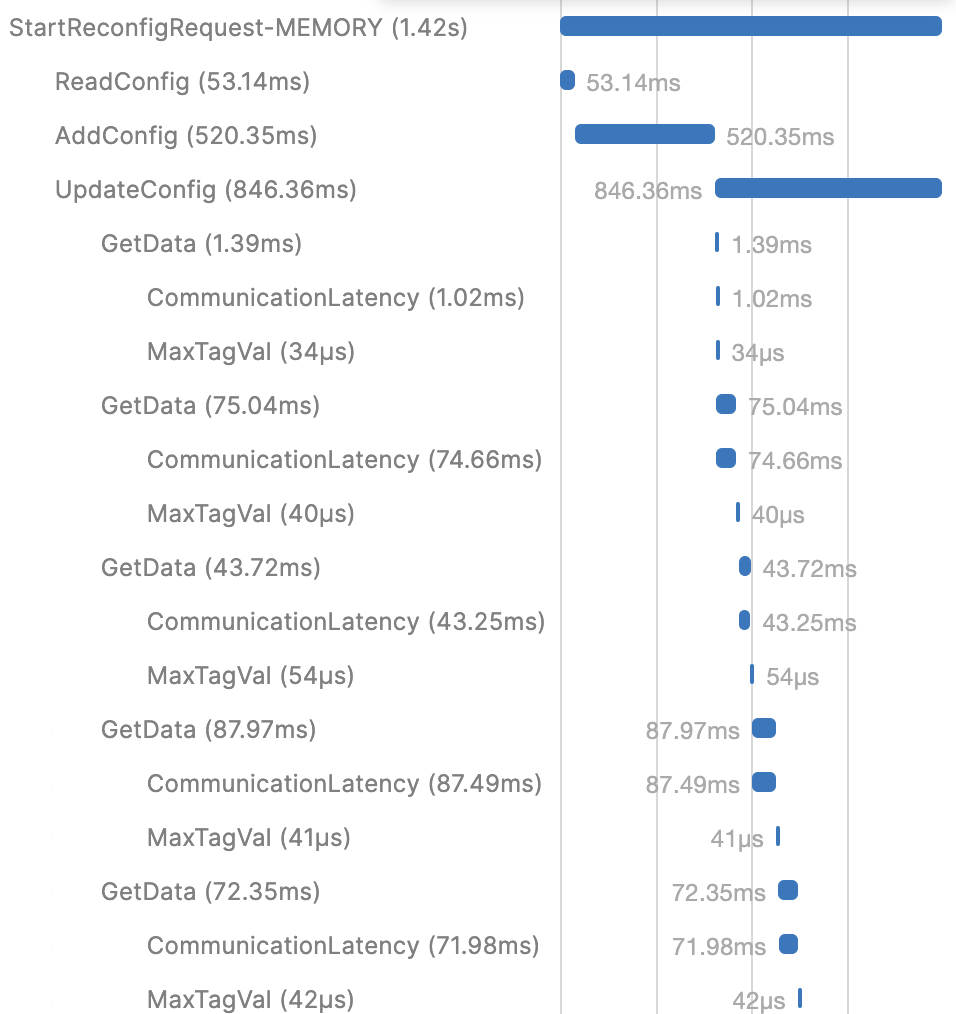}
    \caption{\footnotesize RECON Operation - Algorithm:\fcoARES{}, S:11, W:5, R:15, G=5, fsize:4MB, Debug Level:DSMM}
    \label{fig:lonevity:RECON - G=5 - DSMM}
    \end{minipage}
\end{figure}

\myparagraphitalic{Results:} In Figs.~\ref{fig:lonevity:READ - G=1 - DSMM}-\ref{fig:lonevity:RECON - G=5 - DSMM}, single read/write/recon operations may access various configurations, including both \abd{} and \ec{} algorithms when concurrent with recon operations. Reconfiguration introduces higher delays due to more communication rounds. As the number of reconfigurers increases, all latencies rise, as evident in Figs.~\ref{fig:lonevity:READ - G=1 - DSMM}-\ref{fig:lonevity:READ - G=5 - DSMM}. In \fcoARES{}, recon operations involve multiple $GetData$ and $PutData$ operations, optimizing reconfiguration latency by performing it on the DSMM level for the entire file, gathering and transferring blocks from previous configurations to the new one during a single reconfig.

\subsubsection{Results Summary} A general observation from the tracing evaluation is that configuration discovery adds a stable overhead to the operations, as the client traverses the entire configuration sequence.
A representative sample of our results appears in Fig.~\ref{fig:filesize:READ - ARES_EC & CoARES_EC-F - DSMM} showing the time it takes for two 
read operations to complete on the shared memory (one block in \fARESec{} and the whole object in \ARESec{}).
The tracing mechanism is able to illustrate the various phases and actions in the two operations, and lay down the time it takes for their completion. This allows us to pinpoint potential bottlenecks of the two algorithms.
From Fig.~\ref{fig:filesize:READ - ARES_EC & CoARES_EC-F - DSMM} we observe that the communication and computation latencies vary, while the latency of the $\act{read-config}$ operation incurs a stable overhead regardless of the object size. 
It is worth noting that in this experiment the configuration remains unchanged, and thus time spent for configuration discovery is unnecessary. 
Due to $\act{read-config}$, fragmented algorithms also suffer a stable overhead for each block of the file. This overhead increases when handling 
large objects split into many blocks, ultimately defeating the purpose of fragmentation. 
Lastly, experiments involving concurrent recon operations (cf. Appendix~\ref{appendix:scenarios+results}), demonstrated increased latencies as operations have to access/traverse multiple configurations. This led to the conclusion that latency also suffers as the number of reconfigurations (and thus the number of concurrent configurations) increases.

\section{From \ares{} to \ARESopt{}}
\label{sec:optimizations}
In this section, we introduce a new algorithm called \ARESopt{}, which enhances its predecessor, \ares{}, by addressing identified shortcomings found on Section~\ref{sec:TracingBottlenecks}, and mainly affecting the reconfiguration 
mechanism of the original algorithm. 
We present the following enhancements: 
(i) to expedite configuration discovery we introduce piggy-back data on read/write messages (Section \ref{ssec:optimization:piggy-back}),
(ii) for 
service longevity and to expedite configuration discovery, we introduce a garbage collection mechanism that removes obsolete configurations and updates older configurations with newly established ones (Section \ref{ssec:optimization:GC}), and
(iii) to expedite reconfiguration we introduce a batching mechanism where a single configuration is applied not on a single but multiple objects concurrently (Section \ref{ssec:optimization:reconfig}).
{Below we describe the modifications required for \ares{}, \ecdap{}, and the reconfiguration protocol to support the above enhancements, resulting in \ARESopt{} and \ecdapII{}. We apply similar changes to \abddap{} and implement \abddapII{}, which are evaluated in Section~\ref{sec:Optimized Results}.}
In the pseudocode of the algorithms that follow, 
struck-out text annotates code that has been removed compared to \ares{}.
The colored text annotates the changed code, and the colored box annotates newly added code. Each optimization has a different color: (i) \textcolor{red}{\textbf{red}}, (ii) \textcolor{blue}{\textbf{blue}}, (iii) \textcolor{mydarkgreen}{\textbf{green}}.

\subsection{Optimization 1: Configuration Piggyback} \label{ssec:optimization:piggy-back}

As explained in Section~\ref{sec:ARESnCo}, during the read and write operations in \ares{}, a node learns about new configurations performing $\act{read-config}$ actions. However, as seen in Section~\ref{sec:TracingBottlenecks}, the latency of a $\act{read-config}$ operation produces a stable communication latency overhead, regardless the size of the object. 


\myparagraph{{Description}.}
In this optimization, we 
combine data retrieval with configuration discovery in an attempt to avoid 
the time spent during \act{read-config}. In particular,
in \ARESopt{}, a server piggybacks the $nextC$ variable on every $\act{get-data}$/$\act{put-data}$ reply,
allowing the client to both obtain/update the object value and discover any new configuration through a single message exchange.
This optimization brings changes in the specification of read/write operations, the DAP implementations, and the \myemph{sequence traversal} (cf. Section~\ref{sec:ARESnCo}).

\begin{algorithm*}[!ht]
	\begin{algorithmic}[2]
    \footnotesize
		\begin{multicols}{2}{
  \footnotesize  
    \Part{Write Operation}
				\State at each writer $w_i$ 

				\State {\bf State Variables:}
				\State  $cseq[]~s.t.~cseq[j]\in\confSet\times\{F,P\}$ 
				\State {\bf Initialization:} 
				\State $cseq[0] = \tup{c_0,F}$
				
				\Statex		

				\Operation{write}{$val$}, $val \in V$ 
                \State \PBcolor{\sout{$cseq\gets$\act{read-config}($cseq$)}}  \label{line:writer:p1:lastfin}
				\State $\mu\gets\max(\{i: \status{cseq[i]]} = F\})$\label{line:writer:p1:cseq-lastfin}
    			\State $cs \gets cseq[\mu]$ \label{line:writer:p1:cseq-init}
                \While{\PBcolor{$cs \neq \bot$}} \label{line:writer:p1:whilebegin}
                \State $\tg{c}$\PBcolor{$, Cs$}$ \gets \configPB{cs}.\act{get-tag}()$ \label{line:writer:p1:get-tag}
                \State $\tg{max} \gets \max(\tg{c}, \tg{max})$ 
                \State \PBcolor{$cs,cseq \gets$\act{find-next-config}$(cseq,Cs)$}
				\EndWhile \label{line:writer:p1:whileend}

                \State $\tup{\tg{},v} \gets \tup{ \tup{\tg{max}.ts+1, \wrt_i}, val}$ \label{line:writer:p1:increase}                
                \Statex
       			\State \PBcolor{$\lambda\gets\max(\{i: cseq[i]\neq \bot\})$}
              	\State $cs \gets cseq[\lambda]$  
                \While{$cs \neq \bot$} \label{line:writer:p2:whilebegin}
                \State \PBcolor{$Cs$}$ \gets \configPB{cs}.$\act{put-data}$(\tup{\tg{},v})$\label{line:writer:prop} 
                \State \PBcolor{\sout{$cseq\gets$\act{read-config}($cseq$)}} \label{line:writer:p2:lastfin}
                \State \PBcolor{$ cs,cseq \gets$\act{find-next-config}$(cseq, Cs)$}
				\EndWhile \label{line:writer:p2:whileend}

				\EndOperation
				\EndPart
                \Statex
                \Part{Read Operation}
				\State at each reader $\rdr_i$ 
				\State {\bf State Variables:}
				\State  $cseq[]~s.t.~cseq[j]\in\confSet\times\{F,P\}$ 
				\State {\bf Initialization:} 
				\State $cseq[0] = \tup{c_0,F}$

				\Statex
				
				\Operation{read}{ } 
                \State \PBcolor{\sout{$cseq\gets$\act{read-config}($cseq$)}} \label{line:reader:p1:lastfin}
			     \State $\mu\gets\max(\{j: \status{cseq[j]} = F\})$ \label{line:reader:p1:cseq-lastfin}
                \State $cs \gets cseq[\mu]$ \label{line:reader:p1:cseq-init}
                \While{\PBcolor{$cs \neq \bot$}}\label{line:reader:p1:whilebegin}
                \State $\tup{\tg{c},v_c},$\PBcolor{$ Cs$}$ \gets \configPB{cs}.\act{get-data}()$ \label{line:reader:p1:get-data}
                \State $\tup{\tg{max}\!,\!v_{max}}\!\!\gets\!\!\max(\tup{\tg{c},v_c},\! \tup{\tg{max},v_{max}})$\WRP s.t. $v_c\!\!\neq\!\!\bot$ 
                \label{line:reader:p1:maxpair}
                \State \PBcolor{$ cs,cseq \gets$\act{find-next-config}$(cseq,Cs)$}
                
                \label{line:reader:p1:handleTimerror}
                \EndWhile \label{line:reader:p1:whileend} 
                \Statex\vspace{-.5em}
       			\State
          \PBcolor{$\lambda\gets\max(\{i: cseq[i]\neq \bot\})$}
              	\State $cs \gets cseq[\lambda]$
                \While{$cs \neq \bot$} \label{line:reader:p2:whilebegin}
                \State \PBcolor{$Cs$}$ \gets \configPB{cs}.$\act{put-data}$(\tup{\tg{max},v_{max}})$\label{line:reader:prop} 
                \State \PBcolor{\sout{$cseq\gets$\act{read-config}($cseq$)}} \label{line:reader:p2:lastfin}
                \State \PBcolor{$cs,cseq \gets$\act{find-next-config}$(cseq,Cs)$}            
			\EndWhile \label{line:reader:p2:whileend}

                \State {\bf return} $v_{max}$ \label{line:reader:return}

			\EndOperation

                 \EndPart

  }\end{multicols}	
	\end{algorithmic}
	\caption{Write and Read protocols at the clients for \ARESopt{}.}
	\label{algo:read_writeProtocol}
\end{algorithm*}

\begin{algorithm*}[!ht]
    \begin{algorithmic}[2]
        \footnotesize
		\begin{multicols}{2}	{\footnotesize
                \Procedure{\PBcolor{find-next-config}}{$cseq, Cs$}

                \State $\lambda\gets \max(\{j: cseq[j]\neq \bot\})$	\label{line:find-next-config:final}
                \State $lastC\gets cseq[\lambda]$
                 \State \GCcolor{$Cs_F\gets\{cs: cs \in Cs ~\wedge \status{cs}=F\}$}
          
                \If{$Cs_F\neq\emptyset$}
                    \State \GCbox{$next\lambda\gets \max({\config{cs}.ID: cs\in Cs_F})$} \label{line:find-next-config:maxF}
                    \State $nextC \gets cs \in Cs_F \text{ s.t. } \config{cs}.ID=next\lambda$
				\ElsIf{$\exists cs\in Cs \text{ s.t. }$$cs\neq\bot ~\wedge~ $$ \status{cs} = P$}
					\State $next\lambda\gets \config{cs}.ID$
                    \State $nextC \gets cs$
				\Else
					\State $nextC \gets \bot$   
				\EndIf
                    
				\If{$nextC \neq \bot$}
                    \label{line:find-next-config:update:if1:start}
				    \State $cseq[next\lambda] \gets nextC$ \label{line:readconfig:assign}	
				    \State \act{put-config}$(\config{lastC}, cseq[next\lambda])$\label{line:find-next-config:update:if2:put-config}
				\EndIf 
                \State {\bf return} $cseq, nextC$
				\EndProcedure
				
				\Statex

                \Procedure{read-config}{$cseq$}
				\State $\mu\gets \max(\{j: \status{cseq[j]} = F\})$	
				\State $cs \gets cseq[\mu]$  

                \While{$cs \neq \bot$}
				\State \RECONcolor{$Cs$} $\gets$\act{get-next-config}$(\config{cs})$ 
                \State \RECONcolor{$cseq, cs \gets \act{find-next-config(cseq,Cs)}$}
				\EndWhile
				\State {\bf return} $cseq$
				\EndProcedure
    
				\Statex
				\Procedure{get-next-config}{$c$}
				\State {\bf send} $(\text{{\sc read-config}})$ to each $s\in \servers{c}$
				\State {\bf until} $\exists\quo{},  \quo{}\in\quorums{c}$ s.t. $rec_i$ receives $nextC_s$\WRP from $\forall s\in\quo{}$
                \State\RECONcolor{$Cs\gets\{nextC_s: \text{ received } nextC_s \text{ from each } s\in \quo{}\}$}
				\State \RECONcolor{{\bf return} $Cs$}
				\EndProcedure

                \Statex
				
				\Procedure{put-config}{$c, nextC$}
				\State {\bf send} $(\text{{\sc write-config}}, nextC)$ to each $s\in \servers{c}$
				\State {\bf until} $\exists\quo{},  \quo{}\in\quorums{c}$ s.t. $rec_i$ receives {\sc ack} from $\forall s\in\quo{}$
				\EndProcedure	

                \Statex
                \GCboxParam{1.2}{
                \Procedure{gc-config}{$cseq$}
				\State $\mu\gets\max(\{i: \status{cseq[i]}=F\})$\label{line:gc-config:lastfin}
                \State $CID\gets \{i: cseq[i]\neq\bot\wedge i<\mu\}$

                \For{$id$ in $CID$} \label{line:gc-config:forbegin}
                \State {\bf send} $(\text{{\sc gc-config}}, next)$ to each $s\in \servers{cseq[id]}$
				\State {\bf until} $\exists\quo{},  \quo{}\in\quorums{cseq[id]}$ s.t. $rec_i$ receives {\sc ack}\WRP from $\forall s\in\quo{}$
                \State /* remove the  \{id, cseq[id]\} */
                \State $cseq \gets cseq \backslash~\{id, cseq[id]\}$
                \EndFor \label{line:gc-config:forend}
                \State {{\bf return} $cseq$}             
                \label{line:gc-config:whileend}
                \EndProcedure
                }
		}\end{multicols}

	\end{algorithmic}
	\caption{Sequence traversal at each process $\pr\in\idSet$ of \ARESopt{}.}
	\label{algo:parser}
\end{algorithm*}

\begin{algorithm*}[!htbp]
				\begin{algorithmic}[2]
					{\footnotesize
					\begin{multicols}{2}\footnotesize
							\State{ at each process $\pr_i\in\idSet$}
                            \Statex
							\Procedure{c.get-tag}{}
							\State {\bf send} $(\text{{\sc query-tag}})$ to each  $s\in \servers{c}$
							\State {\bf until}   $\pr_i$ receives $\tup{t_s}$\PBcolor{$,nextC_s$} from $\left\lceil \frac{n + k}{2}\right\rceil$ servers in $\servers{c}$    
                		 \State \PBbox{$Cs \gets \{nextC_s : \text{ received } nextC_s \text{ from } s \}$}       
				        \State $t_{max} \gets \max(\{t_s : \text{ received } t_s \text{ from } s \})$
						\State {\bf return} $t_{max}$\PBcolor{$,Cs$}
						\EndProcedure
							\Statex
							\Procedure{c.get-data}{}
				            \State {\bf send} $(\text{{\sc query-list}}$$)$ to each  $s\in \servers{c}$
							\State {\bf until}    $\pr_i$ receives $List_s$\PBcolor{$,nextC_s$} from $\left\lceil \frac{n + k}{2}\right\rceil$ servers in $\servers{c}$ 
                                   
                                 \State \PBbox{$Cs \gets \{nextC_s : \text{ received } nextC_s \text{ from } s \}$}

                                \State  $Tags_{dec}^{\geq k} =$ set of tags that appears in $k$ Lists 
								\label{line:getdata:max:begin:dec}
                                \State  $t_{max}^{dec} \leftarrow \max(Tags_{dec}^{\geq k})$ \label{line:getdata:max:end}
                                
			                \If{$Tags_{dec}^{\geq k} \neq \emptyset$}
                                     \State \PBbox{$fragments\gets\{e : \tup{\tg,e} \in Lists~\&~\tg=t_{max}^{dec}\}$} \label{line:getdata:fragments}


				                    \If{\PBbox{$\nexists \bot \in fragments$}} \label{line:getdata:if}
			                        \State  \textcolor{black}{$v \leftarrow $ decode value for $t_{max}^{dec}$} \label{line:getdata:decode}  
                                        \Else
                                         \State \PBbox{$v \gets \bot$} \label{line:getdata:bot_v}  
                                        \EndIf
			                        \State {\bf return} $\tup{t_{max}^{dec},v}$\PBcolor{$,Cs$}
                                    
			                \EndIf

							\EndProcedure
							
							\Statex				
							
							\Procedure{c.put-data}{$\tup{\tg{},v})$}\label{line:putdata:begin}
							    
								\State $\Coded = [(\tg{}, e_1), \ldots, (\tg{}, e_n)]$, $e_i = \Phi_i(v)$
								\State {\bf send} 
							$(\text{{\sc PUT-DATA}},
							\tup{\tg{},e_i})$ to each $s_i$ $\in \servers{c}$
								\State {\bf until} $\pr_i$ receives \PBcolor{$nextC_s$} from each server $s\in\srvSet_g$\WRP s.t. $|\srvSet_g|=\left\lceil \frac{n + k}{2}\right\rceil$\\\hspace{1.7cm} and  $\srvSet_g\subset \servers{c}$ 

								
                                 \State \PBboxParam{1.06}{
                                 \State $Cs \gets \{nextC_s\!\!:\!\! \text{ received } nextC_s \text{ from each } s\!\!\in\!\!\srvSet_g\}$
                        	     \State {\bf return} $Cs$ \label{line:putdata:return}
                                }

							\EndProcedure\label{line:putdata:end}

					\end{multicols}
				}
				\end{algorithmic}	
				\caption{\ecdapII{} implementation 
					}\label{algo:casopt}
			\end{algorithm*}
	\begin{algorithm*}[!ht]
	\begin{algorithmic}[2]
		{\footnotesize
		\begin{multicols}{2}\footnotesize
				\State{at each server $s_i \in \srvSet{}$ in configuration $c_k$}
				\State{\bf State Variables:}
					\Statex $List \subseteq  \mathcal{T} \times \mathcal{C}_s$, initially   $\{(t_0, \Phi_i(v_0))\}$
			        \State $nextC\in \confSet\times \{P,F\}$, initially $\tup{\bot,P}$

            \Statex      
            
			\Receive{{\sc query-tag}}{$s_i,c_k$}
                
			\State $\tg{max} \gets \max_{(t,c) \in List}(t)$
			\State {Send} $\tg{max}$\PBcolor{$,nextC$} to $q$
                
			\EndReceive
			\Statex

			\Receive{{\sc query-list}}{$s_i,c_k$}	
                \State 
                \PBboxParam{0.8}{
                \If{$\status{nextC}=F$} \label{line:server:querylist-nextCstatus:start} 
                    \For{$\tg{},v$ in $List$}	\label{line:server:querylist:list':start} 
			    \State $List' \gets List' \cup \{ \tup{\tg{}, \bot}\}$
			    \EndFor \label{line:server:querylist:list':end} 
                    \State Send $List',nextC$ to $q$ \label{line:server:querylist-nextCstatus:true} 
			\EndIf
                }

                \State Send $List$\PBcolor{$,nextC$} to $q$ \label{line:server:querylist:originalreturn} 
			\EndReceive
		    \Statex
			\Receive{{\sc put-data}, $\tup{\tg{},e_i}$}{$s_i,c_k$}
                    \State \PBbox{$fragments\gets\{e: \tup{t,e} \in List\}$} \label{line:putdata:fragments}
				\If{\PBbox{$\nexists \bot \in fragments$}} \label{line:server:ifNotBot}
				\State $List \gets List \cup \{ \tup{\tg{}, e_i}  \}$ \label{line:server:insert}
				\If{$|List| > \delta+1$}\label{line:server:if}
					\State $\tg{min}\gets\min\{t: \tup{t,*}\in List\}$\label{line:server:findmin}
                    \State /* remove the coded value */
					\State $List \gets List \backslash~\{\tup{\tg{},e}: \tg{}=\tg{min} ~\wedge \tup{\tg{},e}$\\\hspace{2.4cm} $\in List\}$\label{line:server:removemin}
     				\State \PBcolor{\sout{$List \gets List  \cup \{  (  \tg{min}, \bot)  \}$}} \label{line:server:bottags}
				\EndIf
                \EndIf
                 \State \PBbox{Send $nextC$ to $q$}
		    \EndReceive

                \Statex
			
			\Receive{{\sc read-config}}{$s_i,c_k$}
			\State send $nextC$ to $q$
			\EndReceive
			
			\Statex
			
			\Receive{{\sc write-config}, $cfgT_{in}$}{$s_i,c_k$}
			\If{$\config{nextC}=\bot~\vee~nextC.status=P$\GCcolor{$~\vee~$\WRP $\config{cfgT_{in}}.ID>\config{nextC}.ID$}} \label{line:server:write-config:if}
			\State $nextC\gets cfgT_{in}$
			\EndIf
			\State send {\sc ack} to $q$
			\EndReceive

            \Statex

            \GCboxParam{1.2}{
			\Receive{{\sc gc-config}, $cfgT_{in}$}{$s_i,c_k$}
            
            \If{\GCcolor{$\config{cfgT_{in}}.ID>\config{nextC}.ID$}} \label{line:server:gc-config:is_lastbegin}
			\State $nextC\gets cfgT_{in}$ \label{line:servers:nextCupdate}
            \EndIf 
            
            \label{line:server:gc-config:is_lastend}
                
                \For{$\tg{},e$ in $List$} \label{line:server:gc-config:forBot:start}		 
                \State $List \gets List \backslash~\{\tup{\tg{},e}$
			\State $List \gets List  \cup \{  (  \tg{}, \bot)  \}$
		    \EndFor \label{line:server:gc-config:forBot:end}	
           
			\State send {\sc ack} to $q$
			\EndReceive
            }
   
				\end{multicols}
			}

	\end{algorithmic}	
	\caption{The response protocols at  any server $s_i \in \srvSet{}$ of \ARESopt{}}\label{algo:server}
\end{algorithm*}

\myparagraph{\textit{Read/Write operations.}}
Algorithm~\ref{algo:read_writeProtocol} specifies the read and write protocols of \ARESopt{}. 


State variable $cseq$ at each reader, writer, and reconfigurer process
is used to hold a sequence of configurations as these are 
discovered at each client. We use the convention that the $cseq[i]$ of a client $\pr$ stores 
a tuple $\tup{c_i, *}$ if $\pr$ discovered the configuration with identifier $i$; otherwise we say that $cseq[i]=\bot$.
Initially,  $cseq$ contains a single element, $\tup{c_0, F}$, which is an initial configuration known to every participant in the service.
The local variable $Cs\subseteq\confSet\times\{P,F\}$ is a set used by readers and writers to collect
the $nextC$ values received by servers, during any DAP action. 

In contrast to \ares{}, both \act{write} and \act{read} operations, do not issue a $\act{read-config}$ action to obtain the latest introduced configuration
(cf. line Alg.~\ref{algo:read_writeProtocol}:\ref{line:writer:p1:lastfin} (resp. line Alg.~\ref{algo:read_writeProtocol}:\ref{line:reader:p1:lastfin}).
Instead, the reader/writer finds the last finalized entry in its local $cseq$ denoted as $\mu$ and starts the read/write operation from $cs=cseq[\mu]$ (line Alg.~\ref{algo:read_writeProtocol}:\ref{line:writer:p1:cseq-init} (res[. line] Alg.~\ref{algo:read_writeProtocol}:\ref{line:reader:p1:cseq-init})). 

Starting from $cs=cseq[\mu]$, a read/write operation then invokes 
\act{get-data}/\act{get-tag} operations respectively.
In the case of a read operation, the reader collects the $\tup{\tg{c}, v_c}$ pairs and discovers the maximum $\tup{\tg{max}, v_{max}}$ pair among them, where $v_c\neq\bot$. Similarly in the case of a write operation, the writer collects only 
the tag values $\tg{c}$ and discovers the maximum $\tg{max}$ among them. At the same 
time, both operations, collect the set of configurations $Cs$, and
discover using the $\act{find-next-config(cseq,Cs)}$ action the next 
available configuration $cs$.



The process above is repeated until the client reaches $cs=\bot$ which indicates that all servers in a quorum responded with $\config{nextC}=\bot$ (lines Alg.~\ref{algo:read_writeProtocol}:\ref{line:writer:p1:whilebegin}--\ref{line:writer:p1:whileend} (resp. lines Alg.~\ref{algo:read_writeProtocol}:\ref{line:reader:p1:whilebegin}--\ref{line:reader:p1:whileend})). Once done, the writer associates 
a new tag with the value to be written, as done in \ares{}
(cf. line Alg.\ref{algo:read_writeProtocol}:\ref{line:writer:p1:increase}).

Then both read/write operations propagate the tag-value pairs to the servers.
The propagation of $\tup{\tg{}, v}$ in \act{write} (lines  Alg.~\ref{algo:read_writeProtocol}:\ref{line:writer:p2:whilebegin}--\ref{line:writer:p2:whileend}) and $\tup{\tg{max}, v_{max}}$ in \act{read} (lines  Alg.~\ref{algo:read_writeProtocol}:\ref{line:reader:p2:whilebegin}--\ref{line:reader:p2:whileend}) follows the same logic as in the first phase, involving this time the $\act{put-data}$ actions. 

It is worth noting that in \ares{}, the discovery of new configuration and the retrieval of the tags/pairs 
occurs in two separate communication rounds: first executing $\act{read-config}$ to update the $cseq$, and then by $\act{get-tag}$ or $\act{get-data}$ to retrieve the tag/value pairs. Same goes for the propagation phase. 


\myparagraph{\textit{Sequence Traversal}.}
\sequencetraversalII{} is presented in Algorithm~\ref{algo:parser}.

{\em $\act{find-next-config}(cseq,Cs)$:}
Given a batch of servers' replies $Cs$ containing their $nextC$ values, it determines whether any of the replies contains $\config{nextC}\neq\bot$ (line  Alg.~\ref{algo:parser}:\ref{line:find-next-config:update:if1:start}).
In cases where there are replies with $\status{nextC}=F$, we need to identify the $nextC$ with the highest $ID$ (line Alg.~\ref{algo:parser}:\ref{line:find-next-config:maxF}), as servers may point to different next finalized configurations due to garbage collection in the reconfig operation (cf. Section~\ref{ssec:optimization:GC}).
After identifying the $nextC$, 
the $\act{put-config}$ action notifies a quorum of servers in $\servers{c}$, where $c$ the current configuration, to update the value of their $nextC$ variable to the found one (lines  Alg.~\ref{algo:parser}:\ref{line:find-next-config:update:if2:put-config}).
Finally, the action returns the updated $cseq$ along with the found $nextC$.
{\em $\act{read-config}(cseq)$:}
The procedure remains the same as that in \ares{}, with modifications aimed at avoiding code duplication. 

\myparagraph{\textit{EC-DAP II Implementation.}}
The piggyback optimization needed also to be supported by the DAP protocols. Here we present only the modifications done in the \ecdap{} 
protocol. Similar modifications are applied to the \abddap{}.
\ecdapII{} is presented in Algorithms~\ref{algo:casopt} and \ref{algo:server}.

Following \cite{ARES}, each server $s_i$ stores a state variable,  $List$,  which is a set of up to $(\delta + 1)$ (tag, coded-element) pairs; $\delta$ is the maximum number of concurrent put-data operations. 
Also in any configuration $c$ that implements \ecdapII{}, 
we assume an $[n,k]$ MDS code~\cite{verapless_book}, where $|c.Servers| = n$ of which no more than $\frac{n-k}{2}$ may crash. The $[n,k]$ MDS encoder slits the a value $v$ into $k$ equal fragments, which then uses to generate $n$ coded elements $e_1, e_2, \ldots, e_n$, denoted as $\Phi(v)$. We denote the projection of $\Phi$ onto the $i^{\text{th}}$ output component as $\Phi_i$, where $e_i = \Phi_i(v)$.  We associate each coded element $e_i$ with server $i$, $1 \leq i \leq n$. Such erasure code algorithm enables recovery of a coded value $v$ using any $k$ out of the $n$ coded elements, each representing a fraction $\frac{1}{k}$ of $v$. 

We now proceed with the description of the three primitives of \ecdapII{}.

{\em Primitive $\dagettag{c}$:}
The difference from the \ecdap{} is that the client, in addition to requesting the servers' maximum $\tg{}$, also queries their $nextC$ within the same request. So, it returns both a list with servers' $\tg{}$ values and the corresponding $Cs$ values.


{\em Primitive $\dagetdata{c}$:}
Similarly with the $\act{get-tag}$ action, the client, in addition to requesting the servers' $List$, also queries their $nextC$ within the same request. 
If the status of $nextC$ is $F$, the server creates a $List'$ which contains all the tags associated with the $\bot$ value, i.e., $\tup{\tg{},\bot}$ (lines Alg.~\ref{algo:server}:\ref{line:server:querylist:list':start}--\ref{line:server:querylist:list':end}).
As by the reconfiguration algorithm, a finalized configuration 
contains the latest data, we achieve communication efficiency by avoiding 
propagating the data from servers that point to a next finalized configuration, i.e. the status of their $nextC$ is finalized.
In this case, the server responds with $List'$ and $nextC$. 
Otherwise, the server returns $List$ and $nextC$.

It is possible that the $List$ contains $\bot$ values, as a result of a garbage collection (GC) of reconfig operation (cf. Section~\ref{ssec:optimization:GC}). 
So, the client needs to check whether all fragments associated with $t_{\text{max}}^{\text{dec}}$ (line Alg.~\ref{algo:casopt}:\ref{line:getdata:fragments}) are not equal to $\bot$.
If $\bot$ does not exist in fragments (Alg.~\ref{algo:casopt}:\ref{line:getdata:if}), the client can decode $v$
Otherwise, it sets the value to $\bot$.


{\em Primitive $c.\act{put-data}(\tup{\tg{}, v})$:}
In line with \ecdap{}, the client computes coded elements and dispatches the pair ($\tg{}$, $\Phi_i(v)$) to the relevant servers ($s_i$).
When server $s_i$ receive a (\text{\sc put-data}, $\tg{}$, $e_i$) message, $s_i$ verifies if its $List$ does not include $e_i = \bot$ (line Alg.~\ref{algo:server}:\ref{line:server:ifNotBot}). If it is, it indicates that the configuration is garbage collected by $GC$ operation in reconfig operation (cf. Section~\ref{ssec:optimization:GC}), and $s_i$ does not need to update its $List$. 
Like in \ecdap{}, $s_i$ trims pairs with tags exceeding length $(\delta+1)$ (line Alg.\ref{algo:server}:\ref{line:server:removemin}), but it does not keep older tags of coded-elements with $\bot$ to reduce return message size (line Alg.\ref{algo:server}:\ref{line:server:bottags}).
Finally, the client returns a $Cs$ list comprising all the $nextC$ values received from the servers.

				
			
			
			

			
            
            
            
           
   


\subsection{Optimization 2: Garbage Collection}
\label{ssec:optimization:GC}


For storage efficiency, longevity, and expediting configuration discovery, we introduce a garbage collection mechanism to eventually remove obsolete configurations. The main idea of this mechanism is to update older configurations to point to more recently established configurations. At the same time we save storage by removing the obsolete content of older configurations.
This optimization brings changes in the specification of recon operations.

\myparagraph{{Description}.}
The Garbage Collection ($\act{gc-config}$) runs in the end of the reconfiguration operation (line Alg.~\ref{algo:reconfigurer}:\ref{gc-config}). 
The main idea is that the entries that appear in $cseq$ before the last finalized entry can be garbage collected and point to the last finalized entry in $cseq$. 
Thus, the $\act{gc-config}$ sends request to servers for each configuration $cs$ in $cseq$ that has smaller configuration id $\config{cs}.ID$ than the last finalized index $\mu$, followed by removing the garbage collected configuration from its local $cseq$ (lines Alg.~\ref{algo:parser}:\ref{line:gc-config:forbegin}--\ref{line:gc-config:forend}). We pass $cseq[\mu]$ as a parameter to the server message.
When a server receives the message, it checks if the index of received configuration is larger than the index of its local $nextC$ (line Alg.~\ref{algo:server}:\ref{line:server:gc-config:is_lastbegin}). If that holds, the server updates the $nextC$ to point to the last finalized configuration (line Alg.~\ref{algo:server}:\ref{line:servers:nextCupdate}). Next, the server sets the values $e$ of all $\tup{\tg{},e}$ in $List$
to $\bot$ (lines Alg.~\ref{algo:server}:\ref{line:server:gc-config:forBot:start}--\ref{line:server:gc-config:forBot:end}).

\subsection{Optimization 3: Reconfiguration Batching}\label{ssec:optimization:reconfig}
\newcommand{\file}{f}
\newcommand{\block}{b}
A reconfiguration operation in \ares{} is applied on a single atomic object. Thus, in systems where we need to manipulate multiple objects, e.g., the blocks of a fragmented object, whenever a reconfigurer wants to move 
the system from a configuration $c$ to a configuration $c'$ needs to execute a series of $\act{recon}$ operations, one for each object. Running multiple recons however, means executing a different instance of consensus and maintaining different configuration sequence per object, resulting in significant overhead and performance degradation. In this section we examine an optimization that suggests the application of the reconfiguration operation over a domain of (one or all) objects. 
{In the implementation below, instead of having an instance of consensus for each configuration (as defined in Section~\ref{sec:ARESnCo}), we use an external consensus mechanism that maintains the values from $\confSet$.}

				
				

			
 
			

\begin{algorithm*}[!ht]
	\begin{algorithmic}[2]
        \footnotesize
		\begin{multicols}{2}{\footnotesize
				\State at each reconfigurer $rec_i$ 
				\State {\bf State Variables:}
				\State  $cseq[]~s.t.~cseq[j]\in\confSet\times\{F,P\}$ 
				\State {\bf Initialization:} 
				\State $cseq[0] = \tup{c_0,F}$				
				\Statex			
				\Operation{reconfig}{c, D} 
				\If {$c \neq \bot$} 		\label{line:install:valid}
				\State $cseq\gets$\act{read-config}$(cseq)$ \label{line:install:readconfig} 
				\State $cseq \gets \text{\act{add-config}}(cseq, c)$  \label{line:install:add} 
				\State $\text{\act{update-config}}(cseq, D)$
				\State $cseq\gets\text{\act{finalize-config}}(cseq)$    			
                \State \GCbox{$cseq\gets\act{gc-config}(cseq)$} \label{gc-config}
				\EndIf 
				\EndOperation
				%
				%
				%
				\Procedure{add-config}{$cseq$, $c$}
				\State \RECONcolor{$\lambda\gets\max(\{j: cseq[j]\neq \bot\})$}
				\State $c' \gets \config{cseq[\lambda]}$
				\State $d\gets Con.propose(\RECONcolor{\lambda+1}, c)$ \label{line:addconfig:consensus}
                \State \RECONcolor{$d.ID \gets \lambda + 1$}
				\State $cseq[\lambda+1]\gets \tup{d,P}$ 				\label{line:addconfig:assign}
				\State $\act{put-config}(c', \tup{d,P})$		\label{line:addconfig:put}
				\State  {\bf return} $cseq$
				\EndProcedure
				\Statex
				\Procedure{update-config}{$cseq$ \RECONcolor{, $D$}} 
				\State $\mu\gets\max(\{j: cseq[j].status = F\})$
				\State \RECONcolor{$\lambda\gets\max(\{j: cseq[j]\neq \bot\})$}
				\State \RECONbox{$M = []$
                    \For{$o$ in $D$}
    				\State $M[o] \gets \emptyset$
                    \EndFor}
				\For{$i=\mu:\lambda$}
                    \For{\RECONcolor{$o$ in $D$}}
    				\State $\tup{t, v},\_  \gets \config{cseq[i]}.\act{get-data}()$ \RECONcolor{for object $o$}
    				\State \RECONbox{$M[o]  \gets M[o] \cup  \{ \tup{\tg{}, v} \}$} \label{line:reconfig:max}
                    \EndFor
				\EndFor
                    \For{\RECONcolor{$o$ in $D$}}
    				\State $\tup{\tg{},v} \gets \max_{t} \{ \tup{t, v}: \tup{t, v} \in M[\RECONcolor{o}]\}$
    				\State $\config{cseq[\lambda]}.\act{put-data}(\tup{\tg{},v})$ \RECONcolor{for object $o$}
                    \EndFor
				\EndProcedure
				\Statex	
				\Procedure{finalize-config}{$cseq$}
    			\State \RECONcolor{$\lambda\gets\max(\{j: cseq[j]\neq \bot\})$}
				\State $cseq[\lambda].status \gets F$	\label{line:status:finalize}
				\State $\act{put-config}(\config{cseq[\lambda-1]}, cseq[\lambda])$
				\State \textbf{return} $cseq$ 
				\EndProcedure
		}\end{multicols}	
	\end{algorithmic}
	\caption{Reconfiguration protocol of  \ARESopt{}.}
	\label{algo:reconfigurer}
\end{algorithm*}

\myparagraph{{Description}.}
We implement reconfiguration batching in \ARESopt{} by changing the 
specification of the reconfiguration protocol (Alg.~\ref{algo:reconfigurer}), and in particular the 
$\act{update-config}$ procedure. The latter procedure is responsible for moving the latest value of an 
object from an older configuration to the one a reconfigurer tries to install. While \ares{} was executing 
this procedure on a single object, \ARESopt{} executes the procedure on a given \textit{domain} (or set) of
one or more objects. 

More precisely, assume that a reconfigurer $rec_i$ wants to change the current configuration $c$ to $c'$ by invoking a $\act{reconfig}(c',D)$. Then, $rec_i$ collects the sequence $cseq$ of established configurations using 
the $\act{read-config}$ procedure (Alg.~\ref{algo:reconfigurer}:\ref{line:install:readconfig}), and then attempts to add the configuration $c'$ at the end of $cseq$ using the consensus service (Alg.~\ref{algo:reconfigurer}:\ref{line:install:add}). 
Once the newly added configuration is established, given a domain $D$,
$rec_i$ invokes the $\act{update-config}$ procedure. 
%
%
%
%
Using the $\act{get-data}$ DAP, $rec_i$, gathers the tag-value pairs for each object $o$ in $D$, from 
every configuration $cseq[i]$, for $\mu \leq i \leq \lambda$ (i.e., the last finalized configuration with index $\mu$ and the configuration with the largest index $\lambda$ in $cseq$). 
Then it discovers the maximum of those pairs and transfers, using the $\act{put-data}$ DAP, the pair for each object in $D$ to the new configuration $c'$. 
For example, if $\tup{t_{max}, v_{max}}$  is the tag value pair corresponding to the highest tag among the responses from all the $\lambda - \mu + 1$ configurations for a specific object $o$, then 
 $\tup{t_{max}, v_{max}}$  is written to the configuration $c'$ via the invocation of $\config{cseq[\lambda]}.\act{put-data}(\tup{\tg{max},v_{max}})$ of object $o$.

\section{Correctness of \ARESopt}
\subsection{Correctness of \ecdapII{}}
\label{appendix:correctness:ecdapII}
As seen in Section~\ref{sec:ARESnCo}, \ares{} relies on three \emph{data access primitives} (DAPs): $(i)$ the \act{get-tag}, 
$(ii)$ the \act{get-data}, and 
$(iii)$ the \act{put-data}($\tup{\tg, v}$). 
For the DAPs to be useful, they need to satisfy a property, referred in~\cite{ARES} as \emph{Property~\ref{property:dap}}. We slightly revised \emph{Property~\ref{property:dap}} to accommodate the fact that $\act{get-data}$ can return a tag associated with either a value from $\valSet$ or $\bot$.   
See the revised \emph{Property~\ref{property:dap}}.
\begin{property}\label{property:dap}  In an execution $\EX$ 
 	we say that a DAP operation in  $\EX$ is complete if both the invocation and the 
 	matching response step  appear in $\EX$. 
 	If~$\Pi$ is the set of complete DAP operations in execution $\EX$ then for any $\phi,\pi\in\Pi$: 
 \begin{enumerate}
 \item[ C1 ]  If $\phi$ is  $\daputdata{c}{\tup{\tg{\phi}, v_\phi}}$, $\tup{\tg{\phi}, v_\phi} \in\tsSet\times\valSet$, 
 and $\pi$ is $\dagettag{c}$ (or  $\dagetdata{c}$) 
 that returns $\tg{\pi} \in \tsSet$ (or $\tup{\tg{\pi}, v_{\pi}} \in \tsSet \times \valSet\cup\{\bot\}$) and $\phi\bef\pi$ in $\EX$, then $\tg{\pi} \geq \tg{\phi}$.
 \item[ C2 ] \sloppy If $\phi$ is a $\dagetdata{c}$ that returns $\tup{\tg{\pi}, v_\pi } \in \tsSet \times \valSet\cup\{\bot\}$, 
 then there exists $\pi$ such that $\pi$ is a $\daputdata{c}{\tup{\tg{\pi}, v_{\pi}}}$ and $\phi$ did not complete before the invocation of $\pi$. 
 If no such $\pi$ exists in $\EX$, then $(\tg{\pi}, v_{\pi})$ is equal to $(t_0, v_0)$.
 \end{enumerate} \label{def:consistency}
 \end{property}
 
Given that \ARESopt{} use DAP operations that satisfy the conditions in Property~\ref{property:dap}, this allows us to show in Section~\ref{sec: ARES Correctness} that  \ARESopt{} satisfies atomicity. In our implementation of \ARESopt{} we use two DAP algorithms: (i) \ecdapII{} (see Section~\ref{ssec:optimization:piggy-back}), and (ii) \abddapII{}. The main differences of the enhanced algorithms compared 
to their original counterparts that appeared in \cite{ARES}, are the following: (i) each of their \act{read-data} operation may return a $\bot$ value, and (ii) each DAP operation includes in its response a list $Cs$ of the $nextC$ values received from the server's replies (piggyback). As Property~\ref{property:dap} focuses on the tags  
and not the values returned, then the proof that \abddapII{} satisfies Property~\ref{property:dap} is almost identical to the proof for \abddap{} in \cite{ARES}; so we refer the reader to \cite{ARES} for that proof. 
The fact that DAP operations may return $\bot$ however, may affect decodability of the 
value (and thus termination of operations) in \ecdapII{}. Hence in the rest of the section we focus in proving that
\ecdapII{} satisfies Property~\ref{property:dap} as well. 
%
In  particular, to prove the correctness of  \ecdapII{}, we need to show that it is \textit{safe}, i.e., it
ensures Property~\ref{property:dap}, and 
\textit{live}, i.e., it allows each operation 
to terminate. 


For the following
proofs we fix the configuration to $c$ as it 
suffices that the DAPs preserve Property~\ref{property:dap} in any single configuration. Also we assume an $[n,k]$ MDS code~\cite{verapless_book}, $|c.Servers| = n$ of which no more 
than $\frac{n-k}{2}$ may crash. 
We refer to $\delta$ as the
maximum number of $\act{put-data}$ operations concurrent with any $\act{get-data}$ operation.

\begin{lemma}[C2]
\label{lem:p1:c2}
Let $\EX$ be an execution of an algorithm $A$ that
uses the \ecdapII{}.
If $\phi$ is a  $\dagettag{c}$ that returns $\tg{\pi} \in \tsSet$ or a $\dagetdata{c}$ that returns $\tup{\tg{\pi}, v_\pi } \in \tsSet \times \valSet\cup\bot$,
 then there exists $\pi$ such that $\pi$ is a $\daputdata{c}{\tup{\tg{\pi}, v_{\pi}}}$ and $\phi$ did not complete before the invocation of $\pi$. 
 If no such $\pi$ exists in $\EX$, then $(\tg{\pi}, v_{\pi})$ is equal to $(t_0, v_0)$ or $(t_0, \bot)$.
\end{lemma}

\begin{proof}
The proof of property $C2$ of \ecdapII{} when $\pi=\dagettag{c}$ or $\pi=\dagetdata{c}$ and the return value is not $\bot$ is identical to that of \ecdap{} (Theorem 2 in \cite{ARES}). This similarity arises because the initial value of the $List$ variable in each server $s$ in $\mathcal{S}$ remains ${(t_0, \Phi_s(v_{0}))}$, and new tags are added to the $List$ exclusively through $\act{put-data}$ operations.
However, $\act{get-data}$ can return a $\bot$ value in two cases: $(i)$ the value of elements in the $List$ can be set to $\bot$ via a subsequent $\act{gc-config}$ operation, or $(ii)$ when the $\status{nextC}$ of a server is $F$, the server creates a new $List'$ during the $\act{get-data}$ operation, which contains all the tags from $List$ associated with the $\bot$ value (lines Alg.~\ref{algo:server}:\ref{line:server:querylist:list':start}--\ref{line:server:querylist:list':end}).
As a result, during a $\act{get-data}$ operation, each server includes all the tags from the $List$ as before, but now some tags may now have associated $\bot$ value. 
If $\phi$ returns $(t_{\phi}, \bot)$, this means that at least one server returns $\bot$ with $t_{\phi}=t_0$ or $t_{\phi} > t_0$. If $t_{\phi} = t_0$, there is nothing to prove. If $t_{\phi} > t_0$, there is a $\act{put-data}(t_{\pi}, v_{\pi})$ operation $\pi$.
The value $v_{\pi}$ of the pair $\tup{t_{\pi}, v_{\pi}}$ later becomes $\bot$ for any of the two reasons mentioned before.
To show $\phi$ cannot complete before $\pi$ for any $\pi$, we argue by contradiction: if $\phi$ completes before $\pi$ begins for every $\pi$, $t_{\pi}$ cannot be returned by $\phi$, contradicting the assumption.
\end{proof}

\begin{lemma}[C1]
\label{lem:p1:c1}
Let $\EX$ be an execution of an algorithm $A$ that
uses the \ecdapII{}.
If $\phi$ is  $\daputdata{c}{\tup{\tg{\phi}, v_\phi}}$, for $c \in \confSet$, $\tup{\tg{\phi}, v_\phi} \in\tsSet\times\valSet$, 
 and $\pi$ is $\dagetdata{c}$ 
 that returns $\tup{\tg{\pi}, v_{\pi}} \in \tsSet \times \valSet\cup\{\bot\}$ or 
 $\pi$ is $\dagettag{c}$ 
 that returns $\tg{\pi} \in \tsSet$
 and $\phi\bef \pi$ in $\EX$, then $\tg{\pi} \geq \tg{\phi}$.
\end{lemma}

\begin{proof}
Let $p_{\phi}$ and $p_{\pi}$ denote the processes that invoke $\phi$ and $\pi$ in $\EX$. Let $S_{\phi} \subset \mathcal{S}$ denote the set of $\left\lceil \frac{n+k}{2} \right \rceil$
servers that respond to $p_{\phi}$, during $\phi$, and by $S_{\pi}$ the set of $\left\lceil \frac{n+k}{2} \right \rceil$ servers that respond to $p_{\pi}$, during $\pi$.
Per Alg.~\ref{algo:server}:\ref{line:server:insert}, every server $s\in S_\phi$, inserts the tag-value pair 
received by $p_{\phi}$ in its local $List$. Note that once a tag-value pair is added to $List$, is removed only when the $List$ exceeds the length $(\delta+1)$ and the tag of the pair is the smallest in the $List$ (Alg.~\ref{algo:server}:\ref{line:server:if}--\ref{line:server:removemin}). 
Except that the server may remove the tag-value pair from the $List$, it can also set the value of the pair to $\bot$ and retain the tag. 
Notice that as $|S_{\phi}| = |S_{\pi}| =\left\lceil \frac{n+k}{2} \right \rceil $,  then $| S_{\phi} \cap S_{\pi} | \geq k$ reply to both $\pi$ and $\phi$. There are two cases to examine: 
(a) the pair 
$\tup{\tg{\phi}, *}\in List$ of at least $k$ servers in
$S_{\pi}$, and 
(b) the $\tup{\tg{\phi}, *}$ appeared in fewer than $k$ servers in $S_{\pi}$.

\case{a} In this case $\tg{\phi}$ was discovered in at least $k$ servers in $S_\pi$.
This happens since there are not enough concurrent write operations to remove the elements corresponding to tag $\tg{\phi}$.
In the case of $\pi$ being a $\act{get-tag}$ operation, the only difference is the inclusion of $nextC$ in every server reply and their aggregation in a set $Cs$ at the client, then with similar reasoning as in Lemma 19 in \cite{ARES}, we can show that the lemma holds for the $\act{get-tag}$ operation.
In the case of $\pi$ being a $\act{get-data}$ operation, we can break this case in two subcases: $(i)$ no server $s\in S_\pi$ returns $\bot$ associated $\tg{\phi}$ and $(ii)$ at least one server $s\in S_\pi$ returns $\bot$ associated $\tg{\phi}$.
In case $(i)$, $\pi$ discovers $\tg{\phi}$ in at least $k$ servers, ensuring that the value associated with $\tg{\phi}$ will be decodable, as there are no $\bot$ values associated with $\tg{\phi}$ (line Alg.~\ref{algo:casopt}:\ref{line:getdata:decode}). Hence, $t^{dec}_{max}\geq\tg{\phi}$ and $\tg{\pi}\geq\tg{\phi}$.
In case $(ii)$, since at least one server $s\in S_{\pi}$ return $\bot$ value, it indicates that either a $\act{gc-config}$ operation has occurred or $nextC=F$. In the first subcase, servers executing $\act{gc-config}$ set $List$ values to $\bot$, while retaining tags. In the latter subcase, servers in $S_{\pi}$ send all the $List$ with the tags associated with $\bot$ values (Alg.\ref{algo:server}:\ref{line:server:querylist:list':start}-\ref{line:server:querylist:list':end}).
In both cases (possibly both applying), $\tg{\phi}$ was discovered by $k$ servers in $S_\pi$, but there is at least one $\bot$ in their associated elements. Thus, according to Alg.\ref{algo:casopt}:\ref{line:getdata:bot_v}, the value is not decodable, and the client returns $\bot$ during $\act{get-data}$. Per Alg.\ref{algo:casopt}:\ref{line:getdata:max:begin:dec}, $t^{dec}_{max}$ is defined as the set of tags that appear in $k$ Lists with values from the set $\valSet\cup\bot$. Therefore, in this case, $t^{dec}{max}\geq\tg{\phi}$, and thus $\tg{\pi}>\tg{\phi}$.

\case{b} 
In this case $\tg{\phi}$ was discovered in 
less than $k$ servers in $S_\pi$. 
A server $s\in S_{\phi} \cap S_{\pi}$
will not include $\tg{\phi}$ iff $|Lists_s| = \delta+1$,
and therefore the local $List$ of $s$ removed $\tg{\phi}$
as the smallest tag in the list. According to our assumption
though, no more than $\delta$ $\act{put-data}$ operations 
may be concurrent with a $\act{get-data}$ operation. 
Thus, at least one of the $\act{put-data}$ operations
that wrote a tag $\tg{}'\in Lists_s$ must have completed
before $\pi$. Since $\tg{}'$ is also written in $|S'| = \frac{n+k}{2}$ servers then $|S_\pi\cap S'|\geq k$. 
Thus, $\pi$ will be able to find the $\tg{}'$; whereas, when $\pi=\act{get-data}$, $\pi$ will be able to decode the value associated with $\tg{}'$ or return $\bot$. Hence $t^{dec}_{max}\geq\tg{}'$ and 
$\tg{\pi}\geq\tg{\phi}$, completing the proof of this lemma.
\end{proof}

\begin{theorem}[Safety]
    Let $\EX$ be an execution of an algorithm $A$ that contains a set $\Pi$ of complete $\act{get-tag}$/$\act{get-data}$ and $\act{put-data}$ operations of Algorithm~\ref{algo:casopt}. 
    Then every pair of operations $\phi,\pi\in \Pi$ satisfy Property~1.
\end{theorem}

\begin{proof}
Follows directly from Lemmas \ref{lem:p1:c1}  and \ref{lem:p1:c2}. 
\end{proof}



\begin{theorem}[Liveness]
    Let $\EX$ be an execution of an algorithm $A$ 
    that utilises the \ecdapII{}. 
     Then any $\act{get-tag}$, $\act{get-data}$ and $\act{put-data}$ $\op$ invoked in $\EX$ will eventually terminate.
\end{theorem}

\begin{proof}
The only difference between the $\act{get-tag}$ and $\act{put-data}$ operations of \ecdapII{} compared to those of \ecdap{} is that they piggyback a list $Cs$ with the servers' $nextC$ in their replies. Thus, our primary focus now shifts to ensuring the completion of the $\act{get-data}$ operation, as its decodability is affected.
Let $p_{\pi}$ be the process executing a $\act{get-data}$ operation $\pi$. Define $S_{\pi}$ as the set of $\left\lceil \frac{n+k}{2} \right \rceil$ servers responding to $p_{\pi}$.
Let $T_1$ denote the earliest time point when $p_{\pi}$ receives all the $\left\lceil \frac{n+k}{2} \right\rceil$ responses.
Additionally, let $\Lambda$ encompass all $\act{put-data}$ operations starting before $T_1$.
Observe that, by algorithm design, the coded-elements corresponding are garbage-collected from the $List$ variable of a server only if more than $\delta$ higher tags are introduced by subsequent writes into the server.  
According to our assumption
though, no more than $\delta$ $\act{put-data}$ operations 
may be concurrent with a $\act{get-data}$ operation. 
Since $List$ variable has length $\delta+1$, at least one of the $\act{put-data}$ operations
that wrote a tag $\tg{}'\in Lists_s$ must have completed
before $T_1$.
Since $\tg{}'$ is written in $|S'| = \frac{n+k}{2}$ servers then $|S_\pi\cap S'|\geq k$ and hence $\pi$ will be able to find the $\tg{}'$ in $k$ Lists.  
Although the $\act{get-data}$ can find the $t_{max}^{dec}$ (the maximum tag $\pi$ discovered in $k$ Lists), there are two subcases to consider for its accosiated values:
$(a)$ $\pi$ does not receive $\bot$ elements in its replies from any server $s\in S_{\pi}$ and $(b)$ there is at least one server $s \in S_{\pi}$  that returns $\bot$ value. In case $(a)$, since $\pi$ does not receive $\bot$ elements, each server $s\in S_{\pi}$ includes $t_{max}^{dec}$ associated with non-empty coded elements in its replies. Consequently, $t_{max}^{dec}$ must be in $Tag_{dec}^{\geq k}$, and its value is decodable.
In case $(b)$, at least one $\bot$ value associated with $t_{max}^{dec}$ received from any $s \in S_{\pi}$ during the operation $\pi$. As seen above, a server can return $\bot$ in two cases: $(i)$ if the $\status{nextC}$ of a server is $F$, or $(ii)$ the value in its $List$ set to $\bot$ via a subsequent $\act{gc-config}$ operation. While in case $(i)$ it is clear that the $\pi$ can detect the $\bot$ value from any server $s \in S_{\pi}$, we should prove it for the case $(ii)$.  
Let $p_{\gamma}$ be the process that invokes the $\act{gc-config}$ operation $\gamma$ . Define $S_{\gamma}$ as the set of $\left\lceil \frac{n+k}{2} \right \rceil$ servers responding to $p_{\gamma}$ in the $\act{gc-config}$ operation during $\gamma$. 
Notably, at execution point $T_0$ in $\EX$, just before the completion of $\pi$, the garbage collection operation $\gamma$ is initiated for the configuration consisting of the set of servers in $S_{\gamma}$.
As per the algorithm design, the elements corresponding to the configuration of $S_{\gamma}$ set the values in their $List$ to $\bot$ during garbage collection (lines Alg.~\ref{algo:server}:\ref{line:server:gc-config:forBot:start}--\ref{line:server:gc-config:forBot:end}). Hence, between execution points $T_0$ and $T_1$ in $\EX$, the elements of every active server $s\in S_{\gamma}$ may undergo garbage collection.
Since $|S_{\gamma}| = |S_{\pi}| = \left\lceil \frac{n+k}{2} \right \rceil $ and $|S_{\gamma} \cap S_{\pi} | \geq k$, a $\bot$ value can be detected by $p_{\pi}$ from at least one server. Since $\act{get-data}$ detects at least one $\bot$ element, it will not attempt to decode the value. Instead, it returns a $\bot$ value (line Alg.~\ref{algo:casopt}:\ref{line:getdata:bot_v}).
\end{proof}

\subsection{Correctness of \ARESopt} 
\label{sec: ARES Correctness}
The correctness of \ares{} (Section 6 of ~\cite{ARES}) highly depends on the way the configuration 
sequence is constructed at each client process.
Also, atomicity is ensured if the DAP implementation in each configuration $c_i$
satisfies Property~\ref{property:dap}.

This work involves modifications on the reconfiguration aspect of \ares{}, and in particular it changes the way the configuration sequence is constructed at each client process. In this section we show that the modifications proposed in \ARESopt{} 
do not violate the correctness of the algorithm. 

The changes in \ARESopt{} have direct implications on the
reconfiguration properties of \ares{} (cf. Section 6.1 of ~\cite{ARES}). 
In \ares{}, the configuration sequence maintained in two processes is either the same or one is the prefix of the other. In \ARESopt{} this changes: the configuration sequence maintained in two processes is either the same or the one is \emph{subsequence} of the other with respect to their indices. 

In the following subsections we first present the configuration properties
and prove their satisfaction by \ARESopt{}, and then we show that given those properties
we can prove the correctness of \ARESopt{}. We proceed by introducing some definitions and notation we use in the proofs.

\begin{table}[!h]
    \centering
    {\small
        \begin{tabular}{p{4cm} p{9cm}}
            \toprule                        
            $\cvec{\pr}{\state}$ & the value of the configuration sequence variable $cseq$ at process $p$ in state $\st$, i.e. a shorthand of $\atT{\pr.cseq}{\state}$\\
            $\cvec{\pr}{\state}[i]$ & the element with the index $i$ in the configuration sequence $\cvec{\pr}{\state}$\\
            $\mu(\cvec{\pr}{\state})$ & last finalized configuration in $\cvec{\pr}{\state}$\\
            $\lambda(\cvec{\pr}{\state})$ & the largest index $i$ in the configuration sequence $\cvec{\pr}{\state}$, s.t. $\config{\cvec{\pr}{\state}[i]}\in\confSet$\\   
            \bottomrule
        \end{tabular}
    }
    \caption{Notations.}
    \label{tab:recon:notation}
\end{table}

Last, we define the notion of subsequence on two configuration sequences.

\begin{definition} [Subsequence]
A configuration sequence $x$ is a \emph{subsequence} of a sequence $y$ 
if $\lambda(x)\leq\lambda(y) $.
Additionally, for any index $j$, if 
$\config{x[j]}\in\confSet$ and $\config{y[j]}\in\confSet$, then $\config{x[j]}=\config{y[j]}$.
\end{definition}

\subsubsection{Reconfiguration Protocol Properties}
In this section we analyze the properties that we can achieve through our reconfiguration algorithm. In high-level, we do show that the following properties are preserved: 
\begin{itemize} 
\item[i] {\bf configuration uniqueness:} the configuration sequences in any two processes have identical configuration at any common index $i$,
\item[ii] {\bf subsequence:} the configuration sequence observed by an operation is a subsequence of the sequence observed by any subsequent operation, and 
\item[iii] {\bf sequence progress:} if the configuration with index $i$ is finalized during an operation, then a configuration $j$, for $j\geq i$, will be finalized
in a succeeding operation.
\end{itemize}

\begin{lemma}
\label{lem:consconf}
	For any reconfigurer $r\in\recSet$ that invokes an $\act{reconfig}(c)$ operation in an execution $\EX$ 
	of the algorithm, If $r$ chooses to install $c$ in index $k$ of its local $r.cseq$ vector, then $r$ invokes 
	the $Cons.propose(k,c)$ and $\lambda(\cvec{r}{\st})=k-1$ where $\st$ the state of $r$
 at the completion of \act{read-config} procedure.
\end{lemma}

\begin{proof}
	This Lemma follows directly from Alg.~\ref{algo:reconfigurer} and the \act{read-config} procedure in Alg.~\ref{algo:parser}. Notice that the reconfigurer traverses to a 
 configuration with index say $k-1$ and then proposes the new configuration to be installed on the next index, i.e. $k$.
\end{proof}

\begin{lemma}
	\label{lem:server:monotonic}
	If a server $s$ sets $s.nextC$ to $\tup{c,*}$ with index $c.ID=i$ at some state $\st$ in an execution $\EX$ 
	of the algorithm, then $s.nextC$ will set to $\tup{c',*}$ with an index $c'.ID\geq i$ for any state $\st'$ that appears after $\st$ in $\EX$.
\end{lemma}

\begin{proof}
Notice that a server $s$ updates its $s.nextC$ variable for some specific configuration $c_k$ in a state $\state$ when it receives: $(i)$ a {\sc gc-config} message or $(ii)$ a {\sc write-confing} message. 
In case $(i)$, the $s.\config{nextC}=c$ of $c_k$ has index $c.ID=i$ and the {\sc gc-config} message received contains a tuple $\tup{c',F}$. This $\act{gc-config}$ action is initiated by a reconfigurer $r$ which wants to propagate $c'$ before state $\st'$. At first, $r$ executes $\act{read-config}$ and among other configurations it detects the $c$ (i.e., $r.cseq[i]$). By Alg~\ref{algo:reconfigurer}:\ref{gc-config}, $r$ sends a {\sc gc-config} message to the servers of all configurations in its $r.cseq$ where their index is smaller than $c'.ID$ (including $r.cseq[i]$) in order to propagate to them the proposed $r.cseq[c'.ID]$.
By Alg.~\ref{algo:server}:\ref{line:server:gc-config:is_lastbegin}, server $s\in c_k$ updates its local $nextC$ only when the  received $c'$ has larger index than the local $nextC$. Thus $c'.ID > i$.
In case $(ii)$, the {\sc write-config} message is either the first one received at $s$ for $c_k$ (and thus $s.nextC=\bot$), or $s.nextC = \tup{c,*}$ and the message received contains a tuple $\tup{c',F}$.
There are only two cases where $s$ can modify its $s.nextC$: $(i)$ its status is finalized by a reconfigurer, or $(ii)$ a $\act{gc-config}$ updates the $s.nextC$ to $c'$ with index $c'.ID>c.ID$ and a DAP operation propagates $c'$ through a {\sc write-config} message to a quorum. In case $(i)$, $c=c'$ and the $\status(c)=P$ hence $c.ID=c'.ID$, while in case $(ii)$ $c'.ID > c.ID$ as shown above and the status of $c$ can be either $P$ or $F$.
\end{proof}

\begin{lemma}[Configuration Uniqueness]
\label{lem:unique}
For any processes $\pr, q\in \idSet$ and any states $\st_1, \st_2$ in an execution $\EX$, it must hold that 
	$\config{\cvec{\pr}{\st_1}[i]}=\config{\cvec{q}{\st_2}[i]}$,  $\forall i$ s.t. 
	$\config{\cvec{\pr}{\st_1}[i]},\config{\cvec{q}{\st_2}[i]}\in \confSet$.
\end{lemma}
\begin{proof}
The lemma holds trivially for index \(i = 0\) such that $\config{\cvec{\pr}{\st_1}[0]}=\config{\cvec{q}{\st_2}[0]}=c_0$. 
	So in the rest of the proof we focus in the case where index $i > 0$. Let us assume 
	w.l.o.g. that $\st_1$ appears before $\st_2$ in $\EX$.

According to our algorithm a process $\pr$ sets $\pr.\config{cseq(i)}$ to a configuration $c$ with index $i$ in two cases: $(a)$ either it received $c$ as the result of invoking a propose operation on index $i$ to the consensus external service (Alg.~\ref{algo:reconfigurer}:~\ref{line:addconfig:consensus}), or $(b)$ $\pr$ receives $\config{s.nextC} = c$ from 
	a server through a DAP operation (Alg.~\ref{algo:read_writeProtocol}:\ref{line:writer:p1:get-tag}\&\ref{line:writer:prop}\&\ref{line:reader:p1:get-data}\&\ref{line:reader:prop}) or a $\act{read-config}$ (Alg.~\ref{algo:reconfigurer}:\ref{line:install:readconfig}).
 Note here that $(a)$ is possible only 
	when $\pr$ is a reconfigurer and attempts to install a new configuration while its latest discovered configuration is $\pr.\config{cseq(i-1)}$, by Lemma~\ref{lem:consconf}. On the 
	other hand $(b)$ may be executed by any process in any operation that reads the $nextC$ of some server $\in\servers{\pr.\config{cseq(k)}}$, where $k\leq i-1$ (i.e., in either $\act{get-tag}$ or $\act{get-data}$ or $\act{put-data}$ or $\act{read-config}$). 
    By Lemma~\ref{lem:server:monotonic}, the $nextC$ at each server is monotonic, thus the index of $s.\config{nextC}$ always increases or remains the same.  
    
    We are going 
	to prove this lemma by induction on the configuration index. 

 \emph{Base case:} The base case of the lemma is when $i=1$.
 Let us first assume that $p$ and $q$ receive $c_p$ and $c_q$, as the result of the consensus at index $i=0$.
 As per Alg.~\ref{algo:reconfigurer}, a reconfigurer proposes configurations at a specific index $i$ by sending a request to a consensus external service and receiving the decided configuration $c_{i+1}$ with index $i+1$ in response (line Alg.~\ref{algo:reconfigurer},~\ref{line:addconfig:consensus}). 
 By Lemma \ref{lem:consconf}, since both processes want to install a configuration 
	in $i=1$, then they have to run the external consensus service on the index $0$.
	Since $\lambda(\cvec{\pr}{\st_1})=\lambda(\cvec{q}{\st_2})=0$, both processes have to run the consensus on the same index $i=0$. Therefore, by the agreement property, they have to decide on the same configuration with the index $i=1$. Consequently, $c_p=c_q=c_1$ and $\config{\cvec{\pr}{\st_1}[1]}=\config{\cvec{q}{\st_2}[1]}=c_1$.
	 
	 Let us examine the case now where $p$ or $q$ 
	assign a configuration $c$ they received from some server $s\in\servers{c_0}$. 
 According to the algorithm, either
	the configuration that has been decided by the consensus instance on 
	index $i=0$ is propagated to the servers in $\servers{c_0}$, or a configuration that is propagated to the $nextC$ of servers in $\servers{c_0}$ by a $\act{gc-config}$ operation.
 In the first case, If $c_1$ is the decided configuration, then 
	$\forall s\in\servers{c_0}$ such that $s.nextC(c_0)\neq\bot$, it holds that $s.nextC(c_0) = \tup{c_1,*}$.
	So if $p$ or $q$ set $\config{\cvec{\pr}{\st_1}[1]}$ or $\config{\cvec{q}{\st_2}[1]}$ to some received configuration, then 
	$\config{\cvec{\pr}{\st_1}[1]}=\config{\cvec{q}{\st_2}[1]}=c_1$ in this case as well. 
In the second case, the $\act{gc-config}$ is executed by a reconfigurer which propagates the finalized proposed configuration $c$ to all the configurations before that. The configuration $c$ can be $c_1$ or a subsequent finalized configuration. Thus if both $\pr$ or $q$ receives $c_1$, it holds that $s.nextC(c_0) = \tup{c_1,F}$, and hence $\config{\cvec{\pr}{\st_1}[1]}=\config{\cvec{q}{\st_2}[1]}=c_1$. However, if one of them receives a subsequent configuration $\neq c_1$, due to a change of $nextC$ to a configuration with a larger index (Lemma~\ref{lem:server:monotonic}), it implies that its $cseq$ sequence at index $1$ is denoted as $\config{cseq[1]}=\bot\notin\confSet$. 
This contracted our hypothesis, which assumes that $\config{\cvec{\pr}{\st_1}[1]},\config{\cvec{q}{\st_2}[1]}\in\confSet$.
	
     \emph{Hypothesis:} We assume  that $\cvec{\pr}{\st_1}[k]=\cvec{q}{\st_2}[k]\neq \bot$  for some $k$, $k \geq 1$.
 
	\emph{Induction Step:} We need to show that the lemma holds for some index $i \geq k+1$, where $i$ is the first index after the index $k$ where $\cvec{\pr}{\st_1}[i]=\cvec{q}{\st_2}[i]\neq\bot$.
 Let's break down the problem into two subcases: $(i)$ $i=k+1$ and $(ii)$ $i>k+1$.
 \emph{Case (i):}
	If both processes retrieve index $i$ with $\config{\cvec{\pr}{\st_1}[k+1]}$ and $\config{\cvec{q}{\st_2}[k+1]}$ respectively through consensus, 
	then both $\pr$ and $q$ run consensus
	on the previous index $k$. Since according to our hypothesis the index is $k$ where
	$\cvec{\pr}{\st_1}[k]=\cvec{q}{\st_2}[k]\neq \bot$ then both processes will receive the same decided value for index $k+1$, say $c_{k+1}$, and hence $\config{\cvec{\pr}{\st_1}[k+1]}=\config{\cvec{q}{\st_2}[k+1]}=c_{k+1}$. Similar to the base case,
	a server in $\servers{c_k}$ only receives the configuration $c_{k+1}$ decided by the consensus run on index $k$. 
	
 \emph{Case (ii):}
Now, consider the case where $\pr$ and $q$ receive a configuration with index $i>k+1$. 
The two processes can receive $i$ through $nextC$ of some server s in $\servers{c_j}$ where $j<i$. This happens since a reconfiguration operation adds the index $i$, then executes the $\act{gc-config}$ operation on every configuration with an index smaller than $i$ and updates the
$nextC$ of their servers to point to the configuration with index $i$ (line Alg.~\ref{algo:server}:\ref{line:servers:nextCupdate}). 
Thus if both processes $\pr$ and $q$ receive $i$, then $\config{\cvec{\pr}{\st_1}[i]}=\config{\cvec{q}{\st_2}[i]}=c_{i}$. However, if one of them does not receive index $i$, i.e. $\config{cseq[i]}=\bot\notin\confSet$ (due to a change of $nextC$ to a configuration with a larger index according to Lemma~\ref{lem:server:monotonic}), it implies that its $cseq$ sequence at index $i$ is denoted as $\config{cseq[i]}=\bot\notin\confSet$. 
This contracted our hypothesis, which assumes that $\config{\cvec{\pr}{\st_1}[i]},\config{\cvec{q}{\st_2}[i]}\in\confSet$.

\end{proof}

Lemma \ref{lem:unique} showed that any two operations store the same configuration in any cell $k$ of their $cseq$ variable. However, whether the two processes discover the same maximum configuration id is still uncertain. In the following lemmas, we will demonstrate that if a process learns about a configuration in a cell $k$ (and it is not its first configuration), it also learns about some configuration ids for some indices 
$i$ such that $0\leq i\leq k-1$. Notably, while the number of configurations may differ between processes, the subsequences must have the same maximum tag or a smaller one.




\begin{lemma}
	\label{lem:config:propagation}
	If at a state $\st$ of an execution $\EX$ of the algorithm $\lambda(\cvec{\pr}{\st}) = k$ 
	for some process $\pr$, then for any element $0\leq j < k$, $\exists Q\in \quorums{\config{\cvec{\pr}{\st}[j]}}$
	such that $\forall s\in Q, s.nextC(\config{\cvec{\pr}{\st}[j]})= \cvec{\pr}{\st}[i]$, for some $i \in [j+1,k]$. 
\end{lemma}
\begin{proof}
    Similar to Lemma 14 in \cite{ARES} we can show that this lemma holds if for every $s\in Q$, $s.nextC(\config{\cvec{\pr}{\st}[j]})= \cvec{\pr}{\st}[j+1]$. 
    As the $\act{gc-config}$ changes the $nextC$ of servers, we need to show that no 
    server $s\in Q$ will have $s.nextC(\config{\cvec{\pr}{\st}[j]})= \cvec{\pr}{\st}[i]$ 
    for $i<j+1$. This follows from the implementation of the $\act{gc-config}$ as well as
    from Lemma \ref{lem:server:monotonic}. 
    In particular, whenever a reconfigurer executes a $\act{gc-config}$ 
    propagates the latest finalized configuration to a quorum of each 
	previous configuration in its sequence (line Alg~\ref{algo:reconfigurer}:\ref{gc-config}).
    By Lemma~\ref{lem:server:monotonic}, a server updates its $nextC$ to point to a configuration
    with a larger index than the one it holds. Therefore, since a server $s$ in a configuration $\config{\cvec{\pr}{\st'}[j]}$ sets $s.nextC(\config{\cvec{\pr}{\st'}[j]})= \cvec{\pr}{\st'}[j+1]$ at some state $\st'$ before $\st$ then by Lemma~\ref{lem:server:monotonic} it
    can only have $s.nextC(\config{\cvec{\pr}{\st}[j]})= \cvec{\pr}{\st}[k]$, for $k\geq j+1$ in state $\st$. This completes the proof.
\end{proof}

\begin{lemma}[Subsequence]
	\label{lem:subsequence}
	Let $\op_1$ and $\op_2$ be two completed read/write/reconfig operations invoked by processes $\pr_1, \pr_2 \in \idSet$ respectively, such that $\op_1 \bef \op_2$ in an execution $\EX$. Let $\st_1$ be the state after the response step of $\op_1$, and $\st_2$ be the state after the response step of $\op_2$. Then $\cvec{\pr_1}{\st_1} \sqsubseteq_p \cvec{\pr_2}{\st_2}$.
\end{lemma}

\begin{proof}
 By Lemma \ref{lem:unique}, for any $i$ such that $\cvec{\pr_1}{\st_1}[i] \neq \bot$ and $\cvec{\pr_2}{\st_2}[i] \neq \bot$, then $\config{\cvec{\pr_1}{\st_1}[i]} = \config{\cvec{\pr_2}{\st_2}[i]}$. So it remains to show that $\lambda_1 \leq \lambda_2$.


Let $\lambda_1 = \lambda(\cvec{\pr_1}{\st_1})$ and $\lambda_2 = \lambda(\cvec{\pr_2}{\st_2})$. Since $\op_1 \bef \op_2$, it follows that $\st_1$ appears before $\st_2$ in $\EX$.
Let $\mu = \mu(\cvec{\pr_2}{\st'})$ be the last finalized element that $\pr_2$ established during operation $\op_2$ at some state $\st'$ before $\st_2$. It is easy to see that $\mu \leq \lambda_2$. If $\lambda_1 \leq \mu$, then $\lambda_1 \leq \lambda_2$, and the lemma follows. Thus, it remains to examine the case where $\mu < \lambda_1$. Notice that since $\op_1 \bef \op_2$, then $\st_1$ appears before $\st'$ in execution $\EX$. By Lemma \ref{lem:config:propagation}, we know that by $\st_1$, there exists $Q\in\quorums{\config{\cvec{\pr_1}{\st_1}[j]}}$ for $0 \leq j < \lambda_1$ and for each $s\in Q$, $s.\text{nextC} = \cvec{\pr_1}{\st_1}[i]$, for some $i \geq j + 1$, possibly different for every $s$. Since $\mu < \lambda_1$, then it must be the case that $\exists Q\in \quorums{\config{\cvec{\pr_1}{\st_1}[\mu]}}$ such that $\forall s\in Q$, $s.\text{nextC} = \cvec{\pr_1}{\st_1}[i]$, for $i \geq \mu + 1$. Let $\pr_2$ 
as the outcome of an action (either \act{get-next-config} or DAP with piggyback) to read the configuration from a quorum $Q'\in\config{\cvec{\pr_2}{*}[\mu]}$ during $\op_2$,at a state $\st''$. Since by Lemma~\ref{lem:unique}, $\config{\cvec{\pr_2}{*}[\mu]}=\config{\cvec{\pr_1}{\st_1}[\mu]}$, the $Q$ and $Q'$ belong to the same configuration and thus by definition $Q'\cap Q\neq \emptyset$. Therefore, there exists a server $s\in Q\cap Q'$ which by Lemma~\ref{lem:server:monotonic}, replies to $\pr_2$ with either $s.nextC = \cvec{\pr_1}{\st_1}[\mu+1]$, or with $s.nextC = \cvec{r}{*}[j]$ for a $j > \mu+1$. If $j<\lambda_1$ then 
by a simple induction we can show that the process will be repeated in every configuration
with index $k<\lambda_1$ until we reach at least $\lambda_1$. In that case $\lambda_2\geq\lambda_1$. In case where $j\geq \lambda_1$ then $\pr_2$ will set 
$\cvec{\pr_2}{\st''}[j] = \cvec{r}{*}[j]$. Since $\st_2$ comes after $\st''$, then $\lambda_2\geq j >\lambda_1$ and this completes the proof.


\end{proof}

\begin{lemma}
	\label{lem:final:monotonic}
	Let $\st$ and $\st'$ two states in an execution $\EX$ such that $\st$ appears before $\st'$ in $\EX$.
 	Then for any process $\pr$, that executes a $\act{read-config}$ or a DAP action, must hold that $\mu(\cvec{\pr}{\st})\leq \mu(\cvec{\pr}{\st'})$.  
\end{lemma}

\begin{proof}
	This lemma follows from the fact that if a configuration $k$ is such that 
	$\status{\cvec{\pr}{\st}[k]}=F$ at a state $\st$, then $\pr$ will start any 
	future $\act{read-config}$ or DAP action from a configuration $\config{\cvec{\pr}{\st'}[j]}$
	such that $j\geq k$. 
    The $\config{\cvec{\pr}{\st'}[j]}$ is the last finalized configuration at $\st'$ which is added by a subsequent reconfiguration. Hence, $\mu(\cvec{\pr}{\st'})\geq \mu(\cvec{\pr}{\st})$.
\end{proof}

\begin{lemma}  [Sequence Progress]
	\label{lem:finalconf}
	Let $\op_1$ and $\op_2$ two 
	completed 
 read/write/reconfig operations invoked by processes $\pr_1, \pr_2\in\idSet$ 
	respectively, such that $\op_1\bef\op_2$ in an execution $\EX$. 
	Let $\st_1$ be the state after the response 
	step of $\op_1$ and $\st_2$ the state after the response step 
	of $\op_2$. Then 
	$\mu(\cvec{\pr_1}{\st_1})\leq\mu(\cvec{\pr_2}{\st_2})$.
\end{lemma}


\begin{proof}
	By Lemma \ref{lem:subsequence} it follows that $\cvec{\pr_1}{\st_1}$ is a subsequence of $\cvec{\pr_2}{\st_2}$.
	Thus, if $\lambda_1 = \lambda(\cvec{\pr_1}{\st_1})$ and $\lambda_2 = \lambda(\cvec{\pr_2}{\st_2})$, $\lambda_1\leq\lambda_2$.
	Let $\mu_1=\mu(\cvec{\pr_1}{\st_1})$, such that $\mu_1\leq\lambda_1$, be the last element in $\cvec{\pr_1}{\st_1}$
	where $\status{\cvec{\pr_1}{\st_1}[\mu_1]} = F$. Let now $\mu_2=\mu(\cvec{\pr_2}{\st'})$, 
	be the last element which $\pr_2$ obtained 
	during $\op_2$ 
	such that $\status{\cvec{\pr_2}{\st'}[\mu_2]} = F$ in some state $\st'$ before $\st_2$. 
	If $\mu_2\geq\mu_1$, and since $\st_2$ is after $\st'$, then by Lemma \ref{lem:final:monotonic} 
	$\mu_2\leq \mu(\cvec{\pr_2}{\st_2})$ and hence $\mu_1\leq \mu(\cvec{\pr_2}{\st_2})$ as well. 
	
	It remains to examine the case where $\mu_2<\mu_1$. Process  $\pr_1$ 
	sets the status of $\cvec{\pr_1}{\st_1}[\mu_1]$ to $F$ in two cases: $(i)$ either when finalizing a reconfiguration, or $(ii)$ when receiving an $s.nextC = \tup{\config{\cvec{\pr_1}{\st_1}[\mu_1]}, F}$ 
	from some server $s$ during a $\act{read-config}$ or a DAP action. 
    
    \case{(i)} In case $(i)$ $\pr_1$ propagates the $\tup{\config{\cvec{\pr_1}{\st_1}[\mu_1]}, F}$ to a quorum of servers in every $\config{\cvec{\pr_1}{\st_1}[j]}$, where $j\leq \mu_1-1$, before completing, using the \act{gc-config}. 
    Thus, a quorum of servers in every $\config{\cvec{\pr_2}{\st_2}[j]}$ ($j\leq \mu_1-1$), say $Q$, 
    receives $\tup{\config{\cvec{\pr_1}{\st_1}[\mu_1]}, F}$ from $\pr_1$.
     We know by Lemma \ref{lem:subsequence} that since $\op_1\bef\op_2$ then $\cvec{\pr_1}{\st_1}$ is a subsequence of the $\cvec{\pr_2}{\st_2}$, it must be the case that $\mu_2 < \mu_1 \leq \lambda_2$. 
 Thus, during $\op_2$, $\pr_2$ starts from the configuration at index $\mu_2$ and in some iteration performs $\act{get-next-config}$ or DAP operation in a configuration $\cvec{\pr_2}{\st_2}[j]$, where $j \leq \mu_1-1$. 
	Since $\op_1$ completed before $\op_2$, then it must be the case that $\st_1$ appears before $\st'$ in $\EX$. However, $\pr_2$ invokes the $\act{get-next-config}$ or DAP in a state $\st''$
	which is either equal to $\st'$ or appears after $\st'$ in $\EX$. Thus, $\st''$ must appear after $\st_1$ in $\EX$.
	From that it follows that when the $\act{get-next-config}$ or DAP is executed by $\pr_2$ there is already 
	a quorum of servers in every $\config{\cvec{\pr_2}{\st_2}[j]}$ ($j\leq \mu_1-1$)
    that received $\tup{\config{\cvec{\pr_1}{\st_1}[\mu_1]}, F}$ from $\pr_1$.
	Since, $\pr_2$ waits from replies from a quorum of servers from the  configuration with index $j$, say $Q'$, then by Lemma~\ref{lem:server:monotonic}
  there is a server $s\in \quo\cap \quo'$, such that $s$ replies to $\pr_2$ with $s.nextC = \tup{\config{\cvec{\pr_1}{\st_1}[\mu_1]}, F}$ or a server $s\in \quo'$ replies to $\pr_2$ with $s.nextC = \tup{\config{\cvec{r}{*}[j]}, F}$ where $r$ a reconfigurer that propagated the configuration in a \act{gc-config} action and $j>\mu_1$. Hence, $\mu(\cvec{\pr_2}{\st_2})\geq \mu_1$ in this case.
  
  \case{(ii)} In this case  $\pr_1$ sets the status of $\cvec{\pr_1}{\st_1}[\mu_1]$ to $F$
  when receiving an $s.nextC = \tup{\config{\cvec{\pr_1}{\st_1}[\mu_1]}, F}$ 
	from some server $s$ during a $\act{read-config}$ or a DAP action. Process $\pr_1$ propagates $\tup{\config{\cvec{\pr_1}{\st_1}[\mu_1]}, F}$ using the action \act{put-config} to a quorum of servers in configuration $\config{\cvec{\pr_1}{\st_1}}$. Similar 
 to Case (i) it holds that $\mu_2 < \mu_1 \leq\lambda_2$. So at $\st_2$, $\cvec{\pr_2}{\st_2}[\lambda_2]$ is the last configuration in $\cvec{\pr_2}{\st_2}$. By our algorithm $\pr_2$ discovers this configuration either by traversing the configurations from index $\mu_2$ to $\lambda_2$, or it discovers $\cvec{\pr_2}{\st_2}[\lambda_2]$ from a configuration $\config{\cvec{\pr_2}{\st_2}[j]}$ for $j<\lambda_2-1$. In the latter case, it must hold that $\cvec{\pr_2}{\st_2}[\lambda_2].status = F$ since the $\act{gc-config}$ action only propagates finalized configurations. Hence, in this case $\mu_2 = \lambda_2 > \mu_1$. By Lemma~\ref{lem:config:propagation}
 every configuration $\cvec{\pr_2}{\st_2}[j]$ has at least one quorum whose servers
 have at least $s.nextC = \cvec{\pr_2}{\st_2}[j+1]$. So in
 the case of traversing the configurations $\pr_2$ will 
 either perform a \act{get-next-config} or DAP action on 
 $\config{\cvec{\pr_2}{\mu_1-1}}$ or it will avoid $\mu_1$ 
 as it will discover a finalized configuration $k > \mu_1$.
 In the latter case $\mu_2 \leq k  >\mu_1$. In the initial
 case $\pr_2$ will access $\config{\cvec{\pr_1}{\st_1}}$ since  by Lemma~\ref{lem:unique}, $\cvec{\pr_1}{\st_1}[\mu_1-1]=\cvec{\pr_2}{\st_2}[\mu_1-1]$ if both are non empty. We know that $\pr_1$ propagated $\tup{\config{\cvec{\pr_1}{\st_1}[\mu_1]}, F}$ to 
 $\config{\cvec{\pr_1}{\st_1}[\mu_1-1]}$ before completing
 (Algo.\ref{algo:parser}:\ref{line:find-next-config:update:if2:put-config}). $\pr_2$ 
 executes \act{get-next-config} or DAP on $\config{\cvec{\pr_1}{\st_1}[\mu_1-1]}$. Thus, some server
 in $Q$ will reply to $\pr_2$ with $s.nextC = \cvec{\pr_1}{\st_1}{\mu_1}$ which is finalized. Thus, $\mu_2\geq\mu_1$ in this case as well.  
\end{proof}

Using the previous Lemmas we can conclude to the main result of this section.

\begin{theorem}
\label{thm:recon:properties}
	Let $\op_1$ and $\op_2$ two 
completed \act{read-config} or DAP actions invoked by processes $\pr_1, \pr_2\in\idSet$ 
respectively, such that $\op_1\bef\op_2$ in an execution $\EX$. 
Let $\st_1$ be the state after the response 
step of $\op_1$ and $\st_2$ the state after the response step of $\op_2$.

Then the following properties hold: 
\begin{enumerate}
\item [$(a)$] 
\textbf{Configuration Consistency}: $\config{\cvec{\pr_2}{\st_2}[i]} = \config{\cvec{\pr_1}{\st_1}[i]}$,  for $ 1 \leq i \leq \nu(\cvec{\pr_1}{\st_1})$,
\item [$(b)$]
 \textbf{Subsequence}: 
 $\cvec{\pr_1}{\st_1}  \sqsubseteq_p \cvec{\pr_2}{\st_2}$, and
\item [$(c)$] 
\textbf{Sequence Progress}:
  $\mu(\cvec{\pr_1}{\st_1}) \leq \mu(\cvec{\pr_2}{\st_2})$
\end{enumerate}
\end{theorem}

\begin{proof}
Statements $(a)$, $(b)$ and $(c)$ follow from Lemmas \ref{lem:unique}, \ref{lem:subsequence}, and 
\ref{lem:finalconf}.
\end{proof}

\subsubsection{Atomicity Property of \ARESopt{}}

\ARESopt{} is correct if it satisfies {\em liveness} (termination) and {\em safety} (i.e., linearizability).
\begin{lemma}
\label{lemma:ARES:termination}
    Every read/write/reconfig operation  terminates in any 
    execution $\EX$ of \ARESopt{}.
\end{lemma}

\begin{proof}
    Termination holds since read, write and reconfig operations on \ARESopt{} always complete given that the DAP completes.
\end{proof}


\begin{lemma} 
\label{lemma:ARES:PB&GC}
     In any execution $\EX$ of \ARESopt{}, if $\op_\rd$ a read operation 
     invoked by some process $\pr$ that appears in $\EX$, then $\op_\rd$ returns
     a value $v\neq\bot$.
     
\end{lemma}

\begin{proof} 
In this proof we need to show that any read operation $\op_\rd$ in $\EX$ will decode and return some value $v\neq\bot$. But the read $\op_\rd$, may 
receive $\bot$, as the returned values of \act{get-data} actions, which 
the read calls repeatedly \act{get-data} until it cannot discover a new configuration. 

Let's assume by contradiction that $\pr_\rd$ may return a value $\bot$.
The value returned by $\pr_rd$ is specified by Line~\ref{line:reader:p1:maxpair} when the reader
discovers the max $\tup{\tg{max},v_{max}}$ pair s.t. $v_{max}\neq\bot$ out of all pairs returned by the $\act{get-data}$ operations it invoked. So in order for $\pr_\rd$ to return $\bot$
then it must be the case that all $\act{get-data}$ actions invoked returned 
a $\bot$ value. 

A $c.\act{get-data}$ action, invoked in a configuration $c$, returns a $\bot$ value only if it receives a $\bot$ value in some server reply from a quorum 
$Q$ in $c.\srvSet$. A server $s\in Q$ in turn, may return a $\bot$ value in two cases:
(i) its variable $s.nextC(c).status = F$ Line~\ref{line:server:querylist-nextCstatus:start}, or (ii) it received a message
from a \act{gc-collect} action Lines Alg.~\ref{algo:server}:\ref{line:server:gc-config:forBot:start}--\ref{line:server:gc-config:forBot:end}. Notice, that in both cases
$s.nextC(c).status = F$, as the \act{gc-collect} action propagates a finalized 
configuration. So, assuming that $c$ has an index $i$, then the $s.nextC(c)=\tup{c', F}$ for some $c'$ with index $j>i$. 

Action $c.\act{get-data}$ will include $c'$ in the set of configurations that 
it returns. Therefore, the set $Cs$ observed by the reader will contain 
at least $c'$, and hence the reader will repeat $c'.\act{get-data}$ in $c'$.
Let w.l.o.g $c'$ be the last finalized configuration in the introduced 
list of configurations $\mathcal{G}_L$. There are two cases to examine: 
(i) there is no other configuration after $c'$, or (ii) there exists 
some pending configuration $c''$ after $c'$. 

In the case where $c'$ is the last configuration then for every server 
$s'\in c'.\srvSet$, $s'.nextC=\bot$. Thus, every $s'$ replies to 
$c'.\act{get-data}$ with a value $v\neq\bot$ ($v$ will be at least $v_0$).
Since we cannot have more than $\delta$ concurrent $\act{put-data}$ 
similar to Lemma~\ref{lem:p1:c1} we will find and return 
a decodable value say $v'$. So the reader will find $v'\neq\bot$ and this 
case contradicts our initial assumption. 

So it remains the case where there exists a pending configuration $c''$
after $c'$. In this case the $Cs$ set at the reader will not be empty and 
it will repeat $c''.\act{get-data}$ operation in $c''$. As $c'$ is the last finalized configuration then no server $s''\in c''.\srvSet$ will have a 
$nextC$ variable with $s''.nextC(c'').status = F$. As $c''$ is still pending
then there exists a concurrent reconfigurer that tries to install $c''$.
So we have two cases when $c''.\act{get-data}$ is invoked: (i) the 
reconfigurer managed to write the latest tag-value pair in a quorum in $c''.Quorums$, or (ii) not. In the first case the $c''.\act{get-data}$ 
action will find the written value in at least $k$ lists and thus 
will decode and return a value $v''$ to the reader. In the second 
case the servers in $c''.\srvSet$ will return at least the initial
value $v_0$ so the action $c''.\act{get-data}$ will be able to decode and 
return at least $\tup{\tg{0}, v_0}$. In any of those cases the reader will 
return either $v'$, $v''$, or $v_0$ based on which tag of those values 
is greater. Since any of those values belong in $\vSet$ they are different
than $\bot$ and they also contradict our initial assumption. This completes 
our proof.

\end{proof}

\begin{theorem}
    In any execution $\EX$ of \ARESopt{}, if in every configuration $c\in G_L$,
 $c.\act{get-data}()$, $c.\act{put-data}()$, and $c.\act{get-tag}()$ satisfy Property 1, then \ARESopt{} satisfies atomicity.
\end{theorem}

\begin{proof}
As shown in \cite{ARES}, \ares{} implements
a linearizable object given that the DAP used satisfy Property~\ref{property:dap}.
In \ARESopt{}, the read and write operations have similar behaviour with that of \ares{}, even though it unites the $\act{read-config}$ operations with the $\act{get-tag}$, $\act{get-data}$ and $\act{put-data}$ operations into one communication round. One main difference introduced by this change in execution is that the $\act{get-tag}$ and $\act{get-data}$ are executed until an empty configuration is found. Additionally, the algorithm transfers only the $Cs$ with next configurations in DAP operations without the tag/data (similar to how the $\act{read-config}$ does) when the status of the next configuration is finalized. In contrast, in \ares{}, after completing the $\act{read-config}$ and reaching $\bot$, it then proceeds with the $\act{get-tag}$/$\act{get-data}$. The main challenge to the proof is to show that \ARESopt{} satisfies the 
linearizability despite, the changes in DAPs (i.e., piggy-back), the new addition of a garbage collection and the optimization in the reconfiguration operation.

    By \ARESopt{}, before a read/write/reconfig operation completes 
it propagates the maximum tag, or the new tag in the case of a write it discovered, by executing the $\act{put-data}$ action in the last 
configuration of its local configuration sequence (Lines Alg.\ref{algo:reconfigurer}:\ref{line:addconfig:put}, Alg.\ref{algo:read_writeProtocol}:\ref{line:writer:prop} \& \ref{line:reader:prop}). 
When a subsequent operation is invoked, it reads the latest configuration 
sequence by beginning from the last finalized configuration in its local sequence and invoking $\act{get-tag}$/$\act{get-data}$ to
all the configurations until the end of that sequence. 

Lemma~\ref{lemma:ARES:PB&GC} shows that a read operation retrieves a tag-value pair, where 
the value is from $\valSet$.
In addition the reconfiguration properties help us show that the consistency of operations is preserved. Finally, the batching mechanism cannot be distinguished from multiple 
    reconfigurations on different objects, hence the correctness of read/write operations is not affected.
\end{proof}



\section{Optimization Results}
\label{sec:Optimized Results}
In this section, we analyze the performance improvements yielded from optimizations in Section~\ref{sec:optimizations}.

\subsection{Piggyback}
In this experiment we use the same scenario as 
in Section~\ref{sec:TracingBottlenecks}.
To demonstrate the effectiveness of the $piggyback$ optimization from Section~\ref{ssec:optimization:piggy-back}, we conducted a comparison between the original algorithms and their optimized counterparts.


\newcolumntype{C}[1]{>{\centering\arraybackslash}p{#1}}
\begin{table}[h]
      \begin{adjustwidth}{-.6cm}{}
    \centering
    \tiny
    \begin{tabular}{|l|l:C{1cm}|l:C{0.75cm}||l:C{1cm}|l:C{1cm}||l:C{1.25cm}|l:C{1cm}|}
    \hline
        alg./$f_{size}$ & \ARESabd{} & \ARESabd{} $PB$ & \ARESec{} & \ARESec{} $PB$ & \coARESabd{} & \coARESabd{} $PB$ & \coARESec{} & \coARESec{} $PB$ & \fARESabd{} & \fARESabd{} $PB$ & \fARESec{} & \fARESec{} $PB$ \\ \hline
        1MB & 343ms & 334ms & 257ms & 251ms & 308ms & 302ms & 136ms & 127ms & 284ms & 278ms & 149ms & 142ms \\ \hline
        256MB & 72s & 72s & 54s & 53s & 71s & 63s & 25s & 27s  & \textcolor{red}{9s} & \textcolor{red}{5s} (44\%) & \textcolor{red}{9.65s} & \textcolor{red}{3.82s} (60\%) \\ \hline
        512MB & 264s & 192s & 109s & 106s & 176s & 164s & 55.6s & 52.4s & \textcolor{red}{21.8s} & \textcolor{red}{15.2s} (30\%) & \textcolor{red}{23.2s} & \textcolor{red}{10.9s} (53\%) \\ \hline

    \end{tabular}
    \end{adjustwidth}
    \caption{READ Operation - File Size - S:11, W:5, R:5:}
    \label{table:filesize:READ - Algorithms with and without Optimizations - USER}
\vspace{-0.5cm}
\end{table}

In Table \ref{table:filesize:READ - Algorithms with and without Optimizations - USER}, a comprehensive comparison between the original algorithms and their optimized counterparts with the $piggyback$ optimization is presented for READ operations with three different object sizes (\SI{1}{\mega\byte}, \SI{256}{\mega\byte}, and \SI{512}{\mega\byte}). 
In non-fragmented algorithms, no improvements are observed. Handling only one relatively medium-sized object, the removal of $\act{read-config}$ which happened only one time does not substantially impact the latencies. 
In contrast, \fcoARES{} exhibits higher drops in the case of \SI{256}{\mega\byte} and \SI{512}{\mega\byte}. 
In this experiment, \fcoARES{} uses 4 communication rounds, while its $PB$-optimized counterpart completes it in $2$ rounds. Due to the DAP optimization (outlined in Section~\ref{sec:ARESnCo}) for $\act{get-data}$ and $\act{put-data}$, often no data transfer occurs (e.g., in \fARESec{}, fast reads average $13.8$ ms, while slow reads take $74.1$ ms. With $PB$, fast reads are $5.84$ s, and slow reads are $67.1$ ms.). Also, as there is only one configuration, $\act{read-config}$ transfers an empty $nextC$. Thus, reducing the rounds to $2$ with $PB$ optimization drastically reduces the latency of fast reads. This demonstrates that when combined with DAP optimization, $PB$ significantly reduces latencies in fragmentation (for the red-highlighted times in Table~\ref{table:filesize:READ - Algorithms with and without Optimizations - USER}, see the improvement percentage).  

\subsection{Garbage Collection}
The scenario below is made to measure the performance of algorithms when we apply the garbage collection optimization.
So, we evaluate the algorithms \ares{}, \coARES{}, and \fcoARES{}, along with their counterparts with $PB$ and $PB$ with $GC$.
We used two different sizes of the object, \SI{1}{\mega\byte} and \SI{64}{\mega\byte}.
The maximum, minimum, and average block sizes (parameters for \emph{rabin fingerprints}) are set to \SI{1}{\mega\byte}, \SI{512}{\kilo\byte}, and \SI{512}{\kilo\byte} respectively. 
We set $|\wSet|=1, |\rdSet|=10, |\recSet|=4, |\srvSet|=11$. For \ecbased{} algorithms we used parity $m=5$ yielding quorum sizes of $9$ and for \abdbased{} algorithms we used quorums of size $6$.
Initially, a writer invokes a write operation,
and subsequently, the $4$ reconfigurers change the configuration using round-robin fashion three times each, with the set of servers remaining the same in all the configurations. The reconfigurers only modify the DAP switching between the \abd{} and the \ec{}, starting from the \ec{}.
Finally, each reader reads the object in serial order.

\begin{table}[h]
\vspace{-0.2cm}
    \centering
    \tiny
    \begin{tabular}{|l|l:l:l|l:l:l|l:l:l|}
    \hline
        alg./$f_{size}$ & \ares{} & \ares{} $PB$  & \ares{} $PB\&GC$ & \coARES{} & \coARES{} $PB$ & \coARES{} $PB\&GC$ & \fcoARES{} & \fcoARES{} $PB$ & \fcoARES{} $PB\&GC$\\ \hline
        \hline
        \multicolumn{10}{|c|}{$11$ Pending Reconfiguration $\&$ $1$ Finalized} \\ \hline     
        1MB & 159ms & 494ms & 107ms & 162ms  & 506ms & 110ms & 181ms & 191ms & 127ms \\ \hline
        64MB & 5.57s & 27.4s & 5.58s & 5.81s & 26.8s & 5.73s & 6.78s & 6.62s & 6.61s \\ \hline
        \multicolumn{10}{|c|}{$12$ Finalized Reconfiguration} \\ \hline  
        1MB & 159ms & 166ms & 119ms & 163ms & 167ms & 122ms & 186ms & 193ms & 135ms \\ \hline
        64MB & 5.80s & 5.76s & 5.71s & 5.88s & 5.98s & 5.82s & 6.92s & 6.73s & 6.74s \\ \hline
    \end{tabular}
    \caption{READ Operation - Rreconfigurations - S:11, W:1, R:10:, G:4}
    \label{table:recon:READ - Algorithms without Optimizations, with PB, and with PB&GC - USER}
\vspace{-0.6cm}
\end{table}

\begin{figure}[t]
    \begin{minipage}{0.45\textwidth}
    \centering
    \includegraphics[width=1\linewidth]{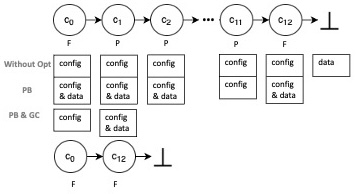}
    \caption{$11$ Pending Reconfigurations $\&$ $1$ Finalized}
    \label{fig:11 Pending Reonfigurations 1 Finalized}
    \end{minipage}
    \begin{minipage}{0.52\textwidth}
    \centering
    \includegraphics[width=1\linewidth]{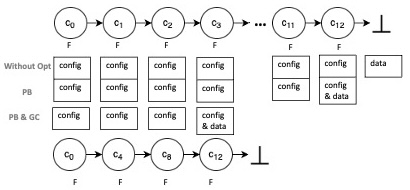}
    \caption{$12$ Finalized Reonfigurations}
    \label{fig:12 Finalized Reonfiguration}
    \end{minipage}
\end{figure}

The Table~\ref{table:recon:READ - Algorithms without Optimizations, with PB, and with PB&GC - USER} is divided into two scenarios. 
In the first scenario, the first $11$ reconfigurations remain pending, and the last reconfiguration is completed as finalized.
As we can see, in this first scenario the version with worst read latency between the three (without optimizations, with $PB$, and with $PB$ and $GC$), is the algorithms that have $PB$. This happens since as we can see in Fig.~\ref{fig:11 Pending Reonfigurations 1 Finalized} the $PB$ version of algorithms transfers the data along with the next configuration ($nextC$) in all round trips. This issue is then solved by the $GC$ optimization of the last reconfiguration, which changes the pointers of all the configurations before the proposed one to point to the finalized configuration. In this way, when the reader
starts its read (from $c_0$) after the last finalized reconfiguration, it finds $c_{12}$ as the next configuration. The read operation in algorithms without optimizations involves performing $12$ $\act{read-config}$ operations followed by $1$ $\act{get-data}$ operation to fetch the data.
This is why the difference between the two algorithms is evident in the smaller object size (\SI{1}{\mega\byte}). However, in the fragmented algorithms, the read operation finds the last finalized configuration during the first block, and subsequent blocks start from that configuration. Thus, the difference between the versions of the fragmented algorithm will only become apparent if a reconfiguration occurs between every block read operation.
Additionally, when (\SI{64}{\mega\byte}), the $\fcoARES{}$ with $PB$ does not outperform the original algorithm as it did in the other cases. This is because it has a larger number of smaller blocks, {which has to migrate to the new configuration.} 

In the second scenario, all $12$ reconfigurations are completed and finalized (Fig.~\ref{fig:12 Finalized Reonfiguration}). The non-fragmented original algorithms and the ones with $PB$ have negligible differences between them, as the only distinction lies in the one extra round trip required by the former to fetch the last finalized configuration, while the latter fetches the data with the last finalized configuration. However, the algorithms with $GC$ reduce the number of communication rounds required to traverse the configuration sequence, as they skip every $4$ configurations. Once again, the benefits of fragmented algorithms with $GC$ are apparent only in the first block.



\section{Conclusions}
In this work, we emphasize the significance of identifying and addressing performance bottlenecks within DSM, specifically with \ares{}, and introduce the use of Distributed Tracing as a methodology for pinpointing these bottlenecks. By injecting checkpoints and monitoring the performance of individual procedures, tracing enables the detection of system inefficiencies. We then turned the identified inefficiencies into optimizations, without jeopardizing correctness,  yielding \ARESopt. 

For future work, it would be interesting to devise strategies on {\em when} new configurations are introduced, such that performance is optimized while prolonging liveness. 
\newpage
\remove{
Our primary contribution includes proposing solutions to eliminate or reduce performance bottlenecks while maintaining correctness. 
To achieve this, we introduce \ARESopt{}, which uses a different approach to manage current configurations, employing a piggy-back technique. This algorithm focuses on initiating a reconfiguration operations for the fragmented object, reducing latency. We also introduce a garbage collection ($GC$) mechanism to enhance storage efficiency and longevity by removing older configurations in the background. 

Additionally, this tracing helps us identify which algorithm to use depending on the specific use case and performance requirements. \ARESec{} may be preferable for small files, while \fARESec{} shines for larger files and scenarios with scalability needs. The block size and \ec{} parameter $k$ should be selected based on the trade-offs between redundancy, performance, and durability. Additionally, consider the impact of reconfigurations when selecting an algorithm for scenarios involving frequent configuration changes.}

\bibliographystyle{ACM-Reference-Format}
\bibliography{references}


\begin{thebibliography}{36}


\ifx \showCODEN    \undefined \def \showCODEN     #1{\unskip}     \fi
\ifx \showDOI      \undefined \def \showDOI       #1{#1}\fi
\ifx \showISBNx    \undefined \def \showISBNx     #1{\unskip}     \fi
\ifx \showISBNxiii \undefined \def \showISBNxiii  #1{\unskip}     \fi
\ifx \showISSN     \undefined \def \showISSN      #1{\unskip}     \fi
\ifx \showLCCN     \undefined \def \showLCCN      #1{\unskip}     \fi
\ifx \shownote     \undefined \def \shownote      #1{#1}          \fi
\ifx \showarticletitle \undefined \def \showarticletitle #1{#1}   \fi
\ifx \showURL      \undefined \def \showURL       {\relax}        \fi
\providecommand\bibfield[2]{#2}
\providecommand\bibinfo[2]{#2}
\providecommand\natexlab[1]{#1}
\providecommand\showeprint[2][]{arXiv:#2}

\bibitem[ref(a)]%
        {ref_url_grafana}
 \bibinfo{year}{{}}\natexlab{a}.
\newblock \bibinfo{title}{{Grafana}}.
\newblock \bibinfo{howpublished}{\url{https://grafana.com}}.
\newblock
\newblock
\shownote{Accessed: [14/02/2024]}.


\bibitem[ref(b)]%
        {ref_url_opentelemetry}
 \bibinfo{year}{{}}\natexlab{b}.
\newblock \bibinfo{title}{{Opentelemetry}}.
\newblock \bibinfo{howpublished}{\url{https://opentelemetry.io}}.
\newblock
\newblock
\shownote{Accessed: [14/02/2024]}.


\bibitem[ope()]%
        {opentelemetry-python}
 \bibinfo{year}{{}}\natexlab{}.
\newblock \bibinfo{title}{OpenTelemetry-Python: A vendor-agnostic framework for distributed tracing, metrics, and logging}.
\newblock \bibinfo{howpublished}{\url{https://github.com/open-telemetry/opentelemetry-python}}.
\newblock


\bibitem[ref(c)]%
        {ref_url_zmq}
 \bibinfo{year}{{}}\natexlab{c}.
\newblock \bibinfo{title}{{ZeroMQ}}.
\newblock \bibinfo{howpublished}{\url{https://zeromq.org}}.
\newblock
\newblock
\shownote{Accessed: [14/02/2024]}.


\bibitem[ref(d)]%
        {ref_url_zipkin}
 \bibinfo{year}{{}}\natexlab{d}.
\newblock \bibinfo{title}{{Zipkin}}.
\newblock \bibinfo{howpublished}{\url{https://zipkin.io}}.
\newblock
\newblock
\shownote{Accessed: [14/02/2024]}.


\bibitem[Aguilera et~al\mbox{.}(2009)]%
        {dynastore}
\bibfield{author}{\bibinfo{person}{M.K. Aguilera}, \bibinfo{person}{I. Keidar}, \bibinfo{person}{D. Malkhi}, {and} \bibinfo{person}{A. Shraer}.} \bibinfo{year}{2009}\natexlab{}.
\newblock \showarticletitle{Dynamic atomic storage without consensus}. In \bibinfo{booktitle}{\emph{Proceedings of the 28th ACM symposium on Principles of distributed computing (PODC '09)}}. \bibinfo{publisher}{ACM}, \bibinfo{address}{New York, NY, USA}, \bibinfo{pages}{17--25}.
\newblock
\showISBNx{978-1-60558-396-9}


\bibitem[Aniszczyk(2012)]%
        {aniszczyk2012zipkin}
\bibfield{author}{\bibinfo{person}{C. Aniszczyk}.} \bibinfo{year}{2012}\natexlab{}.
\newblock \bibinfo{title}{Distributed Systems Tracing with Zipkin}.
\newblock
\newblock
\urldef\tempurl%
\url{https://blog.twitter.com/2012/distributed-systems-tracing-with-zipkin}
\showURL{%
\tempurl}
\newblock
\shownote{Google Scholar}.


\bibitem[Anta et~al\mbox{.}(2021)]%
        {SIROCCO_2021}
\bibfield{author}{\bibinfo{person}{A.F. Anta}, \bibinfo{person}{C. Georgiou}, \bibinfo{person}{T. Hadjistasi}, \bibinfo{person}{E. Stavrakis}, {and} \bibinfo{person}{A. Trigeorgi}.} \bibinfo{year}{2021}\natexlab{}.
\newblock \showarticletitle{{Fragmented Object : Boosting Concurrency of Shared Large Objects}}.
\newblock \bibinfo{journal}{\emph{In Proc.of SIROCCO}} (\bibinfo{year}{2021}), \bibinfo{pages}{1--18}.
\newblock


\bibitem[Attiya et~al\mbox{.}(1995)]%
        {ref_article_ABD}
\bibfield{author}{\bibinfo{person}{H. Attiya}, \bibinfo{person}{A. Bar-Noy}, {and} \bibinfo{person}{D. Dolev}.} \bibinfo{year}{1995}\natexlab{}.
\newblock \showarticletitle{{Sharing Memory Robustly in Message-Passing Systems}}.
\newblock \bibinfo{journal}{\emph{Journal of the ACM (JACM)}} \bibinfo{volume}{42}, \bibinfo{number}{1} (\bibinfo{year}{1995}), \bibinfo{pages}{124--142}.
\newblock
\showISSN{1557735X}


\bibitem[Berger et~al\mbox{.}(2018)]%
        {berger2018integrated}
\bibfield{author}{\bibinfo{person}{A. Berger}, \bibinfo{person}{I. Keidar}, {and} \bibinfo{person}{A. Spiegelman}.} \bibinfo{year}{2018}\natexlab{}.
\newblock \showarticletitle{Integrated Bounds for Disintegrated Storage}. In \bibinfo{booktitle}{\emph{Proceedings of the 32nd International Symposium on Distributed Computing (DISC)}}. Schloss Dagstuhl – Leibniz-Zentrum für Informatik, \bibinfo{pages}{11:1--11:18}.
\newblock
\urldef\tempurl%
\url{https://doi.org/10.4230/LIPIcs.DISC.2018.11}
\showDOI{\tempurl}


\bibitem[Carpen-amarie(2012)]%
        {ref_article_blobseer}
\bibfield{author}{\bibinfo{person}{A. Carpen-amarie}.} \bibinfo{year}{2012}\natexlab{}.
\newblock \showarticletitle{{BlobSeer as a Data-Storage Facility for Clouds: Self-Adaptation, Integration, Evaluation, PhD Thesis, France}}.
\newblock  (\bibinfo{year}{2012}).
\newblock


\bibitem[{Datadog}()]%
        {datadog}
\bibfield{author}{\bibinfo{person}{{Datadog}}.} \bibinfo{year}{{}}\natexlab{}.
\newblock \bibinfo{title}{{Datadog Overview}}.
\newblock
\newblock
\urldef\tempurl%
\url{https://www.datadog.com/overview}
\showURL{%
\tempurl}
\newblock
\shownote{Accessed: [14/02/2024]}.


\bibitem[Dutta et~al\mbox{.}(2004)]%
        {ref_article_fastRead}
\bibfield{author}{\bibinfo{person}{P. Dutta}, \bibinfo{person}{R. Guerraoui}, \bibinfo{person}{R.R. Levy}, {and} \bibinfo{person}{A. Chakraborty}.} \bibinfo{year}{2004}\natexlab{}.
\newblock \showarticletitle{{How Fast can a Distributed Atomic Read be?}}
\newblock \bibinfo{journal}{\emph{In Prof. of PODC}} (\bibinfo{year}{2004}), \bibinfo{pages}{236--245}.
\newblock


\bibitem[Gafni and Malkhi(2015)]%
        {SpSnStore}
\bibfield{author}{\bibinfo{person}{E. Gafni} {and} \bibinfo{person}{D. Malkhi}.} \bibinfo{year}{2015}\natexlab{}.
\newblock \showarticletitle{{Elastic configuration maintenance via a parsimonious speculating snapshot solution}}.
\newblock \bibinfo{journal}{\emph{Lecture Notes in Computer Science (including subseries Lecture Notes in Artificial Intelligence and Lecture Notes in Bioinformatics)}}  \bibinfo{volume}{9363} (\bibinfo{year}{2015}), \bibinfo{pages}{140--153}.
\newblock
\showISBNx{9783662486528}
\showISSN{16113349}
\urldef\tempurl%
\url{https://doi.org/10.1007/978-3-662-48653-5\_10}
\showDOI{\tempurl}


\bibitem[Georgiou et~al\mbox{.}(2019)]%
        {Self-stabilizationOverhead9}
\bibfield{author}{\bibinfo{person}{C. Georgiou}, \bibinfo{person}{R. Gustafsson}, \bibinfo{person}{A. Lindh{\'e}}, {and} \bibinfo{person}{E.~M. Schiller}.} \bibinfo{year}{2019}\natexlab{}.
\newblock \showarticletitle{Self-stabilization Overhead: A Case Study on Coded Atomic Storage}. In \bibinfo{booktitle}{\emph{Networked Systems}}, \bibfield{editor}{\bibinfo{person}{Mohamed~Faouzi Atig} {and} \bibinfo{person}{Alexander~A. Schwarzmann}} (Eds.). \bibinfo{publisher}{Springer International Publishing}, \bibinfo{address}{Cham}, \bibinfo{pages}{131--147}.
\newblock


\bibitem[Georgiou et~al\mbox{.}(2022a)]%
        {ref_article_ERATO}
\bibfield{author}{\bibinfo{person}{C. Georgiou}, \bibinfo{person}{T. Hadjistasi}, \bibinfo{person}{N. Nicolaou}, {and} \bibinfo{person}{A.~A. Schwarzmann}.} \bibinfo{year}{2022}\natexlab{a}.
\newblock \showarticletitle{{Implementing Three Exchange Read Operations for Distributed Atomic Storage}}.
\newblock \bibinfo{journal}{\emph{J. Parallel Distributed Comput.}}  \bibinfo{volume}{163} (\bibinfo{year}{2022}), \bibinfo{pages}{97--113}.
\newblock
\urldef\tempurl%
\url{https://doi.org/10.1016/j.jpdc.2022.01.024}
\showDOI{\tempurl}


\bibitem[Georgiou et~al\mbox{.}(2005)]%
        {do-rambo}
\bibfield{author}{\bibinfo{person}{C. Georgiou}, \bibinfo{person}{P.~M. Musiał}, {and} \bibinfo{person}{A.~A. Shvartsman}.} \bibinfo{year}{2005}\natexlab{}.
\newblock \showarticletitle{{Developing a consistent domain-oriented distributed object service}}.
\newblock \bibinfo{journal}{\emph{Proceedings - Fourth IEEE International Symposium on Network Computing and Applications, NCA 2005}}  \bibinfo{volume}{2005}, \bibinfo{pages}{149--158}.
\newblock
\urldef\tempurl%
\url{https://doi.org/10.1109/NCA.2005.16}
\showDOI{\tempurl}


\bibitem[Georgiou et~al\mbox{.}(2009)]%
        {ref_article_semifast}
\bibfield{author}{\bibinfo{person}{C. Georgiou}, \bibinfo{person}{N. Nicolaou}, {and} \bibinfo{person}{A.A. Shvartsman}.} \bibinfo{year}{2009}\natexlab{}.
\newblock \showarticletitle{{Fault-tolerant Semifast Implementations of Atomic Read/Write Registers}}.
\newblock \bibinfo{journal}{\emph{J. Parallel and Distrib. Comput.}} \bibinfo{volume}{69}, \bibinfo{number}{1} (\bibinfo{year}{2009}), \bibinfo{pages}{62--79}.
\newblock
\showISSN{07437315}


\bibitem[Georgiou et~al\mbox{.}(2022b)]%
        {FragARES}
\bibfield{author}{\bibinfo{person}{C. Georgiou}, \bibinfo{person}{N. Nicolaou}, {and} \bibinfo{person}{A. Trigeorgi}.} \bibinfo{year}{2022}\natexlab{b}.
\newblock \showarticletitle{{Fragmented {ARES}: Dynamic Storage for Large Objects}}. In \bibinfo{booktitle}{\emph{{Proceedings of the 36th International Symposium on Distributed Computing (DISC)}}}. \bibinfo{pages}{25:1--25:24}.
\newblock
\newblock
\shownote{Also at arXiv:2201.13292.}.


\bibitem[Gilbert et~al\mbox{.}(2010)]%
        {rambo}
\bibfield{author}{\bibinfo{person}{S. Gilbert}, \bibinfo{person}{N.~A. Lynch}, {and} \bibinfo{person}{A.~A. Shvartsman}.} \bibinfo{year}{2010}\natexlab{}.
\newblock \showarticletitle{{{RAMBO}: A Robust, Reconfigurable Atomic Memory Service for Dynamic Networks}}.
\newblock \bibinfo{journal}{\emph{Distributed Comput.}} \bibinfo{volume}{23}, \bibinfo{number}{4} (\bibinfo{year}{2010}), \bibinfo{pages}{225--272}.
\newblock
\urldef\tempurl%
\url{https://doi.org/10.1007/s00446-010-0117-1}
\showDOI{\tempurl}


\bibitem[{Google Cloud Trace}()]%
        {googlecloudtrace}
\bibfield{author}{\bibinfo{person}{{Google Cloud Trace}}.} \bibinfo{year}{{}}\natexlab{}.
\newblock \bibinfo{title}{{Google Cloud Trace Overview}}.
\newblock
\newblock
\urldef\tempurl%
\url{https://cloud.google.com/trace}
\showURL{%
\tempurl}
\newblock
\shownote{Accessed: [14/02/2024]}.


\bibitem[Hadjistasi et~al\mbox{.}(2017)]%
        {HNS17}
\bibfield{author}{\bibinfo{person}{T. Hadjistasi}, \bibinfo{person}{N. Nicolaou}, {and} \bibinfo{person}{A.~A. Schwarzmann}.} \bibinfo{year}{2017}\natexlab{}.
\newblock \showarticletitle{Oh-RAM! One and a Half Round Atomic Memory}. In \bibinfo{booktitle}{\emph{Proc. of NETYS 2017}} \emph{(\bibinfo{series}{Lecture Notes in Computer Science}, Vol.~\bibinfo{volume}{10299})}. \bibinfo{pages}{117--132}.
\newblock
\urldef\tempurl%
\url{https://doi.org/10.1007/978-3-319-59647-1\_10}
\showDOI{\tempurl}


\bibitem[Herlihy and Wing(1990)]%
        {HW90}
\bibfield{author}{\bibinfo{person}{M.~P. Herlihy} {and} \bibinfo{person}{J.~M. Wing}.} \bibinfo{year}{1990}\natexlab{}.
\newblock \showarticletitle{{Linearizability: a Correctness Condition for Concurrent Objects}}.
\newblock \bibinfo{journal}{\emph{ACM TOPLAS}} \bibinfo{volume}{12}, \bibinfo{number}{3} (\bibinfo{year}{1990}), \bibinfo{pages}{463--492}.
\newblock
\showISSN{0164-0925}


\bibitem[Huffman and Pless(2003)]%
        {verapless_book}
\bibfield{author}{\bibinfo{person}{W.~C. Huffman} {and} \bibinfo{person}{V. Pless}.} \bibinfo{year}{2003}\natexlab{}.
\newblock \bibinfo{booktitle}{\emph{Fundamentals of error-correcting codes}}.
\newblock \bibinfo{publisher}{Cambridge university press}.
\newblock


\bibitem[Jehl and Meling(2017)]%
        {jehl_et_al:LIPIcs:2017:7100}
\bibfield{author}{\bibinfo{person}{L. Jehl} {and} \bibinfo{person}{H. Meling}.} \bibinfo{year}{2017}\natexlab{}.
\newblock \showarticletitle{{The Case for Reconfiguration without Consensus: Comparing Algorithms for Atomic Storage}}. In \bibinfo{booktitle}{\emph{20th International Conference on Principles of Distributed Systems (OPODIS 2016)}} \emph{(\bibinfo{series}{Leibniz International Proceedings in Informatics (LIPIcs)}, Vol.~\bibinfo{volume}{70})}, \bibfield{editor}{\bibinfo{person}{Panagiota Fatourou}, \bibinfo{person}{Ernesto Jim{\'e}nez}, {and} \bibinfo{person}{Fernando Pedone}} (Eds.). \bibinfo{publisher}{Schloss Dagstuhl--Leibniz-Zentrum fuer Informatik}, \bibinfo{address}{Dagstuhl, Germany}, \bibinfo{pages}{31:1--31:17}.
\newblock
\showISBNx{978-3-95977-031-6}
\showISSN{1868-8969}
\urldef\tempurl%
\url{https://doi.org/10.4230/LIPIcs.OPODIS.2016.31}
\showDOI{\tempurl}


\bibitem[Jehl et~al\mbox{.}(2015)]%
        {smartmerge}
\bibfield{author}{\bibinfo{person}{L. Jehl}, \bibinfo{person}{R. Vitenberg}, {and} \bibinfo{person}{H. Meling}.} \bibinfo{year}{2015}\natexlab{}.
\newblock \showarticletitle{Smartmerge: A new approach to reconfiguration for atomic storage}. In \bibinfo{booktitle}{\emph{International Symposium on Distributed Computing}}. Springer, \bibinfo{pages}{154--169}.
\newblock


\bibitem[Lynch and Shvartsman(1997)]%
        {ref_article_MWMRABD}
\bibfield{author}{\bibinfo{person}{N.A. Lynch} {and} \bibinfo{person}{A.A. Shvartsman}.} \bibinfo{year}{1997}\natexlab{}.
\newblock \showarticletitle{{Robust Emulation of Shared Memory Using Dynamic Quorum-Acknowledged Broadcasts}}.
\newblock \bibinfo{journal}{\emph{In Proc. of FTCS}} (\bibinfo{year}{1997}), \bibinfo{pages}{272--281}.
\newblock
\showISBNx{0818678313}


\bibitem[Nicolaou et~al\mbox{.}(2022)]%
        {ARES}
\bibfield{author}{\bibinfo{person}{N. Nicolaou}, \bibinfo{person}{V. Cadambe}, \bibinfo{person}{N. Prakash}, \bibinfo{person}{A. Trigeorgi}, \bibinfo{person}{K.~M. Konwar}, \bibinfo{person}{M. Medard}, {and} \bibinfo{person}{N. Lynch}.} \bibinfo{year}{2022}\natexlab{}.
\newblock \showarticletitle{{ARES: Adaptive, Reconfigurable, Erasure coded, Atomic Storage}}.
\newblock \bibinfo{journal}{\emph{ACM Transactions on Storage (TOS)}} (\bibinfo{year}{2022}).
\newblock
\newblock
\shownote{Accepted. Also in \url{https://arxiv.org/abs/1805.03727}}.


\bibitem[Nicolaou et~al\mbox{.}(2016)]%
        {NFG16}
\bibfield{author}{\bibinfo{person}{N. Nicolaou}, \bibinfo{person}{A. {Fern{\'a}ndez Anta}}, {and} \bibinfo{person}{C. Georgiou}.} \bibinfo{year}{2016}\natexlab{}.
\newblock \showarticletitle{{Coverability: Consistent Versioning in Asynchronous, Fail-Prone, Message-Passing Environments}}. In \bibinfo{booktitle}{\emph{{Proc. of IEEE NCA 2016}}}. IEEE.
\newblock


\bibitem[Sigelman et~al\mbox{.}(2010)]%
        {sigelman2010dapper}
\bibfield{author}{\bibinfo{person}{B. Sigelman}, \bibinfo{person}{Luiz~A. Barroso}, \bibinfo{person}{M. Burrows}, \bibinfo{person}{P. Stephenson}, \bibinfo{person}{M. Plakal}, \bibinfo{person}{D. Beaver}, \bibinfo{person}{S. Jaspan}, {and} \bibinfo{person}{C. Shanbhag}.} \bibinfo{year}{2010}\natexlab{}.
\newblock \showarticletitle{Dapper, a Large-Scale Distributed Systems Tracing Infrastructure}. In \bibinfo{booktitle}{\emph{Proceedings of the 10th USENIX Symposium on Operating Systems Design and Implementation (OSDI)}}. Google Inc.
\newblock


\bibitem[Sigelman and contributors(2016)]%
        {opentracing}
\bibfield{author}{\bibinfo{person}{B. Sigelman} {and} \bibinfo{person}{contributors}.} \bibinfo{year}{2016}\natexlab{}.
\newblock \bibinfo{title}{{OpenTracing}}.
\newblock
\newblock
\urldef\tempurl%
\url{https://opentracing.io/}
\showURL{%
\tempurl}
\newblock
\shownote{Accessed: [14/02/2024]}.


\bibitem[Spiegelman et~al\mbox{.}(2016)]%
        {spiegelman2016space}
\bibfield{author}{\bibinfo{person}{A. Spiegelman}, \bibinfo{person}{Y. Cassuto}, \bibinfo{person}{I. Keidar}, {and} \bibinfo{person}{G. Chockler}.} \bibinfo{year}{2016}\natexlab{}.
\newblock \showarticletitle{Space Bounds for Reliable Storage: Fundamental Limits of Coding}. In \bibinfo{booktitle}{\emph{Proceedings of the ACM Symposium on Principles of Distributed Computing (PODC)}}. ACM.
\newblock
\urldef\tempurl%
\url{https://doi.org/10.1145/2933057.2933104}
\showDOI{\tempurl}


\bibitem[Spiegelman et~al\mbox{.}(2017)]%
        {spiegelman2017dynamic}
\bibfield{author}{\bibinfo{person}{A. Spiegelman}, \bibinfo{person}{I. Keidar}, {and} \bibinfo{person}{D. Malkhi}.} \bibinfo{year}{2017}\natexlab{}.
\newblock \showarticletitle{Dynamic Reconfiguration: Abstraction and Optimal Asynchronous Solution}. In \bibinfo{booktitle}{\emph{Proceedings of the 31st International Symposium on Distributed Computing (DISC)}}, \bibfield{editor}{\bibinfo{person}{Andréa~W. Richa}} (Ed.). \bibinfo{publisher}{Leibniz International Proceedings in Informatics, Dagstuhl Publishing}, \bibinfo{address}{Germany}, \bibinfo{pages}{40:1--40:15}.
\newblock
\urldef\tempurl%
\url{https://doi.org/10.4230/LIPIcs.DISC.2017.40}
\showDOI{\tempurl}
\newblock
\shownote{A full version of the paper is available at \url{https://alexanderspiegelman.github.io/alexanderspiegelman.github.io/DynamicTasks.pdf}.}.


\bibitem[Trigeorgi et~al\mbox{.}(2022)]%
        {ref_article_ARESvsCommercial}
\bibfield{author}{\bibinfo{person}{A. Trigeorgi}, \bibinfo{person}{N. Nicolaou}, \bibinfo{person}{C. Georgiou}, \bibinfo{person}{T. Hadjistasi}, \bibinfo{person}{E. Stavrakis}, \bibinfo{person}{V. Cadambe}, {and} \bibinfo{person}{B. Urgaonkar}.} \bibinfo{year}{2022}\natexlab{}.
\newblock \showarticletitle{{Invited Paper: Towards Practical Atomic Distributed Shared Memory: An Experimental Evaluation}}. In \bibinfo{booktitle}{\emph{Stabilization, Safety, and Security of Distributed Systems: 24th International Symposium, SSS 2022, Clermont-Ferrand, France, November 15–17, 2022, Proceedings}} (Clermont-Ferrand, France). \bibinfo{publisher}{Springer-Verlag}, \bibinfo{address}{Berlin, Heidelberg}, \bibinfo{pages}{35–50}.
\newblock
\showISBNx{978-3-031-21016-7}
\urldef\tempurl%
\url{https://doi.org/10.1007/978-3-031-21017-4_3}
\showDOI{\tempurl}


\bibitem[{Uber Technologies, Inc.}(2023)]%
        {uberjaeger}
\bibfield{author}{\bibinfo{person}{{Uber Technologies, Inc.}}} \bibinfo{year}{2023}\natexlab{}.
\newblock \bibinfo{title}{{Jaeger Distributed Tracing}}.
\newblock
\newblock
\urldef\tempurl%
\url{https://www.jaegertracing.io/}
\showURL{%
\tempurl}
\newblock
\shownote{Accessed: [14/02/2024]}.


\bibitem[Vukolic(2012)]%
        {QuorumSystems}
\bibfield{author}{\bibinfo{person}{M. Vukolic}.} \bibinfo{year}{2012}\natexlab{}.
\newblock \bibinfo{booktitle}{\emph{Quorum Systems: With Applications to Storage and Consensus}}.
\newblock \bibinfo{publisher}{Morgan {\&} Claypool Publishers}.
\newblock
\urldef\tempurl%
\url{https://doi.org/10.2200/S00402ED1V01Y201202DCT009}
\showDOI{\tempurl}


\end{thebibliography}

\end{document}